\documentstyle[12pt,aaspp4]{article}

\begin{document}
\title{X-ray Observations of Optically-Selected, Radio-Quiet Quasars I:
The {\it ASCA}\ results} 

\author {I.M. George\altaffilmark{1,2}, 
T.J. Turner\altaffilmark{1,3}, 
T. Yaqoob\altaffilmark{1,2},
H. Netzer\altaffilmark{4},
A. Laor\altaffilmark{5},
R.F. Mushotzky\altaffilmark{1},\\ 
K. Nandra\altaffilmark{1,2}, 
T. Takahashi\altaffilmark{6}}

\altaffiltext{1}{Laboratory for High Energy Astrophysics, Code 660,
	NASA/Goddard Space Flight Center,
  	Greenbelt, MD 20771}
\altaffiltext{2}{Universities Space Research Association}
\altaffiltext{3}{University of Maryland, Baltimore County}
\altaffiltext{4}{School of Physics and Astronomy and the Wise Observatory,
        The Beverly and Raymond Sackler Faculty of Exact Sciences,
        Tel Aviv University, Tel Aviv 69978, Israel.}
\altaffiltext{5}{Physics Department, Technion, Haifa 32000, Israel}
\altaffiltext{6}{Institute of Space \& Astronautical Science, 3-1-1 Yoshinodai,
Sagamihara, Kanagawa 229-8510, Japan}

\slugcomment{Resubmitted to {\em The Astrophysical Journal}}

\begin{abstract}
We present the result of 27 {\it ASCA}\ observations of 26 radio-quiet quasars
(RQQs) from the Palomar--Green (PG) survey. The sample is not statistically 
complete, but reasonably representative of RQQs in the PG-survey.             
For many of the sources, the {\it ASCA} data are presented here for the first
time.
All the RQQs were detected except for two objects, both of which contain
broad absorption lines in the optical band.
We find the variability characteristics of the sources to be consistent with 
	Seyfert 1 galaxies. 
A power law offers an acceptable description of the time-averaged
	spectra
	in the 2--10~keV (quasar-frame) band for all but 1 dataset.
	The best-fitting values of photon index vary from object to object
	over the range $1.5 \lesssim \Gamma_{2-10} \lesssim 3$ with 
	a mean $< \Gamma_{2-10} > \simeq 2$ and dispersion 
	$\sigma(\Gamma_{2-10}) \simeq 0.25$. The 
	distribution of $\Gamma_{2-10}$ is therefore similar to 
	that observed 
	in other RQ AGN, and seems unrelated to X-ray luminosity.
No single model adequately describes the full 0.6--10~keV (observed-frame) 
	continuum of all the RQQs.
	Approximately 50\% of the sources can be adequately described by a 
	single power law or a power law with only very subtle deviations.
All but one of the remaining datasets were found to have
	convex spectra (flattening as one moves to higher energies).
	The exception is PG~1411+442, in which a substantial column
	density ($N_{H,z} \sim 2\times10^{23}\ {\rm cm^{-2}}$) obscures
	$\sim$98~\% of the continuum.
We find only 5 (maybe 6) of 14 objects with $z \lesssim 0.25$ to have 
	a 'soft excess' at energies $\lesssim 1$~keV, but find no universal 
	shape for these spectral components.
The spectrum of PG~1244+026 
	contains a rather narrow emission feature
	centered at an energy $\sim 1$~keV (quasar-frame).
The detection rate of absorption due to ionized material 
	in these RQQs is
        lower in that seen in Seyfert 1 galaxies.
	In part, this may be due to selection effects.
        However, when detected, the absorbers in the RQQs 
        exhibit a similar range of column density and ionization 
	parameter as Seyfert 1 galaxies.
We find evidence of Fe $K$-shell emission in at least 8 RQQs.
	These are all low-luminosity objects, and the line parameters are 
	consistent with other low-luminosity RQ AGN. 
	However the construction of the mean data/model ratios for various  
	luminosity ranges
	reveals a trend whereby the profile and strength of the
	Fe $K$-shell emission changes as a
	function of luminosity.
\end{abstract}


\section{INTRODUCTION}
\label{sec:intro}

Quasars typically emit $\sim$50\% of their bolometric luminosity in a spectral
component generally believe to peak in the extreme UV, but encompassing the
optical to soft X-ray bands  (e.g. Sanders et al 1989).
This emission is
believed to be closely related to the accretion flow onto the putative
supermassive black hole powering the copious energy output
($L_{bol} \gtrsim 10^{44}\ {\rm erg\ s^{-1}}$) of quasars
(eg. Zel'dovich \& Novikov 1964).
The remainder of the bolometric luminosity, however, is emitted with almost
constant power per decade of frequency over two other broad ranges 
of frequencies
in the infrared and X-ray band (eg. Carleton et al 1987).
The origin of this emission is less clear, yet it is clearly
related in some way to the presence of the Active Galactic Nucleus (AGN).
The spectral energy distribution (SED) of all quasars in the infrared to hard
X-ray band are somewhat similar.
However  quasars can be divided into two classes
based on the strength of the radio emission relative to that in the optical
band, hence
radio-quiet and radio-loud quasars (hereafter RQQs and RLQs respectively).

The X-ray band (0.1--50~keV) is thought to contain $\sim$10--30\% of the
bolometric luminosity of most quasars. Despite two decades of study we have
only a crude idea of the X-ray spectrum of the average quasar. The
spectrum appears to be dominated by a simple power law continuum at energies
$\gtrsim$few~keV (quasar frame\footnote{Energies are quoted in the 
quasar's frame throughout unless noted}), 
but steepens at energies $\lesssim$1~keV. The
power law is thought to be related to the underlying non-thermal emission 
whilst 
the soft X-ray component to be the high energy tail of the 
(psuedo-thermal ?) UV bump. 
Both components are known to be variable and hence diagnostics of the
ultimate energy release mechanism(s). 
Additional spectral
features such as absorption by material intrinsic to the source, Fe $K$-shell
emission and 'Compton-reflection' components, so useful as diagnostics
in the case of Seyfert galaxies, appear to be rare in quasars
(eg. Reeves et al 1997).
In many respects, our knowledge of the
X-ray characteristics of quasars is reminiscent of our knowledge of the 
lower-luminosity Seyfert 1 galaxies a decade or so ago. 

Most of the studies in the X-ray band conducted to date have been performed
using small or incomplete samples of sources and/or using instruments of low
spectral resolution and limited bandpass.
Nevertheless
many studies have suggested that RLQs have flatter spectra than RQQs
throughout the X-ray band. Modelling the spectra of 33 quasars in the
0.1--3.5~keV (observed frame) band obtained using the {\it Einstein}\ IPC with
a power law (corrected for line-of-sight absorption), Wilkes \& Elvis 
(1987) found
mean values of the photon indices of $<\Gamma> _{RQ} \sim 2.0$  for RQQs and
$<\Gamma> _{RL} \sim 1.5$ for RLQs. Such a trend has been confirmed from other
studies of  a larger number of sources with the IPC 
(e.g. Canizares \& White 1989). 
Similar modelling of the spectra in the
0.1--2.4~keV (observed frame) band obtained using the {\it ROSAT}\ PSPC reveal
steeper still spectral indices, but again with $<\Gamma>_{RQ}$ greater than
$<\Gamma>_{RL}$ (eg. Laor et al 1997 and references therein). Some studies
using {\it EXOSAT}, {\it Ginga} and {\it ASCA}\ have suggested RLQs also have
flatter spectra than RQQs at energies $\gtrsim 2$~keV (e.g. Williams et al
1992; Lawson et al 1992; Reeves et al 1997). However more recent studies have
disputed this, suggesting 
$<\Gamma>_{RQ} \simeq <\Gamma>_{RL} \sim1.7$--1.8 
(Lawson \& Turner 1997; Sambruna, Eracleous \& Mushotzky 1999). It
should be remembered 
that almost all the above studies note a wide dispersion in
spectral indices within both RLQs and RQQs: there appears to be no
generic quasar spectrum in the X-ray band.

Our original motivation of the current project was to observe a well-defined
sample of PG-quasars using a single satellite ({\it ASCA};
Tanaka, Inoue \& Holt 1994) . We planned 
to compare the temporal and spectral characteristics  in the 0.5--10~keV
(observed frame) band to other parameters such as $R_L$, redshift, luminosity
{\it etc} in a sample where at least the obvervational selection effects were
known. 
Unfortunately time was not awarded for the entire project.

Here we present the results of 27 
{\it ASCA}\ observations of 26 RQQs. The sources were selected from 
the PG-quasars
(see \S\ref{Sec:sample}), with our own observations supplemented by
archival data. The resultant 'sample' is by no means statistically
complete. However 
the sample is reasonably representative of the optical and radio 
characteristics of the RQQs in the PG-survey as a whole. 
Thus we may assume the {\it ASCA}\ results provide a 
reasonable representation of the 
X-ray characteristics of the RQQs in the PG-survey.

In this paper we restrict our discussion to the {\it ASCA}\ results,
postponing comparison with previous X-ray observations and 
other multiwaveband characteristics until 
George et al 2000 (hereafter Paper II). 
In \S\ref{Sec:Obs-red} we describe the
observations, data reduction and preliminary analysis. 
In \S\ref{Sec:temporal}
we present a a temporal analysis of the data, and in \S\ref{Sec:spec} our
spectral analysis. We discuss the various results in \S\ref{Sec:disc} and
present a summary and our conclusions in \S\ref{Sec:conclusions}.

\section{OBSERVATIONS, REDUCTION AND PREPARATORY ANALYSIS}
\label{Sec:Obs-red}

\subsection{The sample}
\label{Sec:sample}

The Palomar--Green (PG) survey of stellar--like objects with ultraviolet
excess ($U-B < -0.44$) covers a survey area approximately a quarter of the sky
(Schmidt \& Green 1983; Green, Schmidt, Liebert 1986). The survey contains a
statistically-complete sample of 1715 objects of which 116 are classified as
either quasars or Seyfert galaxies\footnote{There 114 quasars and 
Seyfert galaxies in the complete sample listed in Schmidt \& Green (1983).
Boroson \& Green (1992) pointed out that one of these objects (PG~0119+229)
does not have broad emission lines.
Later, Green et al (1986) added three objects to the complete sample
(PG~1001+291, PG~1203+355 and PG~2349-014). They commented on the former two
sources but not the latter.}.
The subdivision
between quasars and Seyfert galaxies is based solely upon whether the absolute
$B$-magnitude is greater or less than $M_{B} = -23.0$~mag.
In the current work we make no such distinction and, for simplicity, refer
to both types of object as quasars. 
The optical and infrared continua of 
98 of the PG-quasars have been presented in Neugebauer et al (1987).
The properties of the optical emission lines 
of 87 of the 90 PG-quasars with redshifts $z<0.5$
have been presented in Boroson \& Green (1992).
The radio properties of the majority of
the PG-quasars are presented in Kellermann et al (1989, 1994). From a
comparison of the radio and optical flux densities ($f_{\rm 5GHz}$ and $f_{\rm
440nm}$ respectively) it was found that the majority of the sources had $R_L =
f_{\rm 5GHz}/f_{\rm 440nm} \lesssim 1$, but that $\sim 20$\% of the sources
had $R_L \geq 10$. This led to the common classification into 
RQQs and RLQs, with the most
frequently used division at $R_L = 10$.

\begin{figure}
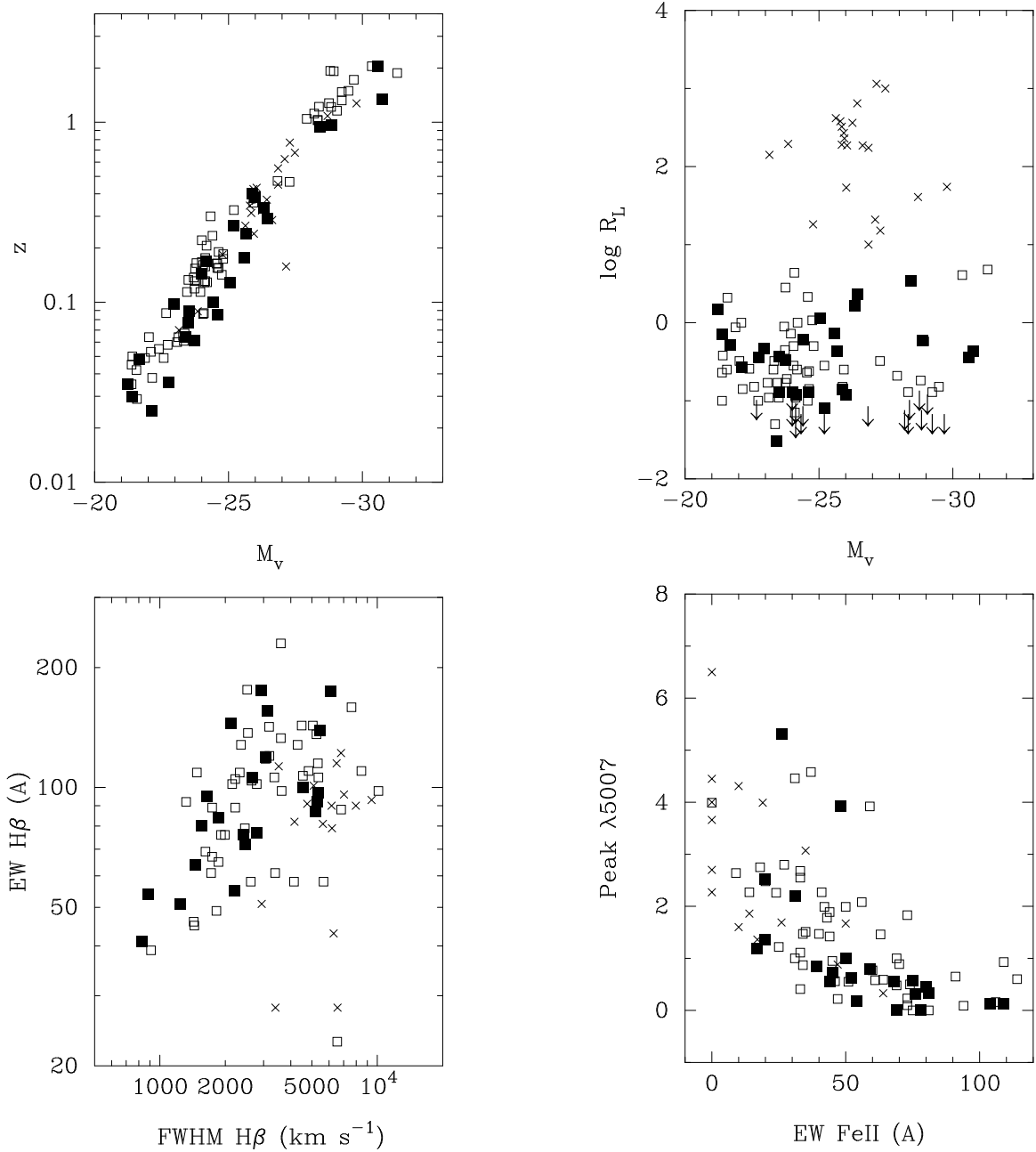

\thispagestyle{empty}
\plotfiddle{fig1a.ps}{2cm}{0}{35}{35}{-250}{0}
\plotfiddle{fig1b.ps}{2cm}{0}{35}{35}{-0}{+75}
\plotfiddle{fig1c.ps}{2cm}{0}{35}{35}{-250}{-100}
\plotfiddle{fig1d.ps}{2cm}{0}{35}{35}{-0}{-25}
\caption{The properties of the RQQs presented here (bold) 
compared to the entire PG-sample.
The data are taken from the references cited in the text. 
Other RQQs in the sample are denoted by open squares, and RLQs by crosses.
The sources presented here are fairly representative of RQQs in 
the entire sample.
\label{fig:rqqs_vs_pgsample}}
\end{figure}

The sample considered in this paper consists of all the radio-quiet 
PG-quasars for which the {\it ASCA}\ data
are available to us (as of 1999 Jul 01). The data are either from our own 
observations, or have been extracted from the {\it ASCA}\ archive at NASA/GSFC.
The sample is listed in Table~\ref{tab:sample} and is comprised of 27
observations of 25 objects. 
(Only four sources, PG~0003+199, PG~0050+124, PG~0921+525 and 
PG~1534+580 are Seyfert galaxies according to the criteria used 
by Schmidt \& Green.)
The sample is therefore approximately a quarter of
the RQQs in the PG sample. 
Table~\ref{tab:sample} also lists the 
source redshift ($z$) from Fabian \& Usher (1996, and references therein),
the absolute visual magnitude\footnote{$H_0 = 50\ {\rm km\ s^{-1}\ Mpc^{-1}}$ 
				and $q_0 = 0.5$ are assumed throughout.} 
   ($M_V$) derived  
	from
        Neugebauer et al (1987) or Boroson \& Green (1992),
the radio-loudness ($R_L$),
and the Galactic column density of neutral material along the line-of-sight 
	($N_{H,0}^{gal}$) as determined from 21~cm measurements.
In Fig.~\ref{fig:rqqs_vs_pgsample} we show 
the distribution of the PG sample in 
$z$, $M_V$ and $R_L$ along with some characteristics of the optical 
emission lines from Boroson \& Green (1992).
These are the equivalent width and FWHM of 
the H$\beta$ line, the equivalent width of the Fe{\sc ii}
complex between $\lambda$4434 and $\lambda$4684, and 
Peak $\lambda$5007 -- the ratio of the peak height of
the \verb+[+O{\sc iii}\verb+]+$\lambda$5007 line to that
of H$\beta$. 
It can be seen that the objects presented here 
are fairly representative of the RQQs in the whole
PG sample. 
The {\it ASCA} observing log is given in 
Table~\ref{tab:asca_obs_log}, along with
various related parameters and 
references to those datasets which have also been presented elsewhere.

\subsection{Data screening}
\label{sec:screening}
\label{Sec:asca-screening}

As described in the {\it ASCA Data Reduction Guide}, 
there are a number of observer--selectable choices for 
observations with the {\it ASCA} instruments. 
All the observations presented here have data available using 
so-called 
'{\tt BRIGHT}-mode' on the SIS detectors. 
Many of the observations also have 
data collected simultaneously using '{\tt FAINT}-mode'. 
However, since this mode is not avaliable for all the datasets, 
here we restrict our analysis to the {\tt BRIGHT}-mode data.
The SIS clocking mode(s) used for each observation is given in 
Table~\ref{tab:asca_obs_log}. In most cases a single clocking mode is used 
throughout an observation, although multiple modes are used occasionally. 
Two of the observations reported here (PG~1416-129 and PG~1634+706)
were performed using two clocking modes. Due to small 
differences in the calibration, the SIS data from each mode were 
analysed independently.
All the data from the GIS detectors 
were obtained using PH mode.
Further details on the instrumentation on {\it ASCA}\ and its performance
can be found in Makishima et al (1996, and references therein).

The raw data for all but one of the observations 
(PG~1543+489) were converted to unscreened event files
using the 'Rev2' version of the 
processing scripts. The event files containing data collected in
all three telemetery modes were combined.
These data were then screened
using the {\tt ascascreen} (v0.39) 
script within the
{\tt FTOOLS} package (v4.0) such as only to include data collected
when:
the spacecraft was outside of the South Atlantic Anomaly;
the radiation belt monitor rate was less than $500\ {\rm ct\ s^{-1}}$;
magnetic cut-off rigidity was $>$6 GeV/c;
angular offset from the the nominal pointing position was $<$36 arcsec;
elevation angle above the Earth's limb was $>5^{\circ}$ for GIS data,
        and $>10^{\circ}$ for SIS data.
In the case of SIS data (only), the following additional
criteria were also applied:
that the Bright Earth angle (elevation angle above the sun-illuminated Earth's
        limb) was $>20^{\circ}$;
that the spacecraft was at least 100 s beyond the day/night terminator.
An appropriate CCD pixel threshold was chosen for each SIS dataset
and 'hot' and 'flickering' pixels were removed using the standard
algorithm. Only SIS 'grades' 0, 2, 3 and 4 were included in the  analysis.
The exposure times for the screened events files are listed in 
Table~\ref{tab:asca_obs_log} along with other observational details.

\subsection{Spatial Analysis and extraction procedure}
\label{Sec:spatial-xtractcells}

After extracting the event files we accumulated and examined the
images for each instrument. 
All but two of the targets 
(PG~0043+030 and PG~1700+518) were detected in all four instruments.
For each detected source, we used 
{\tt XIMAGE} (v2.30) to sum the two SIS images (in sky coordinates) 
and determine the source centroid. 
In all but one case, the X-ray centroids of the detected 
sources were found to be 
$\lesssim 1$~arcmin of the optical position of the host galaxy
as noted in Table~\ref{tab:sample}, and hence
consistent with the current uncertainty in the 
attitude reconstruction of {\it ASCA}\ data 
(Gotthelf \& Ishibashi 1997).
Similar results were obtained using the summed GIS images.
As noted in Nandra et al (1997a), 
an offset $\sim1.5$~arcmin (to the west) was found 
in the case of PG~0003+199. However we are confident that the quasar
is the origin of the X-ray emission as the position of the X-ray source
as determined by {\it ROSAT}\ observations agrees well is that of the 
optical nucleus.
A number of serendipitous sources were evident in the images, and
are noted in the appendix.

Extraction cells were defined for the subsequent temporal and spectral
analysis of the (target) PG-quasar.
In the case of the {\it ASCA}\ SIS data, the source extraction cell
employed was circular of radius $\sim 3.2$~arcmin centered on the target. From
the point spread function (psf) 
of the XRT/SIS instrument, such a cell contains $\sim84$\% of the
total source counts when the cell lies entirely within the active area of the
SIS chip. In a number of cases, however, the pointing of the spacecraft
was such that some part of the source extraction cell falls outside the active
region of the SIS chip, with a corresponding reduction in the fraction of
total source counts contained.
An extraction cell was defined to provide an estimate of the background for
each SIS detector which consisted of the whole of the nominal CCD chip
excluding a circular region of $\sim 4.3$~arcmin centered on the source. In
the case of the GIS data, the source region was circular of radius $\sim
5.2$~arcmin centered on the target. From the psf of the XRT/GIS instrument the
region contains $\sim89$\% of the total source counts. An annulus, centered on
the source and covering $\sim 5.2$--9.8~arcmin was used to provide an estimate
of the background for each GIS detector. For both the SIS and GIS the regions
immediately surrounding the serendipitous sources 
were excluded from the extraction cells.
All fluxes and luminosities quoted below (but {\it not} count rates)
have been corrected for the fraction of the source photons falling outside
the source extraction cells.

\section{TEMPORAL ANALYSIS}
\label{Sec:temporal}

Light curves of the source region were constructed for each observation using
{\tt XRONOS} (v4.02) and several bin sizes ($t_{bin}$). 
We combined the data from 
the separate instrument pairs to form SIS and GIS light curves 
in order to increase the signal-to-noise ratio in our analysis, but only
consider bins which are fully exposed in both instruments (SIS0/SIS1 or
GIS2/GIS3). Following Nandra et al (1997a), we parameterize any variability by
the normalised 'excess variance', $\sigma^2_{rms}$($t_{bin}$), as
prescribed in Turner et al (1999a).
This formulism requires there is a sufficient number of counts ($n$) in
each bin such that Gaussian statistics are applicable, and 
that there is an adquate
number of such bins ($N$). Here we adopt the requirements $n \geq 20$ and $N
\geq 20$ and limit our calculation of $\sigma^2_{rms}$($t_{bin}$) 
to those light
curves in which $< 5$\% of the original bins were excluded due to insufficient
counts. The results for the 15 datasets satisfying these criteria for 
$t_{bin} =$256~s are tabulated in Table~\ref{tab:xs_rms_128}.
The quantity $\sigma^2_{rms}$($t_{bin}$) 
is a measure of the integrated power-density spectrum 
of the source over frequencies 
$f$ (in the quasar frame) in the range  
	$1/t_{dur}   \lesssim  f/(1+z) \leq 1/N\ t_{bin}$, 
where $t_{dur}$ is the total duration of the observation 
(Table~\ref{tab:asca_obs_log}).
For all 15 datasets 
	1--2$\times10^{-5} \lesssim f \lesssim 2\times10^{-4}\ {\rm Hz}$.
Thus $\sigma^2_{rms}$($t_{bin}$) provides a measure of the intrinsic
variability characteristics over a similar range of $f$ in all cases.
From Table~\ref{tab:xs_rms_128} it can be seen that
statistically significant variability (at $\gtrsim 90$\% confidence)
is evident in the majority of the datasets.
Inspection of the light curves for PG1404+226 reveal this source is also
variable, but is too weak to satify our $n$ \& $N$ criteria. Variability of
lower significance may also be present in PG~1116+215 and PG~1211+143. 
The variable sources will be noted as
appropriate within the spectral analysis presented in \S\ref{Sec:spec}, but
general discussion of the variability characteristics is postponed until
\S\ref{Sec:disc-temporal}.

\section{SPECTRAL ANALYSIS}
\label{Sec:spec}
\label{Sec:spec-prelims}
\label{Sec:spec-acceptability}

In the current paper we consider only the time-averaged 
spectra for each observation in order to maximize the signal-to-noise 
ratio
(but bearing in mind some datasets exhibit significant 
variability, \S\ref{Sec:temporal}).
Spectra were extracted from the source and background regions for 
each instrument as defined in \S\ref{Sec:spatial-xtractcells}.
In the case of the SIS data,
the original pulse-height assignment for each event was converted to
a pulse-invariant (PI) scale using {\tt sispi} (v1.1)
and the gain-history file
\verb+sisph2pi_110397.fits+.
In the case of the GIS data 'hard particle flares' were rejected
using the so-called '{\tt HO2}' count rate, and standard 'rise-time' rejection
criteria.
The spectral analysis was performed using the {\tt XSPEC} (v10.00) package
(Arnaud 1996), having first grouped  the raw spectra such that each bin 
contained $\geq 20$ counts permitting us to use $\chi^2$ minimization 
techniques.
In the case of the GIS datasets, the detector redistribution matrices
released on 1995 Mar 06 were used.
In the case of the SIS datasets, appropriate detector redistribution matrices
were generated using {\tt sisrmg} (v1.1).
The effective area appropriate for each dataset was calculated using
{\tt ascaarf}
(v2.72)\footnote{using the calibration files 
\verb+xrt\_eq\_v2\_0.fits+ and \verb+xrt\_psf\_v2\_0.fits+,
and including both the 'Gaussian' and 'filter fudges' to account for problems 
in the calibration of the instruments which have so-far defied physical 
explanation}.
Following common practice, for each observation
we fit the spectra from all four instruments simultaneously, but allowing 
an independent normalization of the model for each 
detector to account differences in the absolute flux calibrations.
All fluxes and luminosities quoted for the sources are derived 
(arbitrarily) using the SIS0 detector.
The corresponding errors represent only the statistical uncertainties.
The luminosities were calculated assuming the preferred model discussed 
below (with the errors derived by allowing only  
the normalizations of the spectral components to vary).
The absolute flux calibration of the instruments on {\it ASCA}\ is believed 
to be uncertain by at least $\sim$5\%.

We use a number of criteria to establish whether we consider a model to
be consistent with a given dataset.
During the spectral analysis we use the $\chi^2$-statistic 
(in conjunction with
the number of degrees-of-freedom, $dof$) to determine the goodness-of-fit over
the energy range used during the analysis.
To this end, we use
the probability, $P(\chi^2)$, that the statistic should be less
than the observed value of $\chi^2$ under the assumption that the model is
indeed the true representation of the data. Thus $P(\chi^2) = 0.5$
corresponds to a reduced--$\chi^2$ value of unity and values of
$P(\chi^2) \sim 1$ indicate that the data are poorly represented by
the model.
For the purposes of this paper, we consider the model to be an adequate
description of the data if
$P(\chi^2)\leq 0.95$

Data from the SIS below 0.6 keV
(observer--frame)
were excluded from the spectral analysis due to uncertainties associated with
the calibration of the SIS detectors below this energy. 
However, 
although the
calibration is suspect at these energies, it is generally considered that the
amplitude of the error is $\lesssim20$\%
(e.g. Orr et al 1998).
Thus, following George et al
(1998a), we make use of this fact by extrapolating our best-fitting models to
energies $<0.6$~keV and calculating
        the ratio of the increase in the $\chi^2$-statistic to the number of
        addition data points
($\frac{\Delta \chi^2_{0.6}}{\Delta N_{0.6}}$),
and
the mean data/model ratio for data points 
below 0.6~keV
($\overline{R_{0.6}}$).
We consider models which give
$\frac{\Delta \chi^2_{0.6}}{\Delta N_{0.6}} > 2$ {\it and}
$\overline{R_{0.6}}$ outside the range 0.8--1.2
disagree with the data at a level unlikely to be the result of any
calibration uncertainties.
Such cases are noted and discussed in the text.

\subsection{The X-ray Continuum}
\label{Sec:spect_all}
\label{Sec:UX-definition}

In this section we consider the form of the X-ray continuum. The goals are 
 to obtain the simplest parametrization of the continuum consistent with the
data.
Thus we exclude the 5--7~keV energy band from the
analysis, postponing the investigation of any Fe $K$-shell emission 
until \S\ref{Sec:Fe-band}.

We first compare the data to three 'standard' spectral 
models in which the underlying continuum is assumed to be 
a single power law.
These and all other models considered in this paper
include Galactic absorption {\it fixed} at the 
effective hydrogen column density ($N_{H,0}^{gal}$)
derived from 21~cm measurements along the appropriate 
line-of-sight as given in Table~\ref{tab:sample}.
We use the abundances and cross-sections given in 
Morrison \& McCammon (1983) and hence assume no 
variations in the abundances of heavy elements within the ISM.
Our simplest model consists of only the power law.
	This model, hereafter refered to as {\bf Model~A},
	has only one interesting parameter: 
	$\Gamma$, the photon index of 
	the power law.
{\bf Model~B} is a single power law, absorbed by a column of neutral 
	material at the redshift of the quasar ($N_{H,z}$) 
	completely covering the cylinder--of--sight.
	This model has two interesting parameters: $\Gamma$ and 
	$N_{H,z}$.
{\bf Model~C} is as for  Model~B, but with the absorbing material 
	is assumed to be photoionized.
	The ionized--absorber models used in this paper were 
	generated 
	using the photoionization code {\tt ION} 
	(Netzer 1993, 1996, version {\tt ION97}). The 
	models are those used in George et al (1998a), except
that here we 
 parameterize the ionization state of the gas
	by a somewhat different ionization parameter 
defined as:
\begin{equation}
	\label{eqn:U_X}
	U_{oxygen}
		 = \int^{10\ {\rm keV}}_{538\ {\rm eV}}
			\frac{(L_E/E)}{4 \pi r^2 n_{H} c} dE
\end{equation}
	where $L_E$ is the monochromatic luminosity  at energy $E$,
	$r$ the distance from the source to the absorbing material, and 
	$n_H$ the number density of the absorber
	(in all cases assumed to be $10^{8}\ {\rm cm^{-3}}$).
  The reasoning for using this version of the ionization parameter is that
the form of the continuum at energies below the oxygen K-edge
(538~eV for O{\sc i}) have only a minor
effect on the strength of the oxygen absorption features 
that are used to
determine the ionization parameter and absorbing column. 
	Model~C therefore has three interesting parameters:
	$\Gamma$, $N_{H,z}$ and $U_{oxygen}$.

As will be discussed in \S\ref{Sec:0.6-10-observed}, 
for many of the datasets, 
Models~A--C do not provide an adequate description of the 
continuum over the full 0.6--10~keV (observed-frame) band.
In most such cases, there appears a gradual steepening 
of the continuum with decreasing energy.
This may not be considered surprizing 
given the evidence for 'soft excesses'
in previous observations of quasars.
Unfortunately however, 
modelling (sometimes rather subtle) deviations from a powerlaw 
due to a convex continuum 
are rather problematic using {\it ASCA} data given the 
limited bandpass and the moderate spectral resolution of the 
instruments.
Thus, it is not clear what functional form 
is most appropriate to model the soft X-ray emission, 
especially since the excesses appear 
to be different in different sources.

Here we have adopted two functional forms to parameterize
such convex continua.
The first is by introducing a second power law. 
In {\bf Model~D} 
the underlying continuum is therefore the sum of 
	two power laws, both absorbed by a column of neutral
        material at the redshift of the quasar and 
	completely covering the cylinder--of--sight.
	The photon indices for the 'soft' and 'hard' power laws 
	are denoted by $\Gamma_s$ and $\Gamma_h$, and the
	relative intensity of the components 
	 by the ratio ($R_{s/h}$) of the soft 
	to hard power laws at 1~keV (quasar-frame).
	This model has four interesting parameters:
	$\Gamma_s$, $\Gamma_h$, $R_{s/h}$ and 
	$N_{H,z}$.
{\bf Model~E} is as for Model~D except the 
absorbing material is photoionized\footnote{For the 
sake of computational expediency, the ionization 
state of the gas was calculated using the power law 
which dominates the continuum in the {\it ASCA}\ 
bandpass. However, in all cases 
the errors associated with this approximation
are less than statistical uncertainties on the derived 
values of $U_{oxygen}$.}, giving a fifth 
interesting parameter, $U_{oxygen}$.

The alternative functional form used to parameterize 
the 'soft excesses' is a Gaussian emission component. 
Thus {\bf Model~F} consists of a single power law 
plus a Gaussian peaking at an energy $E_z$, of width
$\sigma_z$ and total luminosity 
$L_{line}$.
As will be further discussed in \S\ref{Sec:0.6-10-observed}, 
Model~F is the preferred model for a number of datasets 
where the Gaussian component models 
only small deviations from a pure power law in the 0.6--2~keV 
(observed-frame) band. 
In at least some cases this may be an artifact of 
remaining calibration uncertainties or an indicator of 
more subtle spectral complexity (see also \S\ref{Sec:disc-1keV}).

Since the objects in our sample have a range of redshifts, the 
useable 0.6--10~keV {\it ASCA} band covers different energy bands in the 
quasar frame in each object
($\sim$2--20~keV in the most extreme case of PG~1247-267).
Thus in \S\ref{Sec:2-10} we 
consider the form of the continuum 
in the 2--10~keV band, which 
 enables us to make direct comparisons between 
the objects.
Then in \S\ref{Sec:0.6-10-observed}
we consider the form of the 
full 0.6--10~keV (observed-frame) continuum. 
This obviously enables us to study the low-energy spectra of the 
objects with the lower redshifts and the high-energy spectra of those 
with higher redshifts.

We note that the during the spectral analysis 
presented below, different portions 
(in energy-space) of the 
detectors' response functions are used for objects of different redshift.
If the 
sources have simple spectra across the fitted  band, then this is not a 
concern
so long as there is a section of well-constrained data from which to determine
the photon index. If the sources have spectral curvature within the 
fitted band but are modeled using a simple power law, 
then the well-constrained data 
will occur across different parts of the curved spectrum, and could yield a
false correlation between photon index and redshift, and a false distribution
of photon indices.
Bearing in mind these problems, in this paper we have examined the data both
using a fixed rest-frame approach, and then using a fixed observed-frame
approach. We base our conclusions of the sum of information from all of these
fits, and do not find any evidence for instrumentally-biased results.

\subsubsection{Analysis of the 2--10~keV band (quasar-frame)}
\label{Sec:2-10}

In Table~\ref{tab:zpo_ngal_210_noline_fits} we show the results when Model~A
(a single power-law with galactic absorption) 
is applied to the 2--10~keV band, 
excluding the 5--7~keV band.
We find that Model~A provides an adequate description of the data (i.e.
$P(\chi^2) \leq 0.95$) over this restricted energy band for all but one
dataset.
This simple parametrization, is the least dependent on any physical model. 
The exception is PG~1411+442, which has an 'inverted' best-fitting 
photon index, and will be further discussed below. 

In Fig.~\ref{fig:210_ratio} we show the data/model ratios for each of 
the datasets when the best-fitting power law derived from this analysis 
is extrapolated and compared to the 
data in the full {\it ASCA}\ band.
It can be seen that for several of the datasets 
the ratios are close to unity at all energies.
It is important to note that these datasets are not only those 
with low signal to noise ratio, nor do they seem to be 
only the high luminosity sources.
The most noticeable aspect of Fig.~\ref{fig:210_ratio} is, however,
that for the majority of the datasets the data/model ratios exhibit 
large, systematic deviations from unity at one or more energies.
In the case of PG~1114+445 and PG~1543+489, there is 
a deficit of counts $\lesssim 2$~keV (observed-frame).
The remainder of the sources have an excess of counts at 
low and/or high energies compared to the best-fitting power power-law 
2--10~keV continuum.
It should also be noted that several of the datasets contain 
residuals in the 5--7~keV band indicative 
of Fe $K$-shell emission (see \S\ref{Sec:Fe-band}).

\subsubsection{Analysis of the full 0.6--10~keV (observed-frame) Continuum}
\label{Sec:0.6-10-observed}

In this section we report the results from an 
analysis of the 0.6--10~keV (observed-frame) continuum of each 
dataset (again excluding the 5--7~keV band in the quasar's frame).
We have applied Models~A--F to all the datasets, and 
for each defined the model which we 
consider {\it the simplest, acceptable description} of the data.
This is defined as the acceptable model for which the inclusion of 
additional free parameters no longer improves the fit at $>95$\% confidence
(using the $F$-statistic).
The preferred model for each dataset is listed in Table~\ref{tab:fit_summary2}.
The derived spectra and 
corresponding
data/model ratios are shown for a number of the datasets
in Fig.~\ref{fig:ufmodelratio}.

The selection of our preferred model is straightforward for the 
majority of the datasets, and we are confident that the model
provides a fair representation of the time-averaged spectrum.
However, the selection of our preferred model for 
3 datasets (PG~1148+549, PG~1411+442 and PG~1634+706)
warrants some discussion.
Models~A--F do not provide 
a fit which satisfied all our criteria for these datasets. 
In the case of PG~1634+706, our preferred model is an unattenuated power 
law (Model~A).
The lack of an acceptable fit appears to be the result of excess scatter of 
the residuals throughout the 
{\it ASCA}\ bandpass,
and we note $P(\chi^2)$ was only just acceptable in our 2--10~keV fits
(\S\ref{Sec:2-10} and Table~\ref{tab:zpo_ngal_210_noline_fits}).
In the case of PG~1148+549, whilst Model~A provides an acceptable 
description to the fitted data, it does not extrapolate in an acceptable
manner below 0.6~keV ($\overline{R_{0.6}}\simeq1.5$).
This is most likely due to an additional spectral
component dominating the emission below 0.6~keV. However 
since the {\it ASCA}\ data are unable to place useful constraints
on this component, 
Model~A is retained as our preferred model.
PG~1411+442 is particularly interesting. Our preferred model is 
a variant on Model~B in which a fraction $D_f$ of the underlying 
continuum does not suffer attenuation (Model~B$_{\rm pc}$)
whilst the remainder is absorbed by a large column density 
(see Fig.~\ref{fig:ufmodelratio} below).

The best-fitting parameters for those datasets for which 
Models~A--C, D--E and F
are preferred are listed in 
Tables~\ref{tab:prefer_abc}--\ref{tab:prefer_f}.
In general our results are as might be expected from
inspection of Fig.~\ref{fig:210_ratio}. 
In the case of 3 datasets 
(PG~1211+143, PG~1404+026, and PG~1440+356), 
Model~F is preferred with the Gaussian component 
parameterizing the steepening of the spectrum at the lowest 
energies.
A convex continuum is also apparent in 7 other datasets,
but we prefer a model in which it is parameterized
by a second power law (ie. Models~D and E)\footnote{We acknowledge 
        that in many cases the true 
	spectrum is likely to be more complex than either 
	parameterization.
	However, the low spectral resolution afforded by 
	the {\it ASCA} detectors, the relatively low 
	signal-to-noise ratio of the present data, and 
	the remaining calibration uncertainties, 
	make it difficult to make further progress.}.
In the case of PG~0003+199 and PG~1501+106a,b, the second power law 
is a parameterization of a soft X-ray component, with 
$0.1 \lesssim R_{s/h} \lesssim 1$.
In the case of PG~1116+215, the second power law is a parameterization of a 
hard X-ray component dominating the continuum at energies $\gtrsim 5$~keV 
($R_{s/h} \gtrsim 10^2$), with $\Gamma_h$ 'pegging' at the flattest 
value allowed in our analysis ($\Gamma_h =0$)
and $\Gamma_s - \Gamma_h \gtrsim 2.4$.
In the case of PG~0804+761 and PG~1534+580, we find 
$1 \lesssim R_{s/h} \lesssim 10$ with both power laws of similar importance 
over the bulk of the {\it ASCA}\ bandpass.
This is also the case for PG~1244+026 (with $R_{s/h} \sim 10$), but the
spectrum also contains a rather narrow emission feature 
centered at an energy $\sim 1$~keV.
Our preferred model for this source is therefore Model~D plus a 
Gaussian emission line
with $E_z = 0.96^{+0.07}_{-0.31}$~keV, 
$\sigma_z = 0.09^{+0.16}_{-0.09}$~keV and 
$L_{line} \simeq 3\times10^{42}\ {\rm erg\ s^{-1}}$.
This feature has been noted previously 
in this dataset by Fiore et al (1998a) and is further discussed 
in \S\ref{Sec:disc-1keV}. 
Inspection of Fig.~\ref{fig:210_ratio} reveals PG~1216+069 
to have very similar data/model residuals as PG~1244+026
when compared to the best-fitting power law in the 
2--10~keV band
(\S\ref{Sec:2-10}). 
However, since the parameters associated with the putative 
$\sim 1$~keV feature cannot be well-constrained using the current
{\it ASCA}\ data, Model~A remains our preferred model
for PG~1216+069.

For the remaining 3 datasets with convex spectra
(PG~0921+525, PG~0953+415, PG~1416-129), 
our preferred model is Model~F. However in this case 
the Gaussian component represents only a very subtle 
deviation from a pure power law continuum.
This may be indicative of a true, slow curvature in the underlying 
continuum not accounted for in 
Models~A--E), or may be an artifact such as a residual 
calibration problem.
However in all cases the presence of the Gaussian component
does not affect any 
of the remaining spectral parameters more than 
their statistical uncertainty. Thus
we retain Model~F as our preferred parameterization in 
all 4 cases.

To summarize our results regarding the form of the 
continuum in the 0.6--10~keV (observed frame) band
of the 25 datasets (of 24 RQQs detected):
\begin{itemize}
\item 	a single power law continuum (Model~A) is the preferred model 
	for 7 datasets/RQQs.
	Five of these 
	(PG~0050+124, 
	PG~1216+069, 
	PG~1247+267, 
	PG~1407+265, 
	PG~1444+407) satisfy all our criteria
	and 2
	other sources (PG~1148+549 and PG~1634+706) 
	do not.
\item
A single power law continuum with absorption 
	intrinsic to the quasar (Models~B and C) is the preferred model 
	for 4 datasets (PG~0844+349, 
	PG~1114+445, PG~1322+659 and PG~1543+489), 
	with a varient (Model~B$_{\rm pc}$) 
	the preferred model in the case of PG~1411+442.
	The absorption features are, however, weak in the 
	the case of PG~1322+659 and  PG~1543+489 
	(see \S\ref{Sec:disc-absorption}).
\item
A double power law continuum (Models~D and E)
	is preferred for 7 datasets
	(PG~0003+199, 
	PG~0804+761, 
	PG~1116+215, 
	PG~1244+026, 
	PG~1501+106a,b and PG~1534+580).
The second power law represents a soft X-ray component in the case of 
3 datasets (PG~0003+199, PG~1501+106a,b) and a hard X-ray component 
in the case of 1 dataset (PG~1116+215). 
The two power laws are equally 
important within the {\it ASCA}\ band for the remaining 3 datasets
(see Fig.~\ref{fig:ufmodelratio}).
PG~1244+026 also contains 
a 'narrow' emission feature
at $\sim 1$~keV.
\item
Model~F is the preferred model for the remaining 6 datasets/RQQs.
For 3 of these (PG~1211+143, PG~1404+226, PG~1440+356)
the Gaussian component parameterizes a soft X-ray component.
In the remaining 3 
(PG~0921+525, PG~0953-415, and 
PG~1416-129), the Gaussian component parameterizes only very 
subtle curvature in the continuum.
\end{itemize}

\subsection{The Iron K-band}
\label{Sec:Fe-band}

Using the parameterization of the continuum obtained above, we now introduce
the 5--7~keV energy band into the analysis. In
Table~\ref{tab:feline} we list the number of additional spectral bins ($\Delta
N_{pts}$) for each dataset included in the analysis. We also list the
fractional increase in the $\chi^2$-statistic per additional bin ($\Delta
\chi^2$/ $\Delta N_{pts}$) when the preferred model from
\S\ref{Sec:0.6-10-observed} is applied to the data in this band. 
In order to better constrain any Fe $K$-shell emission we have fixed the
parameters associated with the continuum at the best-fitting values derived
above. A broad Gaussian emission line was added to the spectral model. We
constrain the energy of the line to lie in the range $4 \leq E_z \leq 8$~keV
and its width $\sigma_z \leq 2$~keV (both in the quasar's frame).
The intensity of the line is parameterized by its 
luminosity, $L$(Fe-$K$), and its equivalent 
width, $EW$(Fe-$K$), compared to the underlying continuum at $E_z$.

Constraints and upper limits on the emission line are listed in 
Table~\ref{tab:feline}.
We find the addition of the 3 additional parameters associated with the 
emission line significantly ($F$-statistic $\geq 3$) 
improves the fit for 11 datasets.
However, the data/model ratios for 
the two datasets with the the largest values of $EW$(Fe-$K$), 
PG~0050+124 and PG~1444+407, reveals features
more reminisent of an unmodelled continuum component at energies 
$\gtrsim 5$~keV rather than a Fe $K$-shell emission
(Fig.~\ref{fig:ufmodelratio}). 
Further observations are required to distinguish between these 
possibilities. 

We postpone further discussion of the Fe $K$-shell emission 
until \S\ref{Sec:disc-Fe}.

\section{DISCUSSION}
\label{Sec:disc}

To recap, in the previous sections we have presented the results 
from 27 {\it ASCA}\ observations of 26 RQQs
(PG~1501+106 was observed twice).

\subsection{The objects not detected}
\label{Sec:disc-nondet}

Two objects (PG~0043+039 and PG~1700+518)
were not detected during the {\it ASCA}\ observations. 
Both are known to contain broad absorption line 
(BAL) systems in the optical band 
(eg. Turnshek et al 1997) suggesting 
the lack of X-rays observed is most likely due to 
attenuation by material along the line of sight.
The lack of detection of these sources by the 
{\it ROSAT}\ PSPC (with a bandpass up to $\sim$2.5~keV in 
the observer's frame) requires 
column densities $\gtrsim 2\times10^{22}\ {\rm cm^{-2}}$
(eg. Green \& Mathur 1996).
The lack of detection by {\it ASCA}\ (bandpass up to 10~keV)
increases the required column densities to 
$\gg 10^{23}\ {\rm cm^{-2}}$.
The results of {\it ASCA}\ observations of a number BAL 
quasars (including the above two sources)
have been discussed recently 
by Gallagher et al (1999).

\subsection{Time variability}
\label{Sec:disc-temporal}

In \S\ref{Sec:temporal} we found 15 of the datasets 
(14 out of the 24 RQQs detected)
to have a sufficient
number of counts that we were able to perform an analysis of the variability
characteristics.
From an analysis using the normalised
excess variance $\sigma^2_{rms}$($t_{bin}$) with time bins of 
$t_{bin} =$256~s, 
we found statistically significant
variability (at $\gtrsim 90$\% confidence) in 10 datasets
(9 RQQs)
and suspect variability in 3 others.

From an analysis of a hetergeneous sample of Seyfert 1--1.5 galaxies observed
early in the {\it ASCA}\ mission, Nandra et al (1997a) found
a trend whereby sources of higher luminosity exhibit 
a lower degree of variability. This finding confirmed the results 
made by previous satellites (Barr \& Mushotzky 1986; 
Lawrence \& Papadakis 1993; Green et al 1993).
The anti-correlation between variability and luminosity 
has been further explored by Turner et al (1999a), 
who considered 
$L$(2--10~keV) versus $\sigma_{rms}$(256~s)
for a larger, but still heterogenous, collection of Seyfert 1 
Galaxies from the {\it ASCA}\ archive.
In Fig.~\ref{fig:correlations_xray_rms}a
we show 
$L$(2--10~keV) versus $\sigma_{rms}$(256~s) 
for the RQQs presented here along with the other sources from Turner et al
(excluding the 3 RL objects in the latter).
As expected, the RQQs behave in a similar way to other AGN despite 
covering only a limited range in luminosity.
All 5 RQQs with 
  	$L$(2--10~keV) $\gtrsim 10^{44}\ {\rm erg\ s^{-1}}$
have $\sigma_{rms}$(256~s) $\lesssim 10^{-2}$.
The lack of variability in the higher luminosity datasets 
is not a simple case of them having a lower signal to noise ratio. For 
example, $\gtrsim 4$ times as many source counts were obtained 
from PG~0804+761 
($L$(2--10~keV) $\simeq 4\times10^{44}\ {\rm erg\ s^{-1}}$, 
$\sigma_{rms}$(256~s) $\lesssim0.3\times10^{-2}$)
than from the second most variable source in our sample
(PG~0050+124: $L$(2--10~keV) $\simeq 5\times10^{43}\ {\rm erg\ s^{-1}}$,
$\sigma_{rms}$(256~s) $\simeq 4\times10^{-2}$).

From a study of the {\it ROSAT}\ data from 6 PG--RQQs,
including PG~1440+356 and PG~1216+069,  Fiore et al (1998b) found a trend 
whereby the objects with steeper spectra in the 
observed 0.1--2~keV band
are more variable on timescales of 2--20~days than the objects with flatter 
soft X-ray spectra.
Similar behavior has been suggested from the 
{\it EXOSAT}\ and 
{\it ROSAT}\ data 
of low-luminosity AGN 
(e.g. Green, McHardy \& Lehto 1993; K\"{o}nig et al 1997;
Boller et al 1996).
Interestingly, the RQQs presented here reveal no clear correlation between 
$\sigma_{rms}$(256~s) and $\Gamma_{2-10}$, although the 
number of objects is admittedly small.
Turner et al (1999a) also present 
values for $\Gamma_{2-10}$ from a similar analysis 
as presented in \S\ref{Sec:2-10}.
From Fig.~\ref{fig:correlations_xray_rms}b
it can be seen that the RQQs 
occupy similar regions of the $\sigma_{rms}$(256~s)--$\Gamma_{2-10}$ 
plane as the other sources in Turner et al (1999a). 

Discussion of the correlation between $\sigma_{rms}$ and 
parameters from other wavebands, especially the width of the 
optical emission lines, is postponed until 
Paper~II.

\subsection{The 2--10~keV (quasar-frame) continuum}
\label{Sec:disc-slope210}

In \S\ref{Sec:2-10} we found a single power law (Model~A) provided 
an adequate description of the time-averaged spectra
in the 2--10~keV band for 
all the detected RQQs except PG~1411+442. 
Significant variations are seen between the best-fitting values of the 
photon index $\Gamma_{2-10}$ between the datasets, with a range 
	$1.5 \lesssim \Gamma_{2-10} \lesssim 3$.
Using the method of Maccacaro et al (1988), we find a mean 
$< \Gamma_{2-10} > = 1.97^{+0.08}_{-0.09}$ 
and dispersion 
$\sigma(\Gamma_{2-10}) = 0.24^{+0.08}_{-0.05}$.
The latter values are consistent with the nearby Seyfert 1 Galaxies.
For example, excluding the RL objects and the RQQs 
presented here, the Turner et al (1999a) 
sample has
$< \Gamma_{2-10} > = 1.85^{+0.07}_{-0.06}$ and
$\sigma(\Gamma_{2-10}) = 0.21^{+0.04}_{-0.02}$. 
These results are somewhat biased by the presence of multiple observations 
of a single object (e.g. Fairall-9, 
NGC~3783, NGC~5548 \& Mrk~509), however
its clear from 
	Fig.~\ref{fig:correlations_xray_gamma}a
that the RQQs presented here occupy a similar region of the 
$L$(2--10~keV)--$\Gamma_{2-10}$ plane.
Discussion of the results found prior to {\it ASCA}\ 
is postponed until the next section.

\subsection{The 0.6--10~keV (observed-frame) continuum}
\label{Sec:disc-0.6-10}

In \S\ref{Sec:0.6-10-observed} we presented the results from 
an analysis of the the full 0.6--10~keV (observed-frame) continuum
of each dataset. We considered a relatively small number of 
spectral models, and presented that which we felt best described the 
time-averaged continuum of each dataset. 
No single model provides 
an adquate description of the data from all the objects.
The variety of spectral forms exhibited by the RQQs is illustrated in 
Fig.~\ref{fig:sample_asca_nulnu}.
We find no obvious trends with luminosity or redshift.

The continua of 11 out of 25 datasets 
(11 out of the 24 RQQs detected) can be 
adequately described by a single power law 
(i.e. those datasets for which Model~A, B or C was preferred).
Best-fitting continua with only very subtle deviations from a 
power law were the preferred model for an additional 
3 datasets (3 RQQs).
Of the remaining 11 datasets, we found 10 to 
have convex continua (flattening as one moves to 
higher energies). 
In the case of 7 datasets (6 RQQs) our preferred model 
parameterized the continuum as the sum of two power laws 
(Models~D and E), 
and for 3 datasets (3 RQQs) as a power law 
plus a Gaussian component at low energies (Model~F).
The preferred model for 
PG~1411+442 involves absorption only partially covering 
the cylinder-of-sight, and discussion is postponed 
to \S\ref{Sec:disc-absorption}.

For 4 of the datasets for which Models~D and E are 
preferred
(PG~0804+761, PG~1116+215, PG~1244+026 and PG~1534+580), 
the two power laws contribute similar fractions 
to the observed continuum in the {\it ASCA}\ bandpass
(eg. see Fig.~\ref{fig:ufmodelratio}).
For the remaining 21 datasets a single power law dominates 
the observed continuum in the {\it ASCA}\ bandpass.
For these we find\footnote{Using $\Gamma_h$ 
	in the case of PG~0003+199, PG~1501+106a,b.}
a mean spectral index
	$< \Gamma > = 1.99^{+0.08}_{-0.08}$ 
and dispersion 
	$\sigma(\Gamma) = 0.20^{+0.07}_{-0.04}$.
These values are consistent with most previous results from RQQs obtained
using {\it EXOSAT}\ and {\it Ginga}\ in the 2--10~keV (observed-frame) band 
which find $< \Gamma > \simeq 1.8$--2.0 and $\sigma(\Gamma) \simeq 0.2$--0.3.
(e.g. Comastri et al 1992; Lawson et al 1992; Williams et al 1992; 
Lawson \& Turner 1997). 
Our results are also consistent with those from {\it ASCA}\ observations 
in the 0.6--10~keV (observed) band of RQQs (Reeves et al 1997)
and with the general distribution of $\Gamma$ 
exhibited by Seyfert 1 Galaxies. 
This is illustrated in 
Fig.~\ref{fig:correlations_xray_gamma}b in which we 
show $L$(2--10~keV) versus $\Gamma$ from the {\it ASCA}\ observations
of the RQQs presented here, 
and a selection of other RQ AGN published in the literature.
The RQQs presented here show a range of $\Gamma$ coinsistent with the other 
objects, with no obvious relationship between $\Gamma$ and $L$(2--10~keV).
Vignali et al (1999) have recently presented results from
{\it ASCA}\ observations of 5 high-redshift ($z>2$) RQQs, 
all of which have $L$(2--10~keV)$\gtrsim 10^{45}\ {\rm erg\ s^{-1}}$
(none of which are reported here). 
They found $< \Gamma > = 1.7\pm0.1$
and $\sigma(\Gamma) \sim 0.1$ and suggested higher luminosity RQQs possess 
flatter X-ray continua.                        
The 3 highest luminosity RQQs in our sample 
($L$(2--10~keV)$\sim 10^{45}\ {\rm erg\ s^{-1}}$) all have 
$\Gamma \simeq 2$ arguing against such an hypothesis.

Finally we note that {\it Einstein} IPC observations in 
0.2--3.5~keV (observed) band reveal RQQs to have $< \Gamma > \sim 2$ 
(e.g. Wilkes \& Elvis 1987). 
This is consistent with the 
results presented here, but somewhat flatter than 
{\it ROSAT} PSPC observations 
(e.g. $< \Gamma > \sim 2.7\pm0.1$, Laor et al. 1997).

\subsubsection{The Presence of 'Hard-Tails'}
\label{Sec:disc-CR}

Our preferred model in the case of PG~1116+215 consists of 
a 'normal' continuum at low energies (with $\Gamma_s \simeq 2.4$), 
but then flattens to $\Gamma_h = 0$ 
	(the flattest value allowed in the analysis)
at energies $\gtrsim 5$~keV.
Less dramatic 'hard-tails' are present in our preferred model for PG~0804+761, 
and PG~1244+026.
In addition the data/model residuals for several other sources (eg.
PG~0050+124, PG~1216+069 and PG~1444+407) seem systematically 
greater than unity at energies $\gtrsim 5$~keV
 (see Fig.~\ref{fig:ufmodelratio}).
Such features may be indicative of a real up-turn in the spectra 
-- for example a Compton-reflection component (e.g. George \& Fabian 1991)
or an additional spectral component absorbed by Compton-thick material 
($N_{H,z} \gtrsim 10^{24}\ {\rm cm^{-2}}$).
Alternatively such an up-turn may arise as the result of a 
unknown problem with the analysis/calibration of these datasets, 
or may be an artifact of spectral variability within the observation
(variability was apparent in many of the above datasets).
Future observations of these sources at energies $\gtrsim 10$~keV
will be valuable.

\subsubsection{Soft X-ray Excesses \& 1~keV Emission Lines}
\label{Sec:disc-SXS}
\label{Sec:disc-1keV}

For the majority of RQ AGN, extrapolation of the 1--10~keV  
continuum underpredicts the emission observed in the optical/UV.
An up-turn in the emitted spectrum is therefore required 
somewhere between $\sim$1~keV and the Lyman 
edge (13.6~eV). Many models for the so-called 'big, blue bump' 
place this up-turn in the range 0.1--1~keV, and indeed 
observations in this bandpass (eg. by the 
{\it Einstein} IPC and {\it ROSAT}\ PSPC)
typically reveal steeper spectra than observed above 1~keV.

Our preferred model for 6 datasets (5 RQQs)
contains what is often referred to as a 'soft excess' -- a 
separate spectral component dominating at energies 
$\lesssim 1$~keV.
This component is represented by a Gaussian emission feature in our 
preferred model for PG~1211+143, PG~1404+026, and PG~1440+356, 
and as a second power law for PG~0003+199 and PG~1501+106a,b.
A soft excess is also implied in the case of the 
PG~1148+549 ($z=0.969$) just below the {\it ASCA}\ bandpass
(see \S\ref{Sec:0.6-10-observed}). 
We repeat the remarks made in \S\ref{Sec:spect_all}: 
given the limited bandpass, the moderate spectral resolution and calibration 
uncertainties of the instruments on {\it ASCA}, 
we consider these functional forms to be simply {\it parameterizations}
of the true soft X-ray emission.
Other functional forms are undoubtedly also consistent with the {\it ASCA}\ 
data. 
Indeed a combination of the 
numerous emission lines and edges expected from ionized material
both within and outside the immediate circumnuclear region
can also give rise to a soft excess.
Furthermore, variability is detected/suspected 
in the majority of these RQQs. The possibility of 
spectral variability within the observation therefore 
adds to the uncertainty 
on the parameters derived for the soft X-ray component.

Given the limitations of the current {\it ASCA}\ data 
plus the imminent launch of 
the {\it Chandra X-ray Observatory}, 
{\it XMM} and {\it Astro-E}, we do not dwell 
upon these soft X-ray components here. 
Nevertheless we do note 
that we find evidence for an additional soft component in 
only 5 of the 18 detected RQQs with $z\leq 0.25$ 
(i.e sources in which the {\it ASCA}\ bandpass allows a realistic 
possibility of detecting features $\lesssim 1$~keV in the quasar-frame).
It is clear that such components are not universal within 
the {\it ASCA}\ bandpass, and 
even when present, vary in both spectral shape and luminosity.

The {\it ASCA}\ spectrum of PG~1244+026 contains a rather 
narrow emission feature centered at an energy $\sim 1$~keV. 
This feature was first discovered using the same {\it ASCA}\ 
data by Fiore et al (1998a).
Interestingly, a similar feature has recently been noted in the 
{\it ASCA}\ data of the Narrow-line Seyfert 1 Galaxies 
	Ton~S180 (Turner et al 1998)
and 
	Akn~564 (Turner et al 1999b).
Inspection of Fig.~\ref{fig:210_ratio} reveals PG~1216+069 
may also contain a emission feature at $\sim 1$~keV. 
Unfortunately the current {\it ASCA}\ data from this object do not allow 
the parameters associated with the emission to be well-constrained.

Observations with instruments of higher-resolution and bandpasses 
stretching to lower energies  are required to reveal the true nature of 
the soft X-ray excesses and $\sim 1$~keV features.

\subsection{Intrinisic absorbers}
\label{Sec:disc-absorption}

The results of the spectral analysis presented in \S\ref{Sec:spect_all} were
obtained by fixing the Galactic column density at the value ($N_{H,0}^{gal}$)
derived from H{\sc i} measurements (Table~\ref{tab:sample}). All our preferred
models (except for Model~A) contain an additional column density of absorbing
material ($N_{H,z}$) which we assume to be intrinsic to the quasar.
There are two sources of uncertainty which could give to a false 
indication of intrinsic absorption.
First there may be variations in the 
abundances of the heavy elements within the ISM
resulting in an error in the attenuation curve applied to the 
{\it ASCA}\ spectra (by a factor proportional to the 
C, N and O abundance relative to H{\sc i}).
Second, despite us excluding the (worse-affected) SIS data below 0.6~keV 
from the analysis, there are still 
uncertainties in the calibration of both the SIS \& GIS 
below 1~keV (particularly in observations towards the end of the mission).
The combination of these uncertainties means that 
any apparently intrinsic column densities with
$\lesssim$few$\times10^{20}\ {\rm cm^{-2}}$
derived using {\it ASCA}\ data
should be treated with caution.
With this in mind we now discuss our results.

Our preferred model suggests intrinisic absorption 
in 16 datasets (15 out of the 24 RQQs detected).
However, the column density of any such material, 
$N_{H,z}$, is inconsistent with zero at $>90$\% confidence for 
only 9 datasets (8 RQQs).
Furthermore, only 
in the case of 5 of these 9 datasets (PG~0844+349, PG~1114+445, 
PG~1411+422, and both observations of PG~1501+106) 
are we confident that some
intrinisic absorption is indeed present. 
For the remaining 4 datasets 
(PG~0804+761, PG~1116+215, PG~1322+659 and PG~1534+580)
the situation is less clear. 
(As noted in \S\ref{Sec:disc-nondet}, intrinisic absorption also offers 
an explanation of the non-detection of the 2 BAL quasars.)

First we discuss the 4 RQQs for which we are confident 
there is intrinsic absorption.
Ionized material along the line-of-sight to PG~1114+445 
	(with $N_{H,z} \sim 2\times10^{22}\ {\rm cm^{-2}}$, 
	$U_{oxygen} \simeq 10^{-2}$)
has been discussed by George et al (1997). 
A column density of ionized material along the line-of-sight to 
PG~1501+106 
	($N_{H,z} \sim 2\times10^{21}\ {\rm cm^{-2}}$, 
	$U_{oxygen} \simeq 10^{-3}$)
has been discussed by George et al (1998a). 
Our preferred model for PG~1411+442 contains a substantial column 
density ($N_{H,z} \sim 2\times10^{23}\ {\rm cm^{-2}}$) obscurring
$\sim$98~\% of the continuum, 
and offers an explanation of the unusual optical--to--X-ray spectrum 
observed for this object (see Paper~II). 
A relatively low column density of neutral material 
($N_{H,z} \lesssim 10^{21}\ {\rm cm^{-2}}$) is suggested by our 
preferred model for PG~0844+349.

Ionized material is suggested along the the line-of-sight to 
the remaining 4 datasets. 
However, the derived parameters of the ionized material in 
PG~1322+659 are rather extreme. The implied column density is
$N_{H,z} \sim 10^{24}\ {\rm cm^{-2}}$, but the material is 
in such a high ionization state ($U_{oxygen} \sim 1$)
that only relatively weak absorption features are imprinted 
on the spectrum (Fig.~\ref{fig:ufmodelratio}).
However, this observation was performed towards the end of the 
{\it ASCA}\ mission when the calibration of the instruments 
is becoming increasingly uncertain.
The remaining 3 objects have rather complex spectra 
(Fig.~\ref{fig:ufmodelratio}). The apparent absorption 
could therefore be an artifact of incorrectly modeling 
the emission.
Confirmation is required before we can be certain that these 
4 objects indeed contain absorbing material along the line--of--sight.

\subsubsection{The Detection Rate of Ionized Gas}

With the bandpass and spectral resolution of {\it ASCA}, ionized material 
along the line--of--sight can be detected primarily by bound-free absorption 
edges of O{\sc vii} (739~eV) and O{\sc viii} (871~eV) imprinted in the 
continuum. 
Recent observations have shown that a large fraction of
low-luminosity AGN contain such material
with implied column densities in the range
$10^{21} \lesssim N_{H,z} \lesssim 10^{23}\ {\rm cm^{-2}}$
(eg. Reynolds 1997; George et al 1998a).
For example, 
of the 12 RQ objects with 
	$L$(2--10~keV)$< 10^{44}\ {\rm erg\ s^{-1}}$
considered by George et al (1998a), 11 were found to contain 
evidence for ionized material.

In additional to that presented in \S\ref{Sec:0.6-10-observed}, we have 
performed a spectral analysis of each dataset 
in which the model includes both O{\sc vii} and O{\sc viii} edges.
In each case these edges were added to the preferred continuum
(having first set $N_{H,z}$ to zero in the case of Models~C and E).
The results are listed in Table~\ref{tab:2edges}.
Fig~\ref{fig:2edge}a shows the sum of optical depths 
in these edges ($\tau$(O{\sc vii}$)+\tau$(O{\sc viii}))
as a function of signal to noise ratio (S/N).
Only upper limits on the optical depths 
are obtained for the majority of the RQQs.
A similar analysis was also carried out of the objects in the 
George et al sample, and the results are also shown 
in Fig~\ref{fig:2edge}a.
There appear to be no significant variations among the 
distribution of $\tau$(O{\sc vii})$+\tau$(O{\sc viii})
between the two samples.
As expected, the data from many of the 
RQQs are of relatively low signal to noise ratio. 
In the case of PG~1114+445, the low S/N is the 
result of the strong absorption edges in the spectrum. Nevertheless 
useful constraints on the the depth of the edges can be obtained
for this source. 
As noted above, the presence of ionized material is suggested 
in PG~1322+659, but with a very high ionization parameter.
Indeed the material is so highly ionized that 
oxygen is almost fully stripped, resulting in 
only an upper limit on $\tau$(O{\sc vii})$+\tau$(O{\sc viii}).

In Fig~\ref{fig:2edge}b we plot 
$\tau$(O{\sc vii})$+\tau$(O{\sc viii}) versus 
luminosity for the RQQs and George et al sample.
Only objects with a S/N $>25$ 
(hence excluding the RQQs with S/N less than that for any of the 
sources in the George et al sample).
are shown, plus PG~1114+445.
In the luminosity range 
$10^{43} < L$(2--10~keV)$\leq 10^{44}\ {\rm erg\ s^{-1}}$
we find evidence for significant 
O{\sc vii} and O{\sc viii} edges in all 8 RQ AGN 
from the George et al sample, yet in 
only 2 of the 7 RQQs
(4 of 10 RQQs  if no threshold is imposed on S/N).
A study of a larger, well-defined sample is required to determine whether
this is a selection effect.
At luminosities $L$(2--10~keV)$> 10^{44}\ {\rm erg\ s^{-1}}$
we find 2 out of the 3 RQQs with S/N $>25$
contain evidence of significant O{\sc vii} and O{\sc viii} edges
(3 out of 9 RQQs if no threshold is imposed on S/N).
The George et al sample only contains 3 objects with such luminosities, 
one of which has significant O{\sc vii} and O{\sc viii} edges.
A study of a larger number of objects covering a wider 
range of luminosity is required to determine whether 
the detection rate of ionized material changes as a function 
of luminosity.

Finally
in Fig~\ref{fig:nh_vs_ionox} we show the location of the 
ionized material in the $N_{H,z}$--$U_{oxygen}$ plane
for the RQQs and the other objects in the George et al sample.
We find no significant difference between the 
two samples.

\subsubsection{An Absorption Line in PG~1404+226 ?}

Using the same {\it ASCA}\ data reported here, 
Leighly et al (1997) found PG~1404+226 to contain an absorption feature at 
$\sim 1$~keV (observed-frame) which they interpreted as due to oxygen in a 
highly relativistic outflow. 
Ulrich et al (1999) also report the results of the same 
dataset and find an absorption feature at 
$\sim 1$~keV, but prefer alternative interpretations.

With our preferred model we find no requirement for 
such features in this object. For example adding 2 narrow absorption lines to 
our model at the energies found by Leighly et al we obtain a reduction in 
$\chi^2$ of only 2.2. 
This lack of a significant detection is in part due to the different 
parameterizations of the soft X-ray emission in this source used here and 
by Leighly et al. 
We assume a Gaussian component whilst Leighly et al 
use a multiple black-body spectrum
(Ulrich et al assume single-temperature black-body spectrum).
These functional forms all intersect the 
underlying power law close to 1~keV. 
Since the significance of the absorption features is dependent on the form of 
the soft X-ray spectrum, future observations with higher spectral resolution 
are required to verify their presence.
Nevertheless, we do find upper limits on the equivalent widths of the 
putative lines of 
	$EW$(1.12~keV) $\lesssim 64$~eV and 
	$EW$(1.24~keV) $\lesssim 15$~eV
(at 90\% confidence) consistent with the results of Leighly et al.

\subsection{The Fe-K$\alpha$ line}
\label{Sec:disc-Fe}

In \S\ref{Sec:Fe-band} we found the inclusion of 
a broad, Gaussian emission line significantly 
improved our preferred model in the Fe $K$-shell regime for 11 datasets
(10 out of the 24 RQQs detected). 
However some doubt was however cast on the reality of such a feature 
in PG~0050+124 and PG~1444+407.
Unfortunately the parameters associated with the 
Gaussian are very poorly constrained in most of the individual 
datasets, preventing detailed searches for trends with 
other parameters.
For the 9 datasets (8 RQQs) in which we are confident of the presence of 
Fe $K$-shell emission, we find a mean equivalent width
$< EW$(Fe-$K$) $> = 220^{+70}_{-60}$~eV and mean 
line energy of $< E_z > = 6.34^{+0.13}_{-0.12}$~keV. 
The dispersions of both distributions are ill-determined
with upper limits of 
$\sigma$($EW$(Fe-$K$)) $\lesssim 450$~eV
and 
$\sigma(E_z) < 0.21$~keV 
at 90\% confidence.
These results are consistent with those of Nandra et al (1997b) for their 
sample of Seyfert 1 Galaxies when they employed a Gaussian emission 
component to parameterize any Fe $K$-shell emission
($< EW$(Fe-$K$) $> = 160\pm30$~eV, 
	$< E_z > = 6.34\pm0.04$~keV
and 
	$\sigma$($EW$(Fe-$K$)) $\lesssim 90$~eV, 
	$\sigma(E_z) < 0.09$~keV 
at 90\% confidence).

All 9 RQQs in which we detect Fe $K$-shell emission and 
all the objects reported in Nandra et al (1997b) 
have $L$(2--10~keV)$<$few$\times 10^{44}\ {\rm erg\ s^{-1}}$. 
Based on a hetergeneous sample of Seyfert 1 Galaxies and 
(RL \& RQ) quasars observed during the first 2 years of the 
{\it ASCA}\ mission, 
Nandra et al (1997c) have presented evidence of a change in the 
profile and strength of the Fe $K$-shell emission as a 
function of luminosity. 
They found $EW$(Fe-$K$) decreased with increasing $L$(2--10~keV) 
confirming such a suggestion based on {\it Ginga}\ observations 
(Iwasawa \& Taniguchi 1993).
Nandra et al also found evidence that the blue-wing of the 
line profile was stronger than the red-wing in 
sources with 
$L$(2--10~keV) $\sim 10^{44}$--$10^{46}\ {\rm erg\ s^{-1}}$, 
contrary to case in the lower luminosity sources. 
This can be interpreted as the result of Fe becoming more ionized 
in higher luminosity sources. 

In Fig.~\ref{fig:correlations_xray_feline}a we compare 
$EW$(Fe-$K$) vesus $L$(2--10~keV) for 
the RQQs presented here to a selection of other RQ AGN 
(Nandra et al 1997b; Vaughan et al 1999; Reeves et al 1997; 
Vignali et al 1999).
In Fig.~\ref{fig:correlations_xray_feline}b we show 
$E_z$ vesus 
$L$(2--10~keV) for those sources with detected Fe $K$-shell emission.
No significant trends 
in the Fe $K$-shell emission as a function of $L$(2--10~keV)
are obvious from either plot.
However in Fig.~\ref{fig:fe-profile} we show the mean data/model ratios 
for all the RQQs presented here for different ranges of 
$L$(2--10~keV). 
The plots were created using the SIS data (only) and our preferred model 
for the continuum in each dataset (\S\ref{Sec:0.6-10-observed}). 
The resultant data/model ratios were then corrected to the quasar frame, 
and co-added and rebinned for clarity.
The mean profile for the RQQs with 
$L$(2--10~keV) $<10^{44}\ {\rm erg\ s^{-1}}$ is very similar in 
its intensity and profile to that often seen in 
Seyfert 1 Galaxies (eg cf Fig~1 in Nandra et al 1997c), 
with a narrow peak at 6.4~keV (corresponding to Fe{\sc i}--{\sc xii})
and a broader base.
The presence of a red-wing at energies below 6.4~keV is of 
particular interest as this is commonly interpreted as 
the result of relativistic effects close to the putative 
supermassive black hole (eg. Tanaka et al 1995).
Unfortunately the low signal to noise ratios prevents an meaningful 
constraints being placed on such 'diskline' models 
using the individual datasets not already studied by Nandra et al (1997b).

We find no significant change in the intensity or profile for the RQQs with 
	$L$(2--10~keV) $= 10^{44-44.5}\ {\rm erg\ s^{-1}}$ 
although it should be noted that there are only 3 objects in this sub-set.
In \S\ref{Sec:Fe-band} we found the inclusion of Fe $K$-shell emission did not 
significantly improve the fits to any of the RQQs with 
	$L$(2--10~keV) $>10^{44.5}\ {\rm erg\ s^{-1}}$. 
However Fig.~\ref{fig:fe-profile} reveals evidence for such emission in the 
6 datasets with 
	$L$(2--10~keV) $= 10^{44.5-45}\ {\rm erg\ s^{-1}}$. 
Furthermore the peak in the mean profile is consistent with being at a higher 
energy. 
However we note that the limits on the intensity of the line 
in one of the datasets in this sub-set, PG~0953+415
	($EW$(Fe-$K$) $<60$~eV), 
is incompatible with that of the mean profile
and should serve as a reminder that there will undoubtedly be 
object-to-object variations within any luminosity class.
Finally the mean data/model ratio for the 4 datasets with 
	$L$(2--10~keV) $>10^{45}\ {\rm erg\ s^{-1}}$
is consistent with unity at all energies. 
Thus the results for the RQQs presented here confirm the results of 
Iwasawa \& Taniguchi (1993) and Nandra et al (1997c).

\section{SUMMARY \& CONCLUSIONS}
\label{Sec:conclusions}

We have presented the results from 25 {\it ASCA}\ observations of 24 RQQs
from the PG sample. Our main findings are summarized as follows:
\begin{enumerate} 
\item	
All but two objects (PG~0043+039 \& PG~1700+518) were detected.

\item	
We found 15 of the datasets to have a sufficient
number of counts that we were able to perform an analysis of the variability
characteristics.
Significant variability was found in 10 of these datasets 
(9 RQQs) at 
$\gtrsim 90$\% confidence.
Visual inspection of the light curves suggests 3 additional sources
also exhibit variability, but the signal to noise ratios 
are insufficient to allow statistical testing and/or it being quantified.
The variability characteristics of 
these RQQs were consistent with the trend found by 
Nandra et al (1997a) and Turner et al (1999) 
whereby sources of higher luminosity sources exhibit
a lower degree of variability. 

\item
We found 
a power law offers an acceptable description of the time-averaged spectra
in the 2--10~keV band for all but 1 of the detected sources
(PG~1411+442).
The best-fitting values of photon index vary from object to object 
over the range $1.5 \lesssim \Gamma_{2-10} \lesssim 3$.
The distribution of $\Gamma_{2-10}$ is similar to that observed in other 
RQ AGN, and seems unrelated to X-ray luminosity.
For the RQQs presented here we find a mean 
$< \Gamma_{2-10} > \simeq 2$ but with a significant 
dispersion $\sigma(\Gamma_{2-10}) \simeq 0.25$.

\item
No single model adequately describes the full 0.6--10~keV (observed-frame) 
continuum of all the RQQs.
We found the continua of 14 RQQs could be
adequately described by a single power law
(11 RQQs)
or a power law with only very subtle deviations 
(3 RQQs). 
The preferred model for PG~1411+442 has 
$\sim$2\% of the continuum escaping 
the nucleus directly, 
but the remainder attenuated by a substantial 
column density of material.
All the remaining objects were found to have 
have convex spectra (flattening as one moves to higher energies).

\item	We find evidence for 
	a flattening of the continuum at energies $\gtrsim 5$~keV
	in 5 RQQs. 
	Although
	there are physical explanations for such a spectral component, 
	the {\it ASCA}\ data are unable to exclude 
	other possibilities. 

\item
Six of the 10 datasets (5 out of 9 RQQs) 
with convex spectra contain
a 'soft excess' at energies $\lesssim$1~keV.
This component was {\it parameterized} by 
a Gaussian emission feature in the case of 3 datasets 
(3 RQQs)
and as a second power law for the remainder. 
A sixth RQQ, 
PG~1148+549, probably contains such a component just below 
the {\it ASCA}\ bandpass. 
There is no universal soft X-ray spectrum for RQQs.
For instance, out of 14 objects with $z \lesssim 0.25$
(such that there is a possibility of detecting an upturn in
the spectrum $\lesssim$1~keV), only 
5 contain a 'soft excess'.

\item
The spectrum of PG~1244+026 
contains a rather narrow emission feature
centered at an energy $\sim 1$~keV.
A similar feature is present in the
spectra of at least two Narrow-line Seyfert 1 Galaxies.

\item
We find evidence for 
absorption by ionized material in 6 RQQs. 
However we are only confident of its presence in 2 objects:
PG~1114+445 (with $N_{H,z} \sim 2\times10^{22}\ {\rm cm^{-2}}$)
and PG~1501+106 ($N_{H,z} \sim 2\times10^{21}\ {\rm cm^{-2}}$).
Three of the remaining RQQs 
have rather complex spectra raising the possibility that the 
presence of ionized absorber is an artifact of incorrectly 
modelling the soft X-ray spectrum.
The derived parameters of the ionized material in 
PG~1322+659 rather extreme, but require confirmation.
There are indications 
that the detection-rate of ionized material is lower
in RQQs compared to Seyfert galaxies.
However it is premature to 
make any definitive statements.

\item
Our preferred model for PG~1411+442, in which a substantial column
density ($N_{H,z} \sim 2\times10^{23}\ {\rm cm^{-2}}$) obscures
$\sim$98~\% of the continuum, 
and offers an explanation of the unusual optical--to--X-ray spectrum
observed for this object (Paper~II).

\item
We detect significant Fe $K$-shell emission 
in 10 RQQs (although there is 
some doubt in the case of 2 of these),
all of which have 
$L$(2--10~keV)$<$few$\times 10^{44}\ {\rm erg\ s^{-1}}$.
We find the strength and energy of the emission 
to be consistent with that seen in other RQ AGN.
However construction of the mean data/model ratios
using all 25 datasets, but for different ranges of $L$(2--10~keV), 
reveals a trend whereby the profile and strength of the 
Fe $K$-shell emission changes as a
function of luminosity, consistent with previous suggestions 
(Iwasawa \& Taniguchi 1993; Nandra et al 1997c).

\end{enumerate}

\acknowledgements

We thank Fabrizio Fiore for useful discussions and an anonymous referee
for suggested improvements to the paper.
We acknowledge the financial support of the
Universities Space Research Association (IMG, KN, TY),
LTSA grant NAG 5-7385 (TJT), 
and the Israel Science Foundation (HN).
This research has made use of the Simbad database, operated by 
the Centre de Donn\'{e}es astronimiques de Strasbourg (CDS);
the {\tt VizieR} Service for Astronomical Catalogues,
developed by CDS and ESA/ESRIN;
and of data obtained through the HEASARC on-line service, provided by
NASA/GSFC.

\appendix
\section{APPENDIX: SERENDIPTOUS SOURCES WITHIN THE FIELDS OF VIEW}

Besides the target, a number of additional sources were 
evident with the fields of view (fov) of the {\it ASCA}\ detectors. 
The useable portion of the GIS detectors is circular
with a radius of $\sim$20~arcmin. 
The celestial coordinates 
and GIS2 count rates of the serendipitous sources 
detected are listed in Table~\ref{tab:asca_serendip}, along with 
a tentative identification and any references discussing 
X-ray observations.
The FWHM of the point spread function (psf) 
of the GIS detectors is $\sim$3~arcmin at 1~keV, and the 
uncertainties on the positions listed in Table~\ref{tab:asca_serendip}
are $\sim 20$~arcsec.
Each SIS detector consists of a 2$\times$2 array of 
CCDs, with each CCD having a $\sim$11.2$\times$11.2~arcmin
fov.
The psf of the SIS detectors is dominated by that of the 
focussing opticals, consisting of a relatively sharp core 
(FWHM $\sim 0.8$~arcmin), but broad, azimuthally-dependent 
wings (extending out to several arcmin). 
All 4 CCDs on each SIS were active only for one of the observations 
reported here (Table~\ref{tab:asca_obs_log}).
The subsequent reduction in the SIS fov for 
the majority of the observations resulted in 
an accurate position and count rate for only one of the 
serendipitous sources:
that in the PG~1247+267 field, with an SIS0 count rate 
of $3.2\pm0.3\times10^{-3}\ {\rm ct\ s^{-1}}$ in the 
0.6--10~keV band (observer's frame).

\newpage
\typeout{REFS}

\clearpage
\typeout{Figures}

\clearpage
\begin{figure}
\thispagestyle{empty}
\plotfiddle{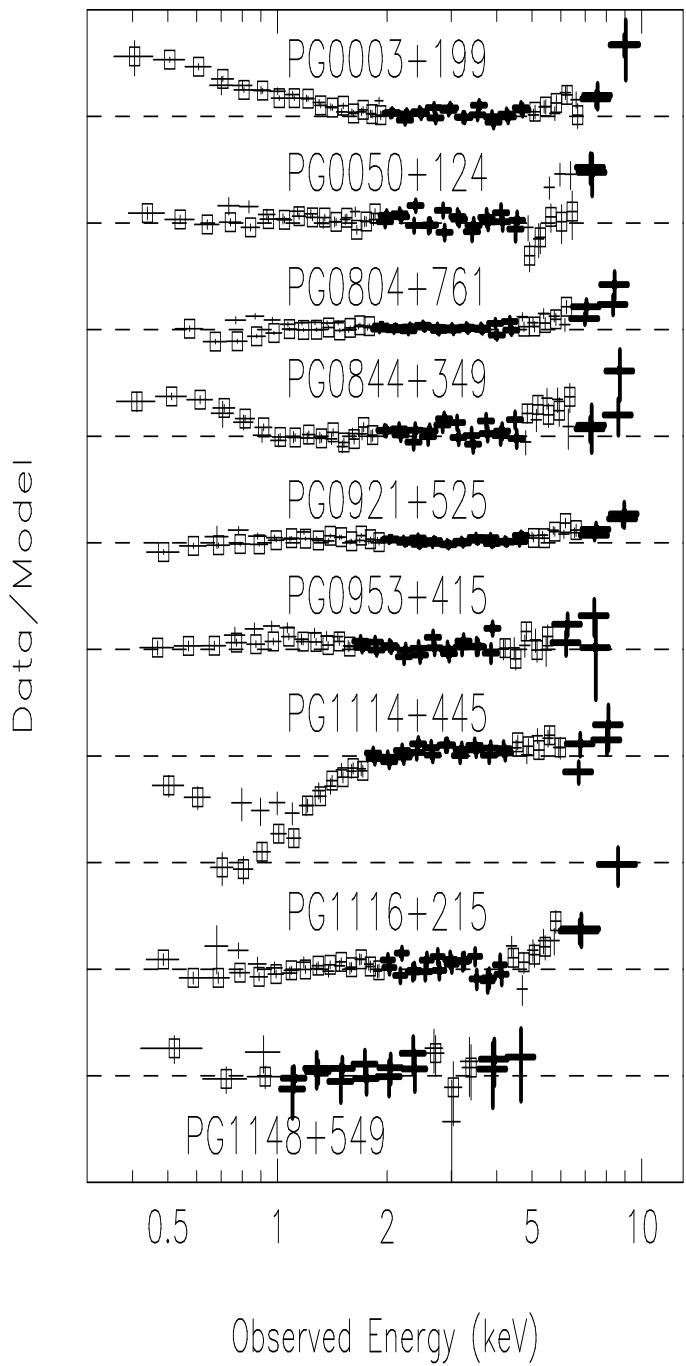}{13cm}{0}{50}{90}{-300}{-200}
\plotfiddle{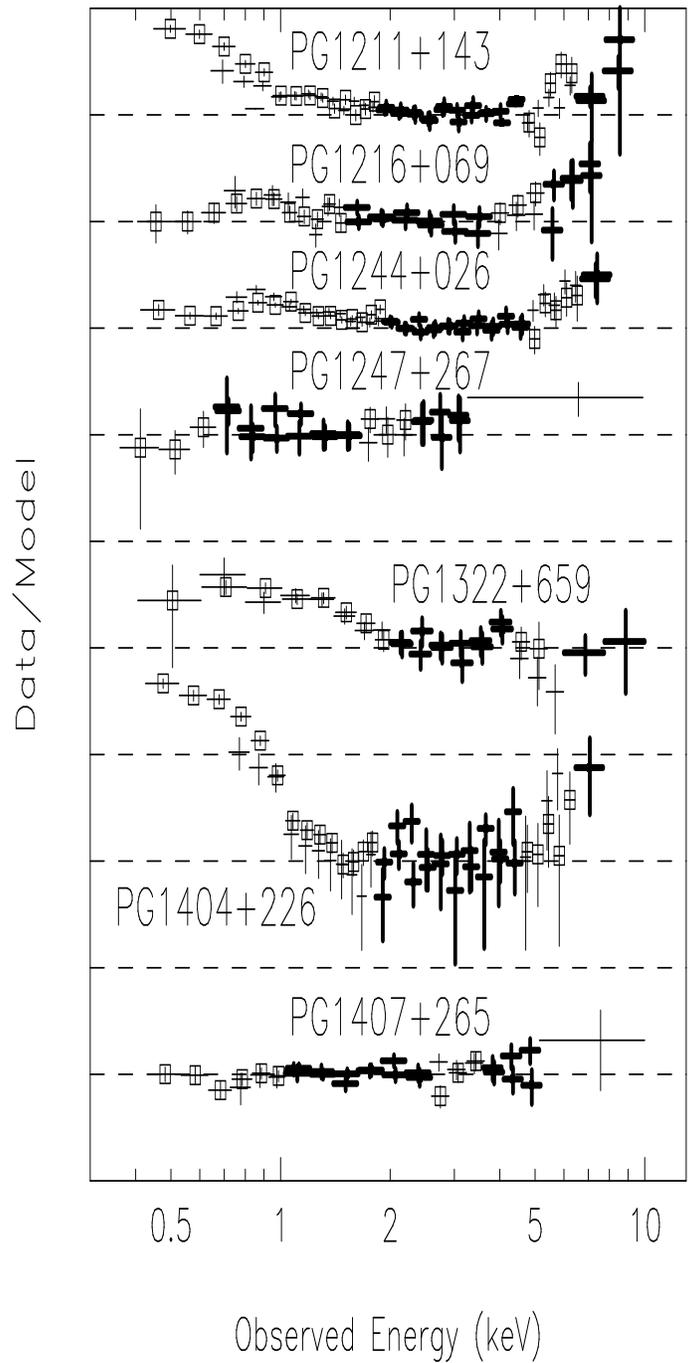}{0cm}{0}{50}{90}{-0}{-165}
\caption{Data/model ratios obtained 
when the best-fitting power law to the 2--10~keV 
continuum (in quasar's frame; data shown in bold) is extrapolated through the 
rest of the {\it ASCA}\ data.
The data are the averages of each instrument pair (with the SIS data 
denoted by the boxes), and have been rebinned in energy-space for clarity.
The y-axis is logarithmic, and the dashed lines separated by
a factor of four.
The model provides an 
adequate description of the data fitted in all cases 
except PG~1411+442.
\label{fig:210_ratio}}
\end{figure}

\clearpage
\begin{figure}
\addtocounter{figure}{-1}
\thispagestyle{empty}
\plotfiddle{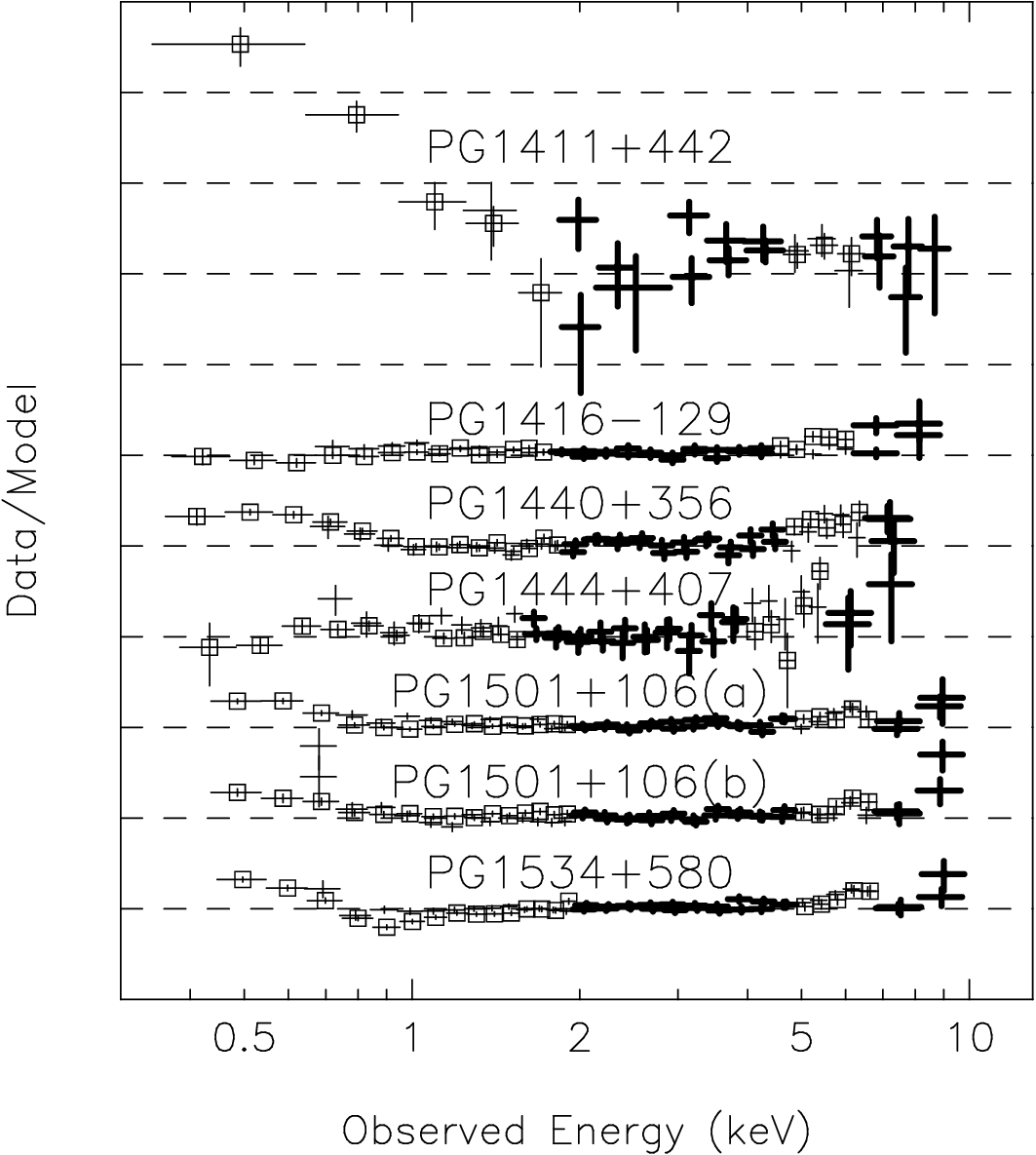}{13cm}{0}{50}{90}{-300}{-200}
\plotfiddle{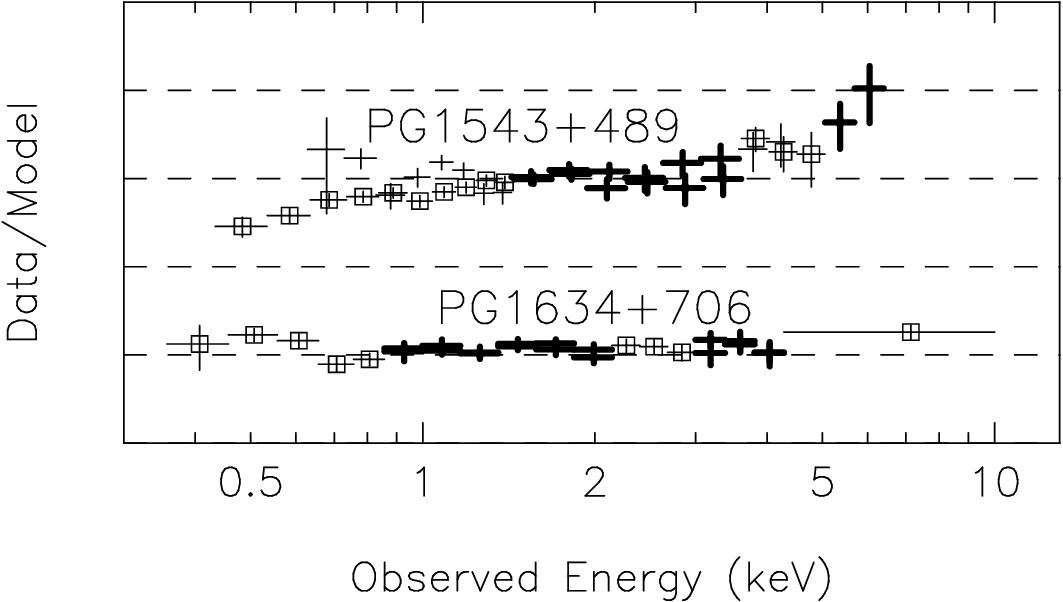}{0cm}{0}{50}{90}{-0}{-165}
\caption{\it continued}
\end{figure}

\clearpage
\begin{figure}
\thispagestyle{empty}
\plotfiddle{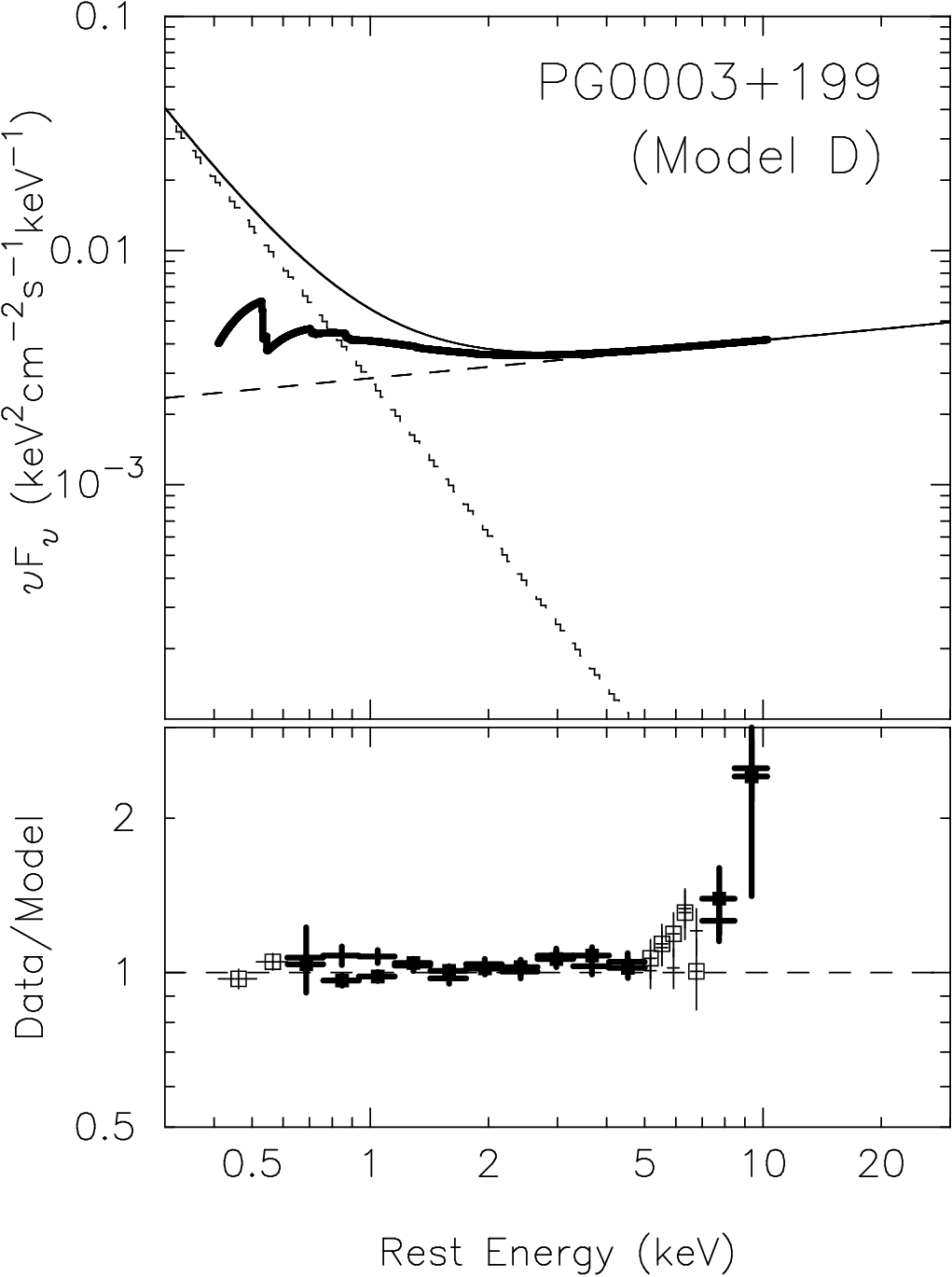}{2cm}{0}{45}{20}{-250}{0}
\plotfiddle{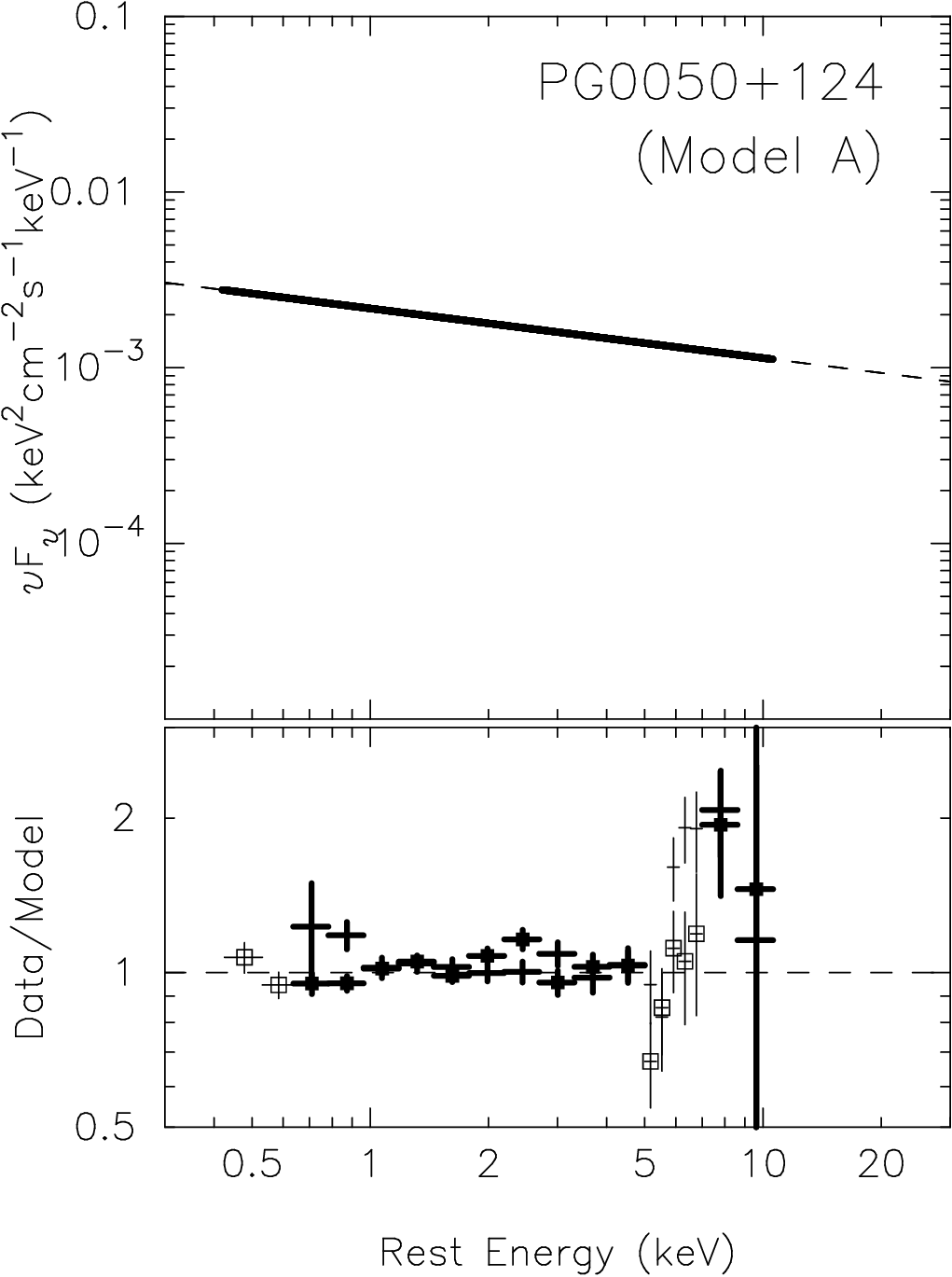}{2cm}{0}{45}{20}{-0}{+75}
\plotfiddle{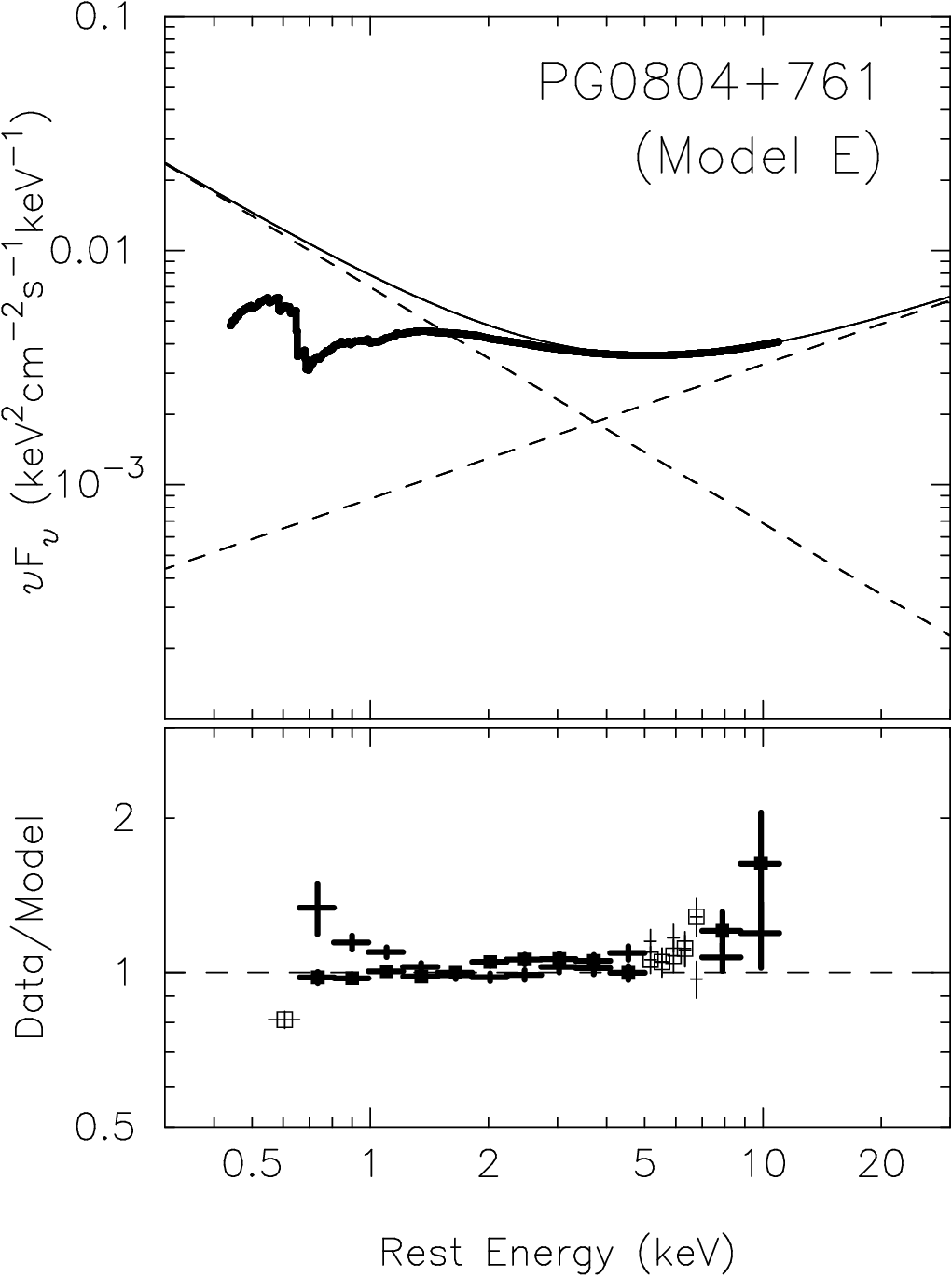}{2cm}{0}{45}{20}{-250}{0}
\plotfiddle{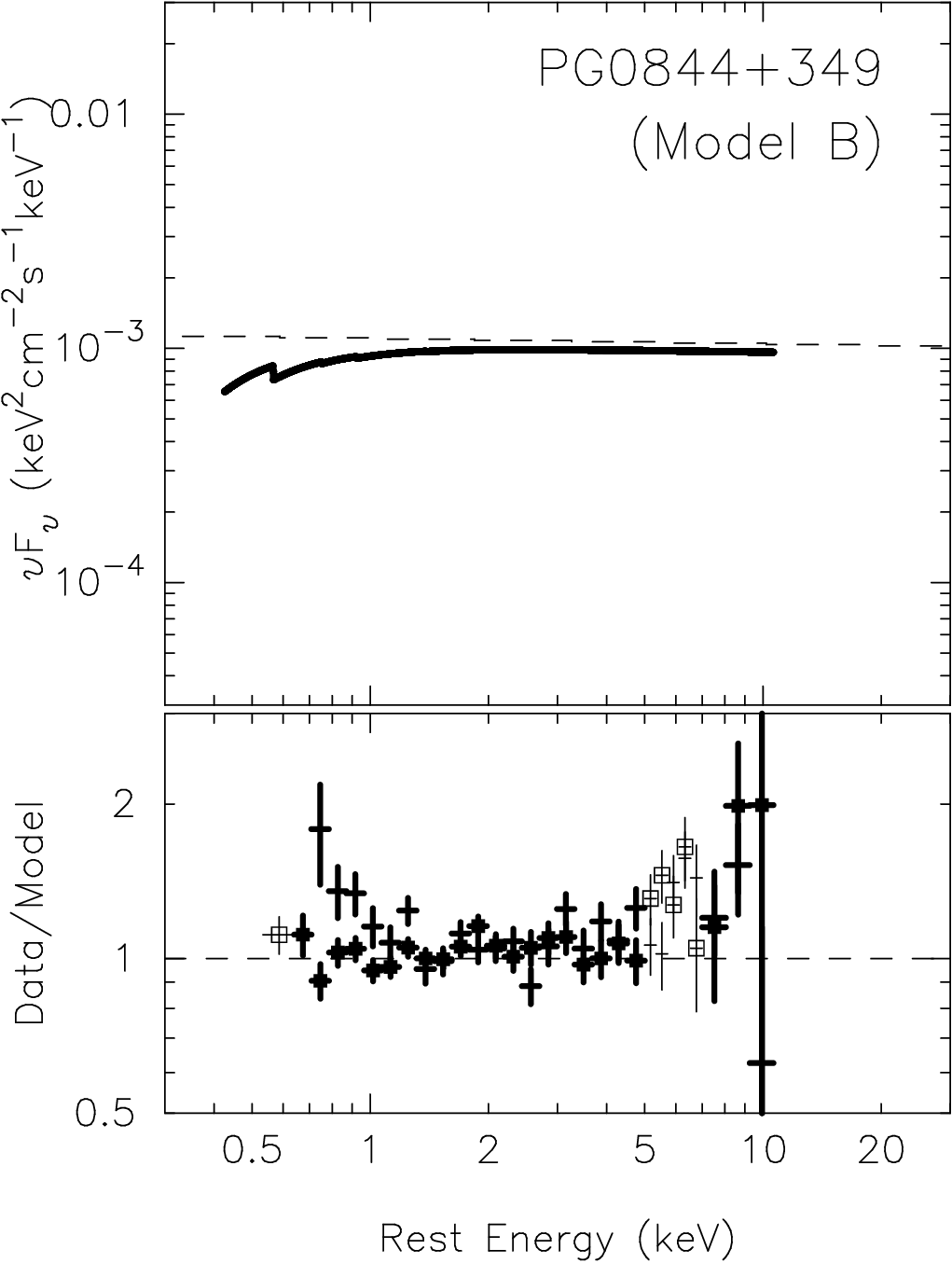}{2cm}{0}{45}{20}{-0}{+75}
\plotfiddle{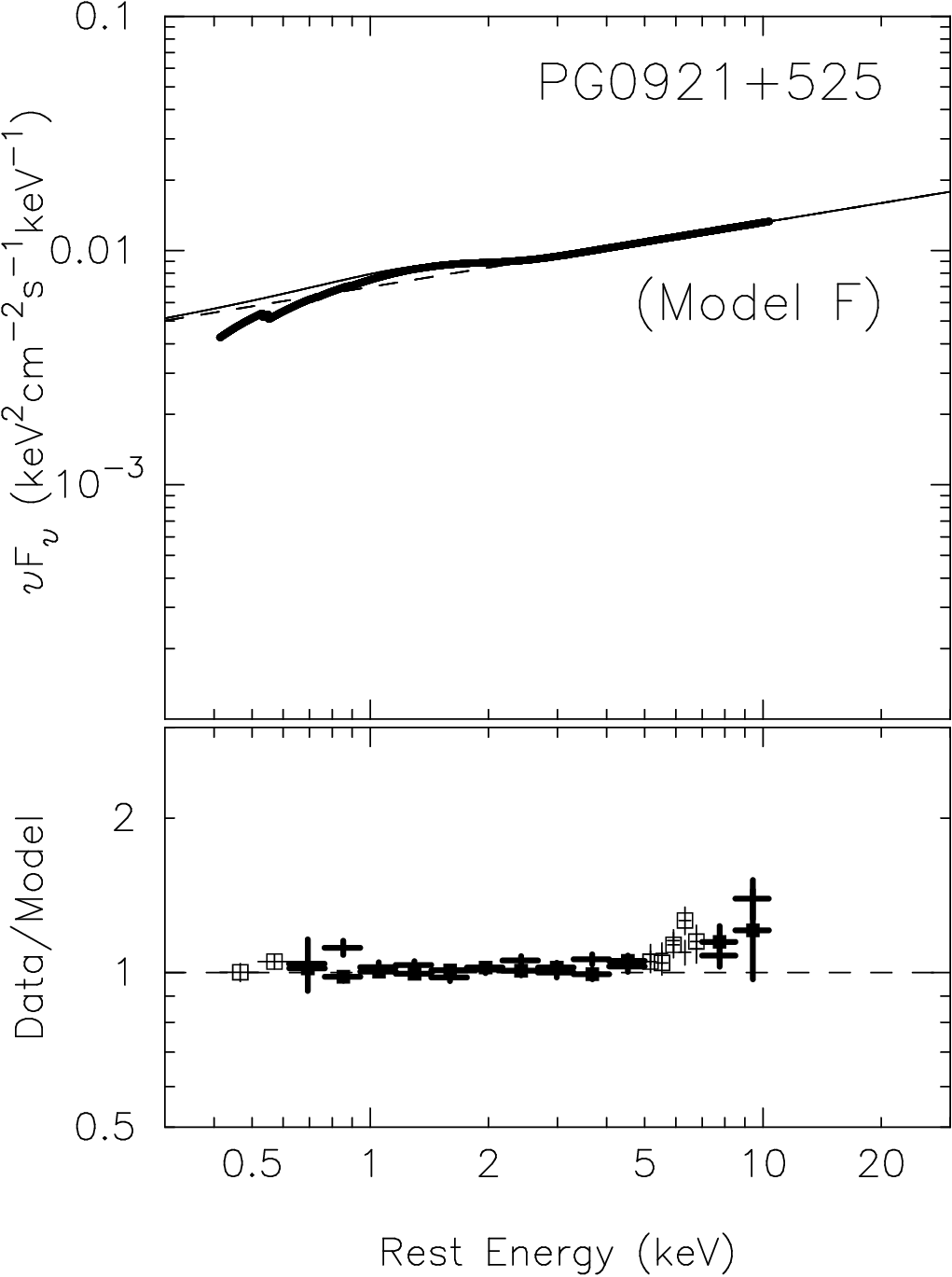}{2cm}{0}{45}{20}{-250}{0}
\plotfiddle{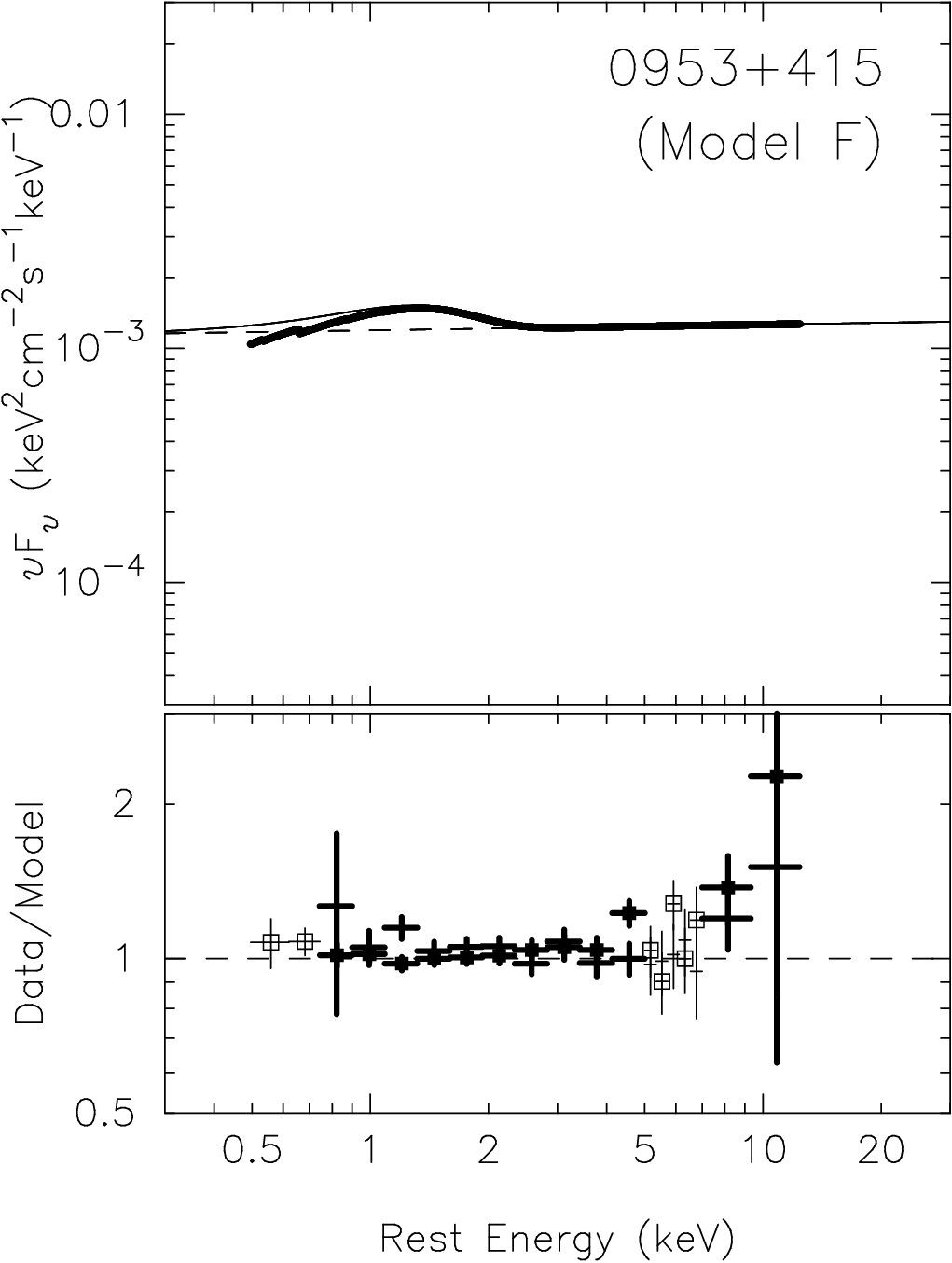}{2cm}{0}{45}{20}{-0}{+75}
\plotfiddle{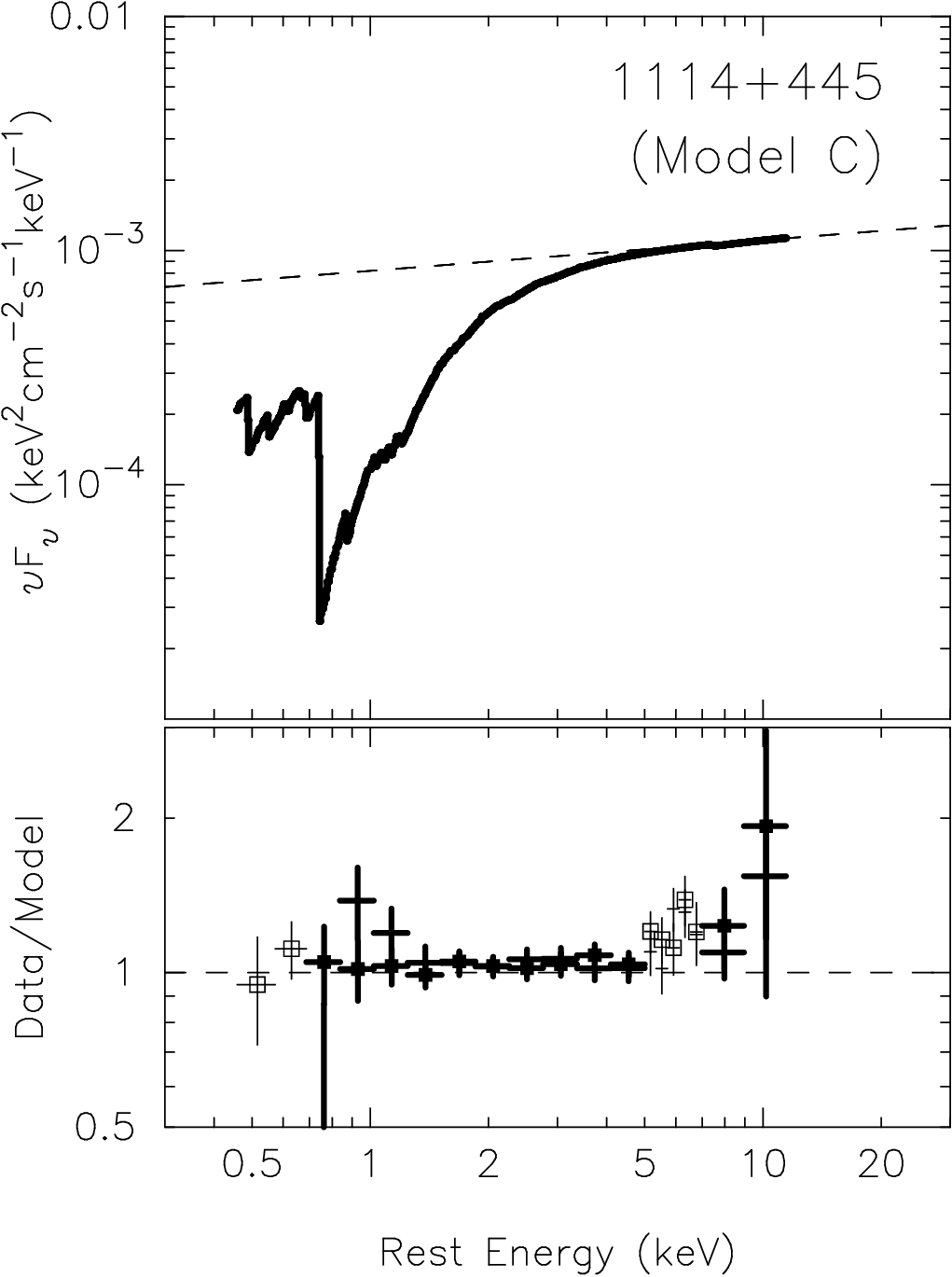}{0cm}{0}{45}{20}{-250}{-50}
\plotfiddle{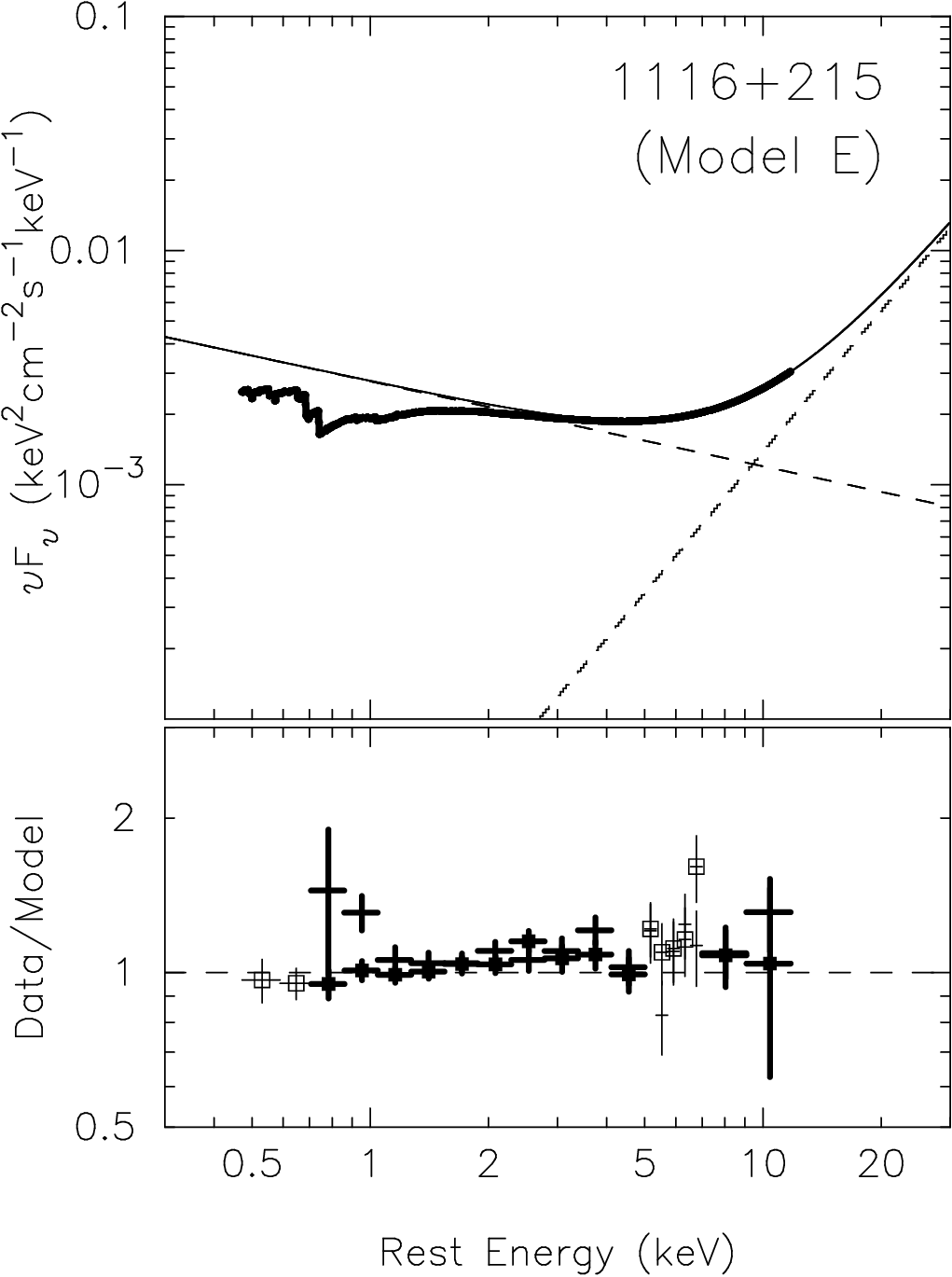}{0cm}{0}{45}{20}{-0}{-20}
\caption{The best-fitting continua (excluding the Fe $K$-shell band) for each
dataset (see \S4.1.2). In each case the {\it bold line} in the upper panel
shows the best-fitting model over the {\it ASCA} bandpass after correcting
for Galactic absorption, the {\it dashed line} shows the underlying power law
component(s), and (for models D--F only) the {\it thin solid line} shows total
underlying spectrum. The lower panel shows the mean data/model ratios as for
Fig.2, except here the 0.6--10~keV (observed frame) band is used in the fits,
and the range of the logarithmic y-axis is 0.5--3.0. \label{fig:ufmodelratio}}
\end{figure}
\clearpage

\addtocounter{figure}{-1}
\begin{figure}
\thispagestyle{empty}
\plotfiddle{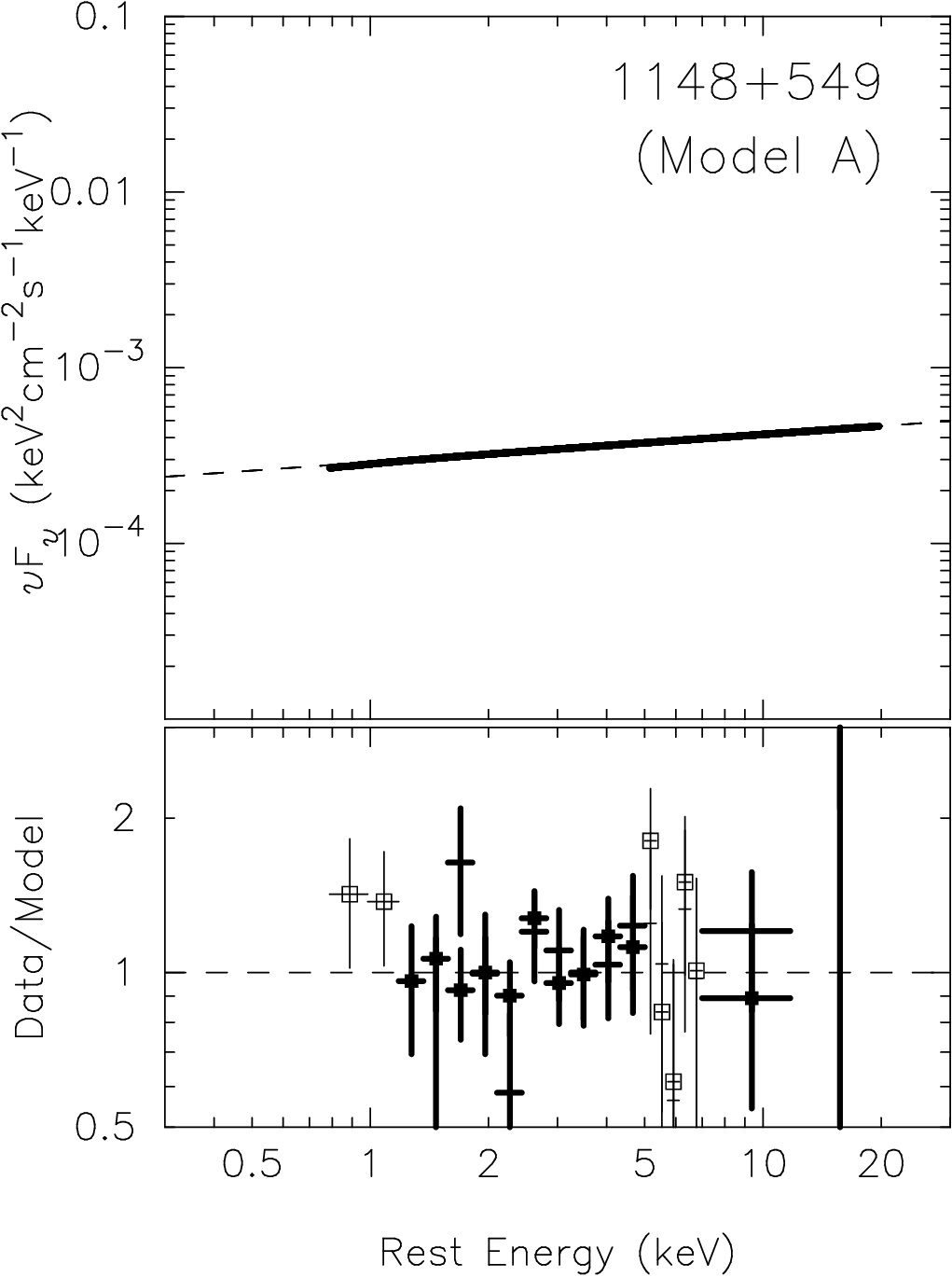}{2cm}{0}{45}{20}{-250}{0}
\plotfiddle{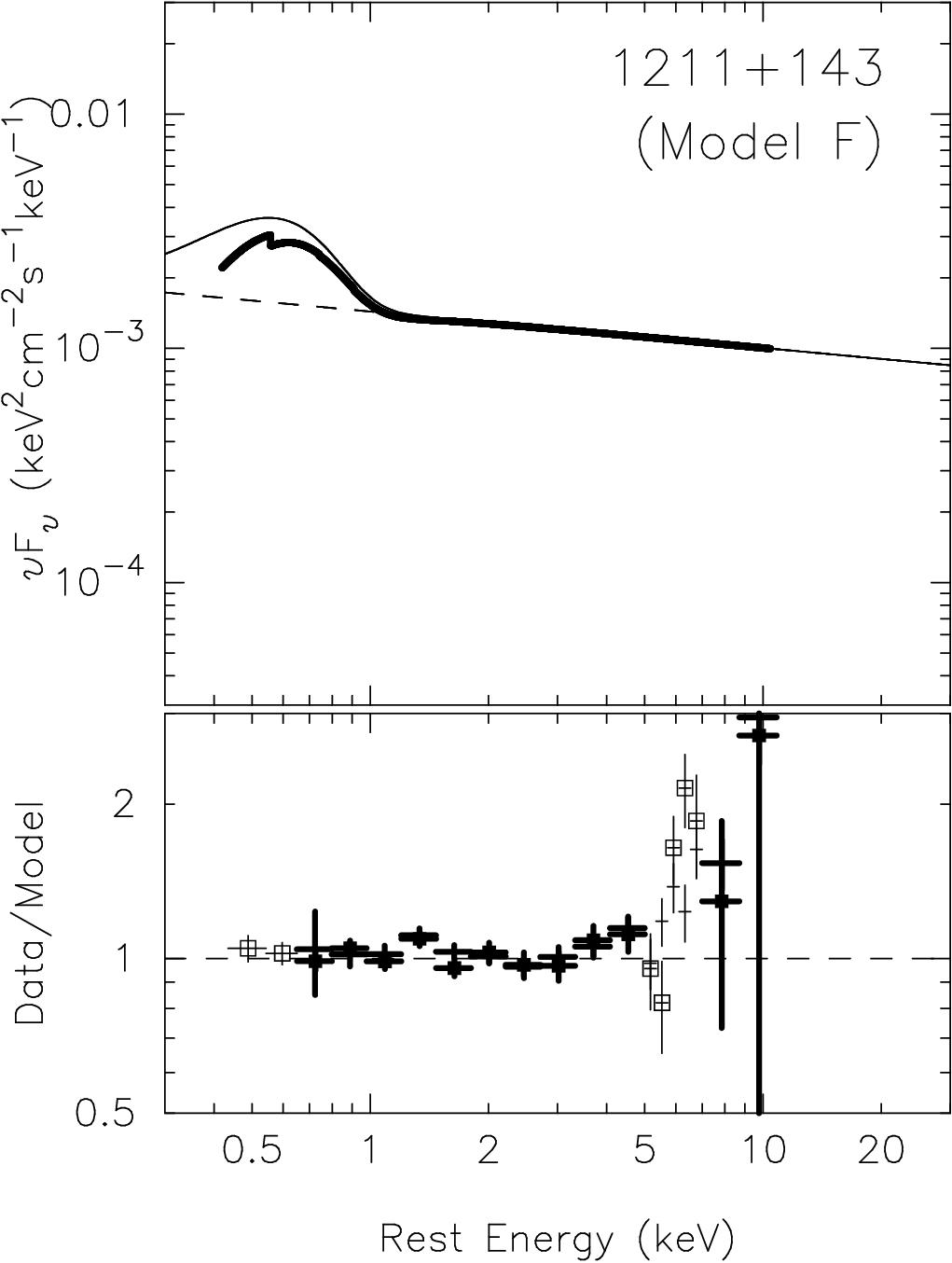}{2cm}{0}{45}{20}{-0}{+75}
\plotfiddle{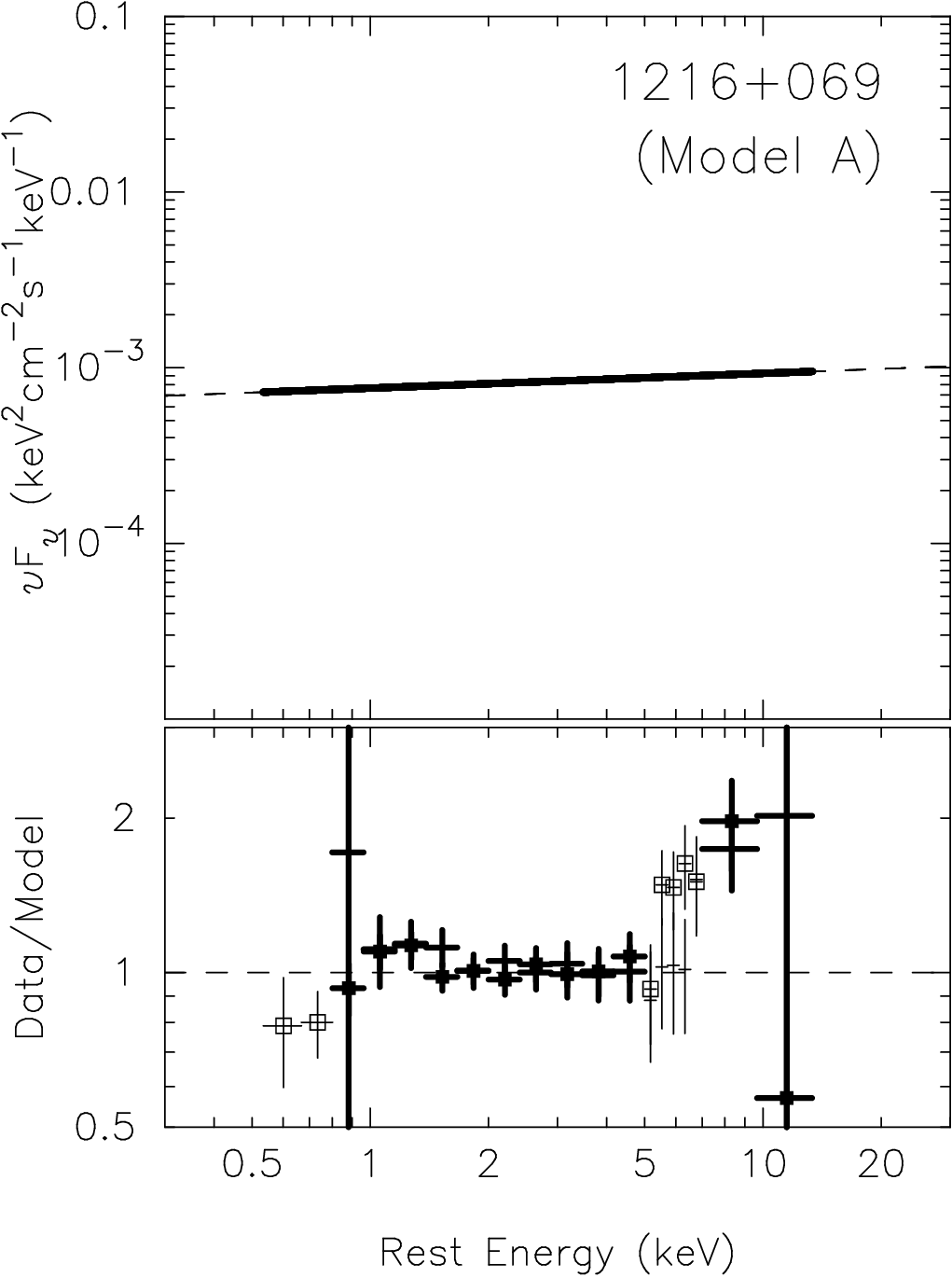}{2cm}{0}{45}{20}{-250}{0}
\plotfiddle{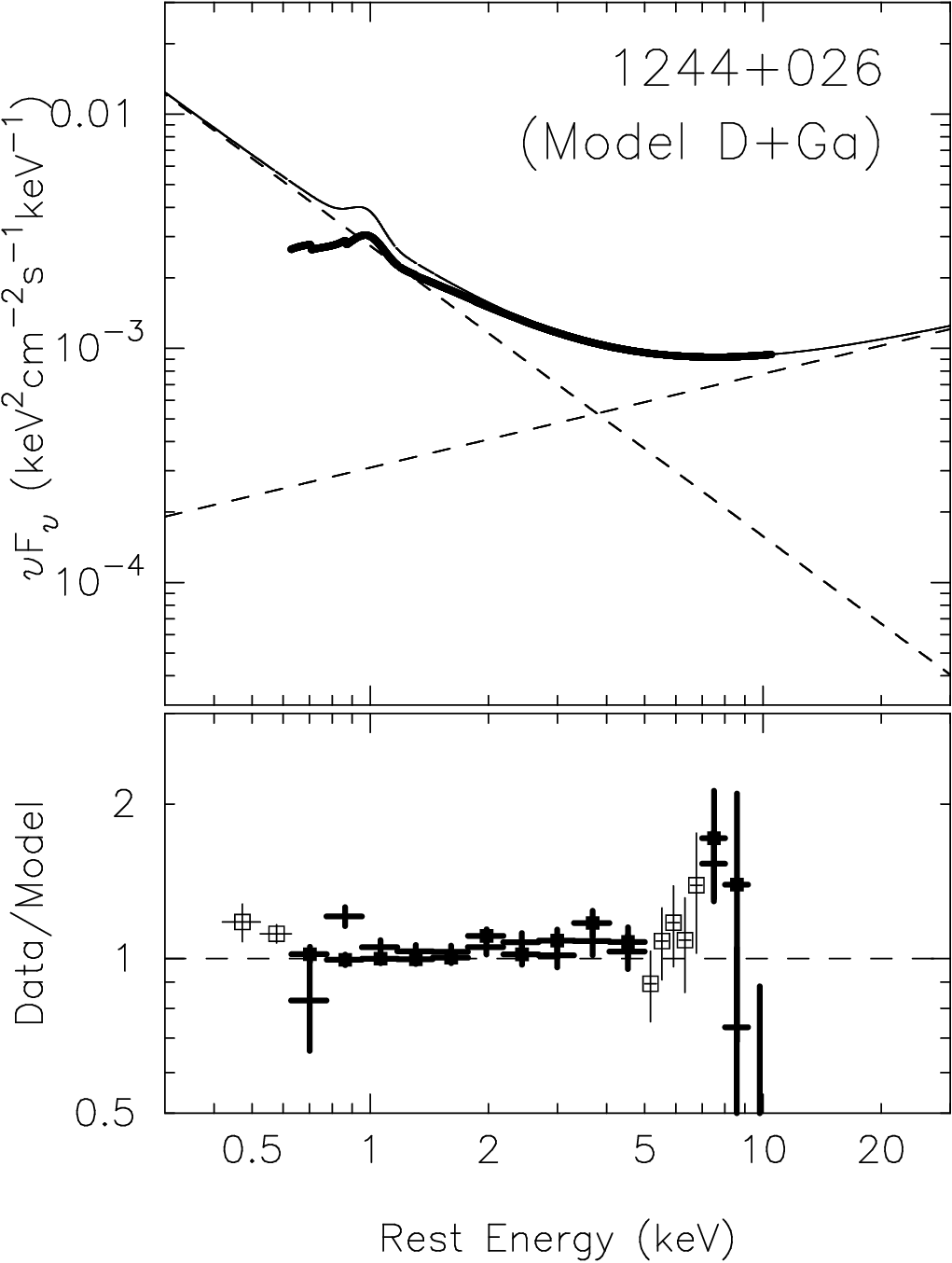}{2cm}{0}{45}{20}{-0}{+75}
\plotfiddle{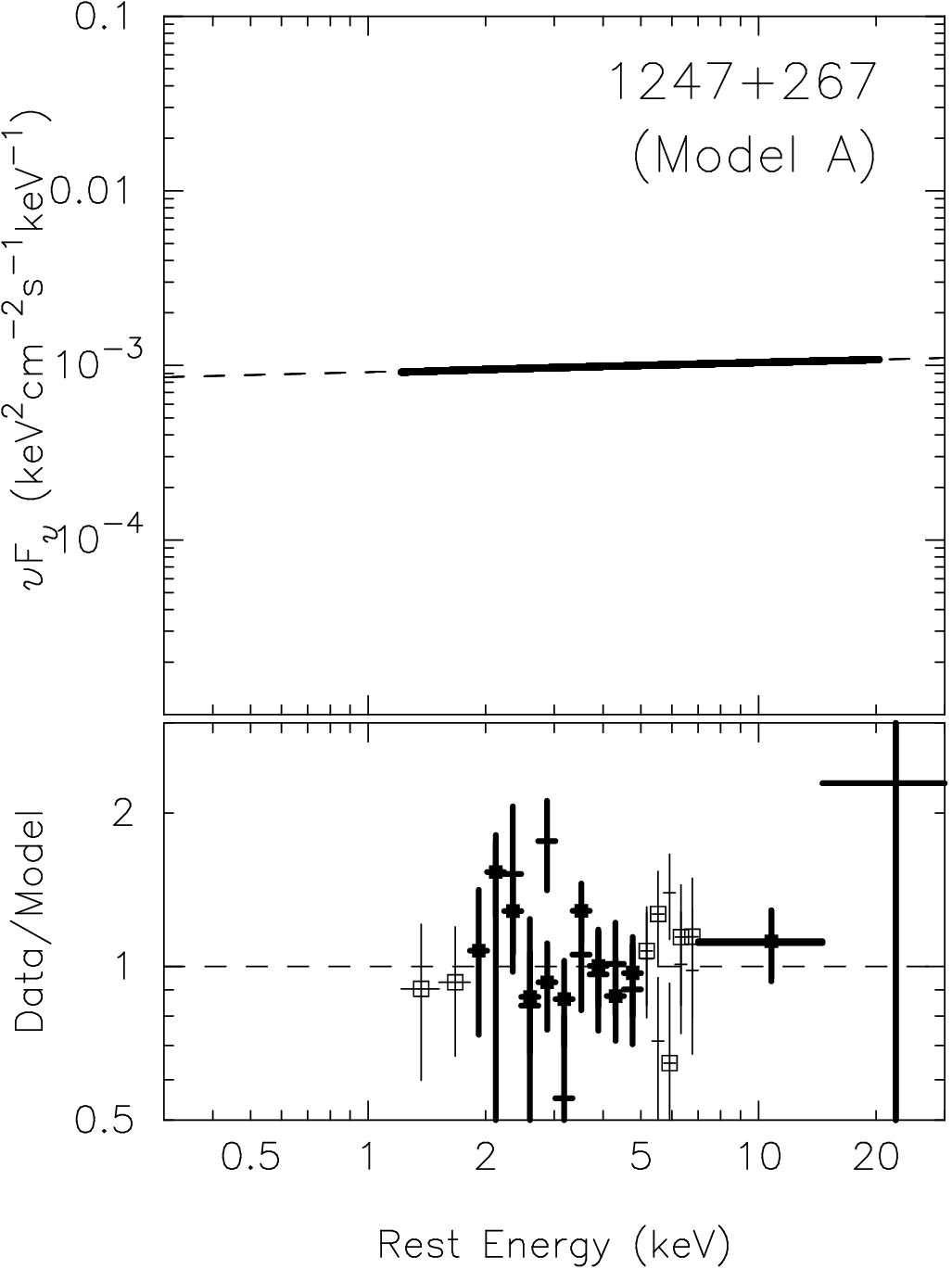}{2cm}{0}{45}{20}{-250}{0}
\plotfiddle{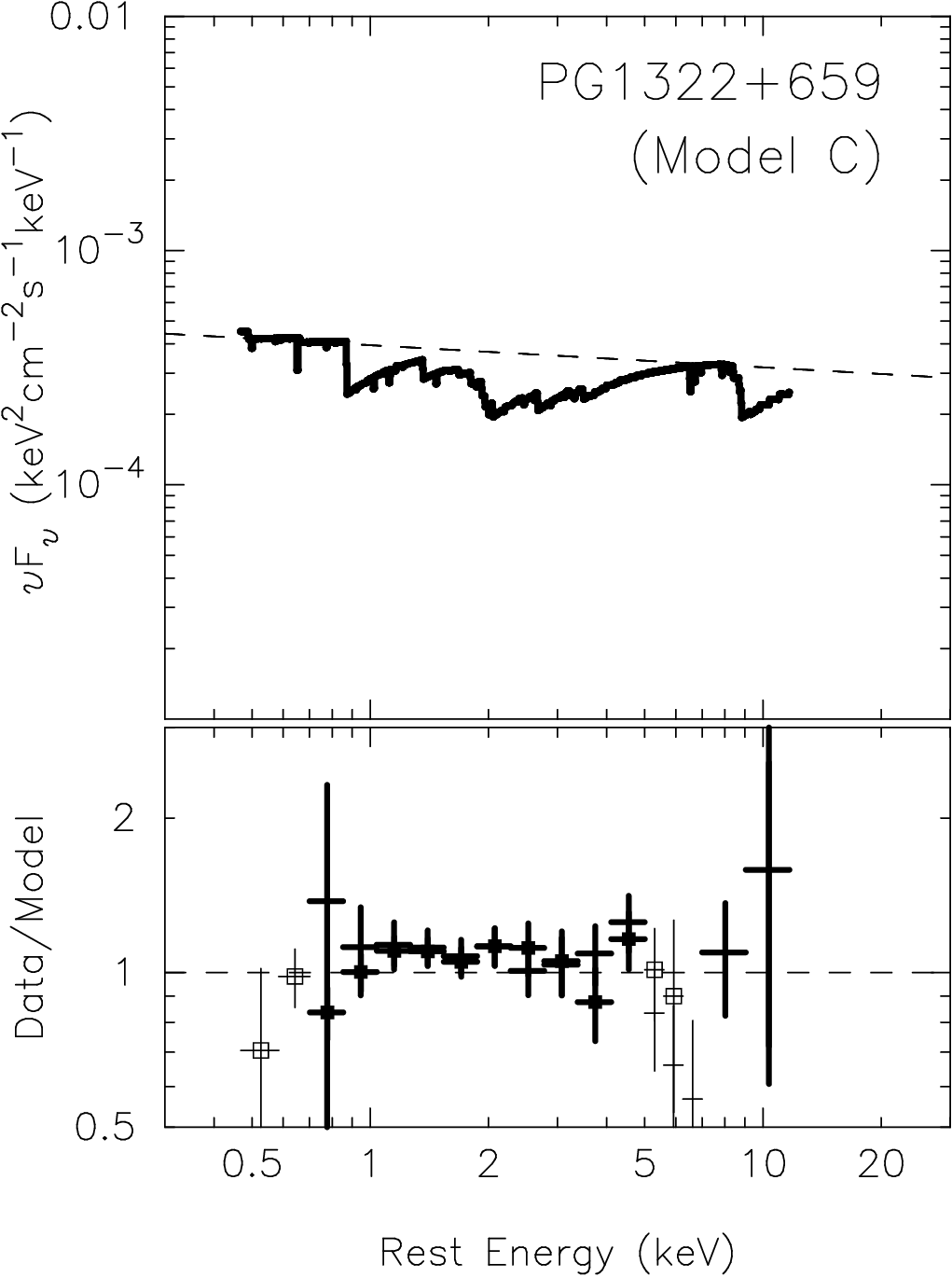}{2cm}{0}{45}{20}{-0}{+75}
\plotfiddle{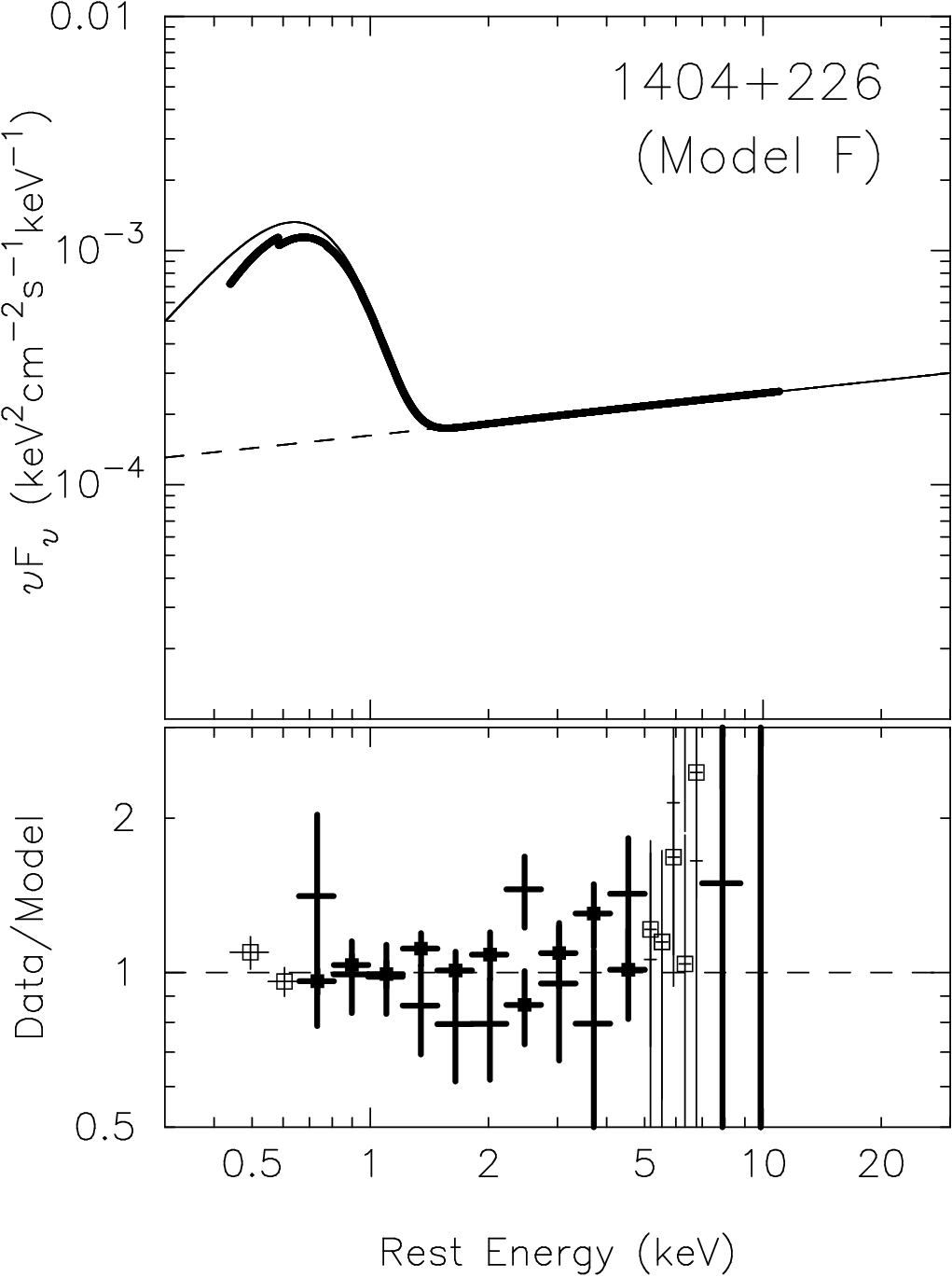}{2cm}{0}{45}{20}{-250}{0}
\plotfiddle{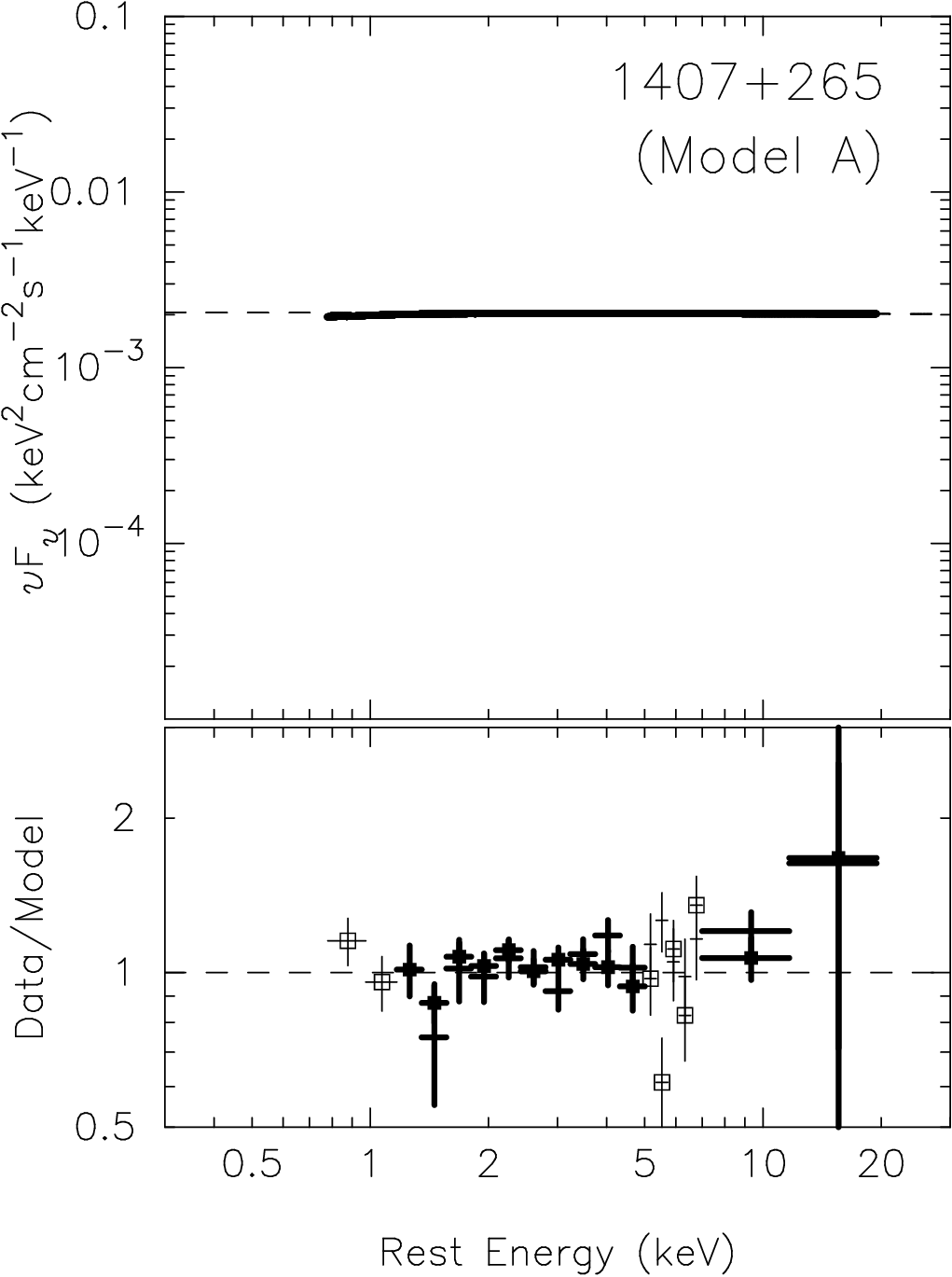}{2cm}{0}{45}{20}{-0}{+75}
\caption{\it continued}
\end{figure}

\addtocounter{figure}{-1}
\begin{figure}
\thispagestyle{empty}
\plotfiddle{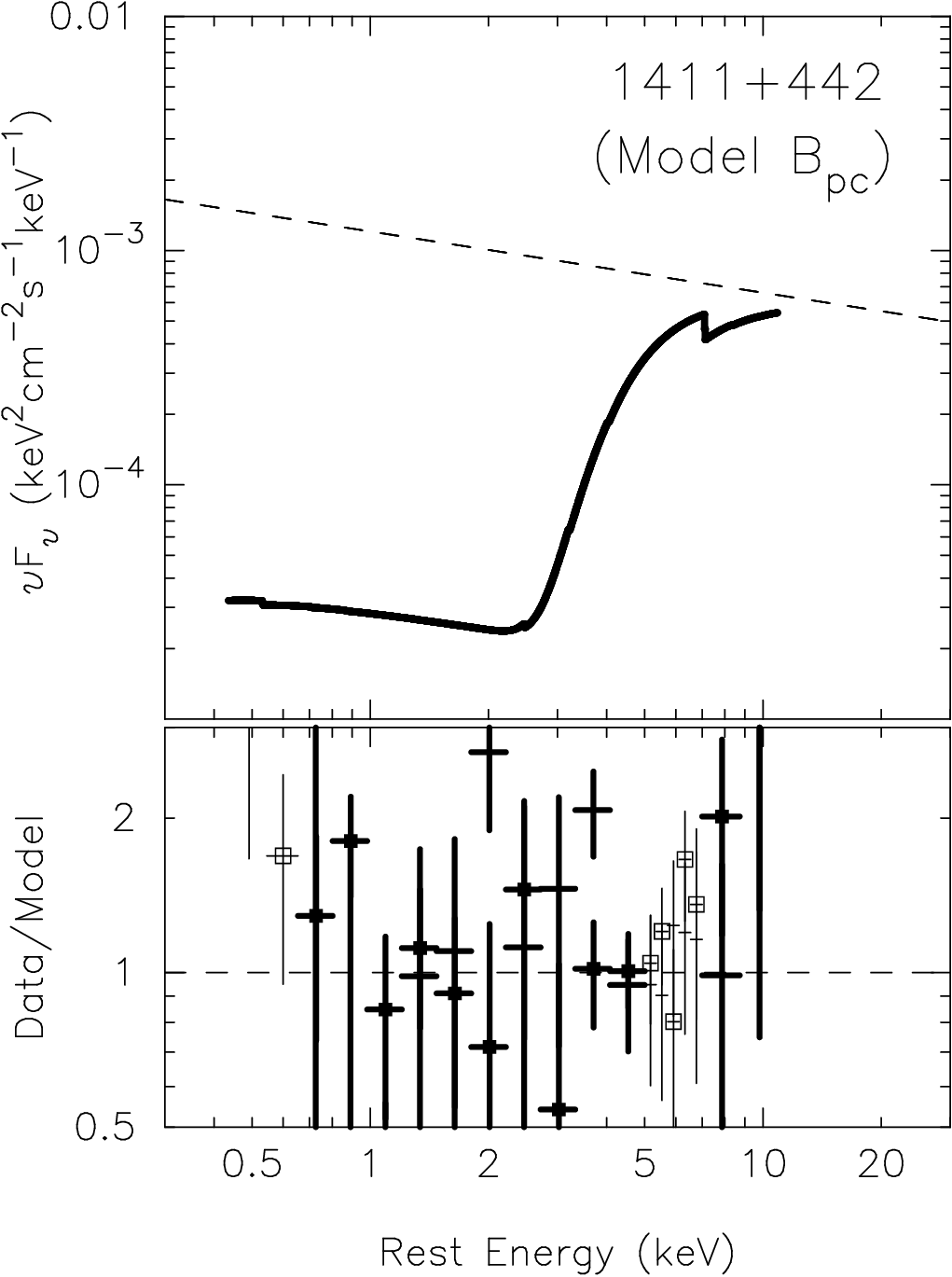}{2cm}{0}{45}{20}{-250}{0}
\plotfiddle{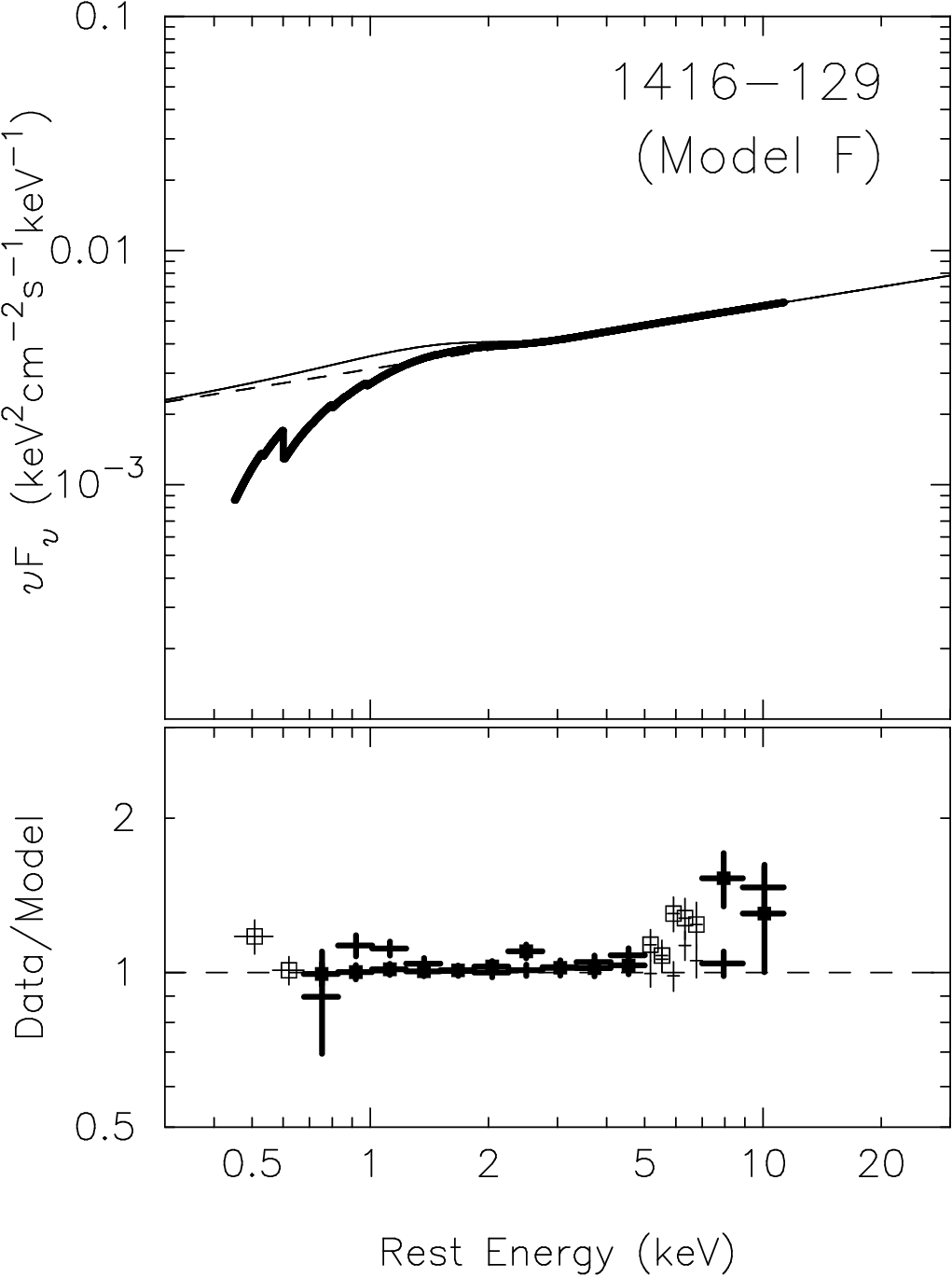}{2cm}{0}{45}{20}{-0}{+75}
\plotfiddle{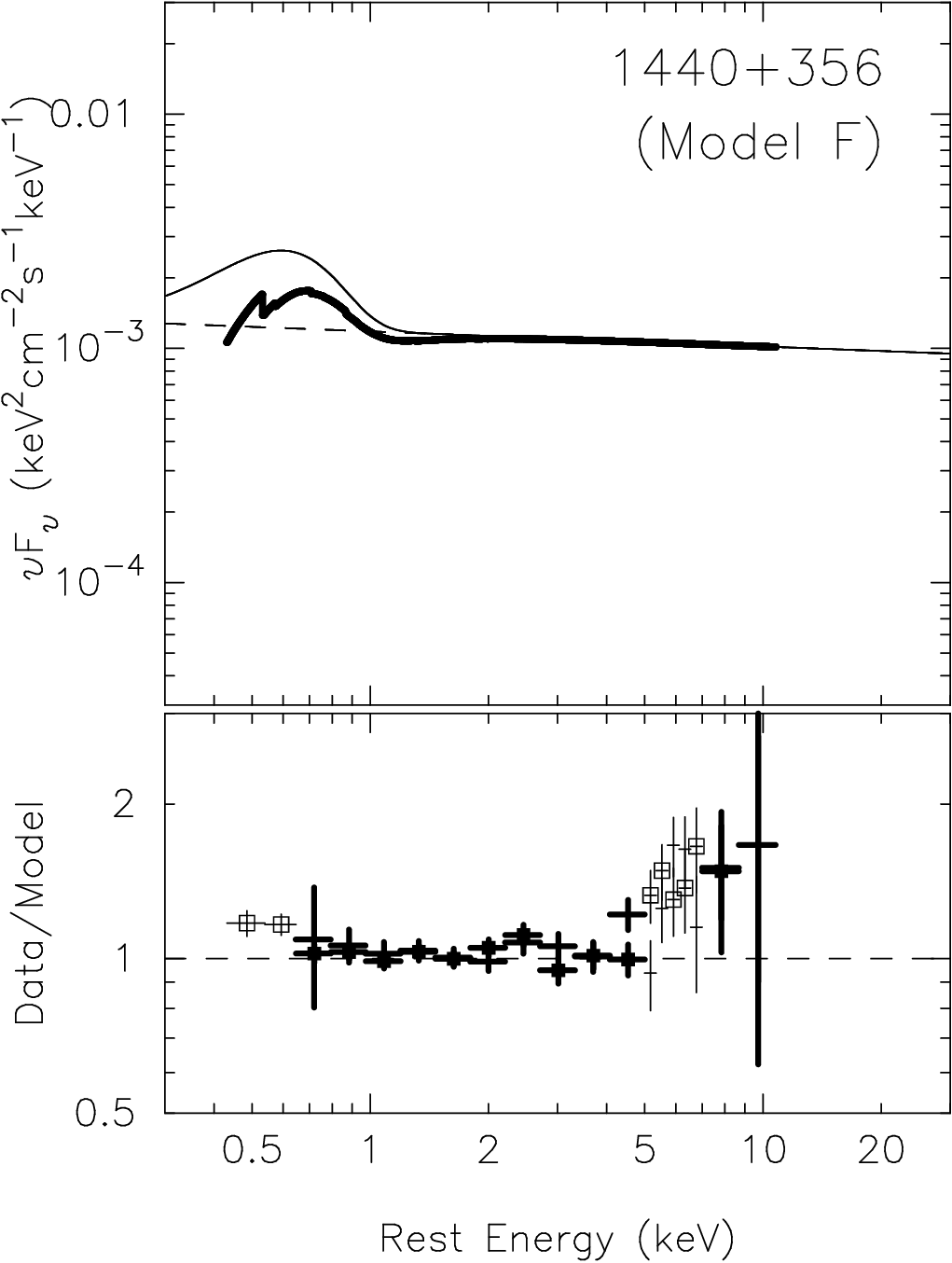}{2cm}{0}{45}{20}{-250}{0}
\plotfiddle{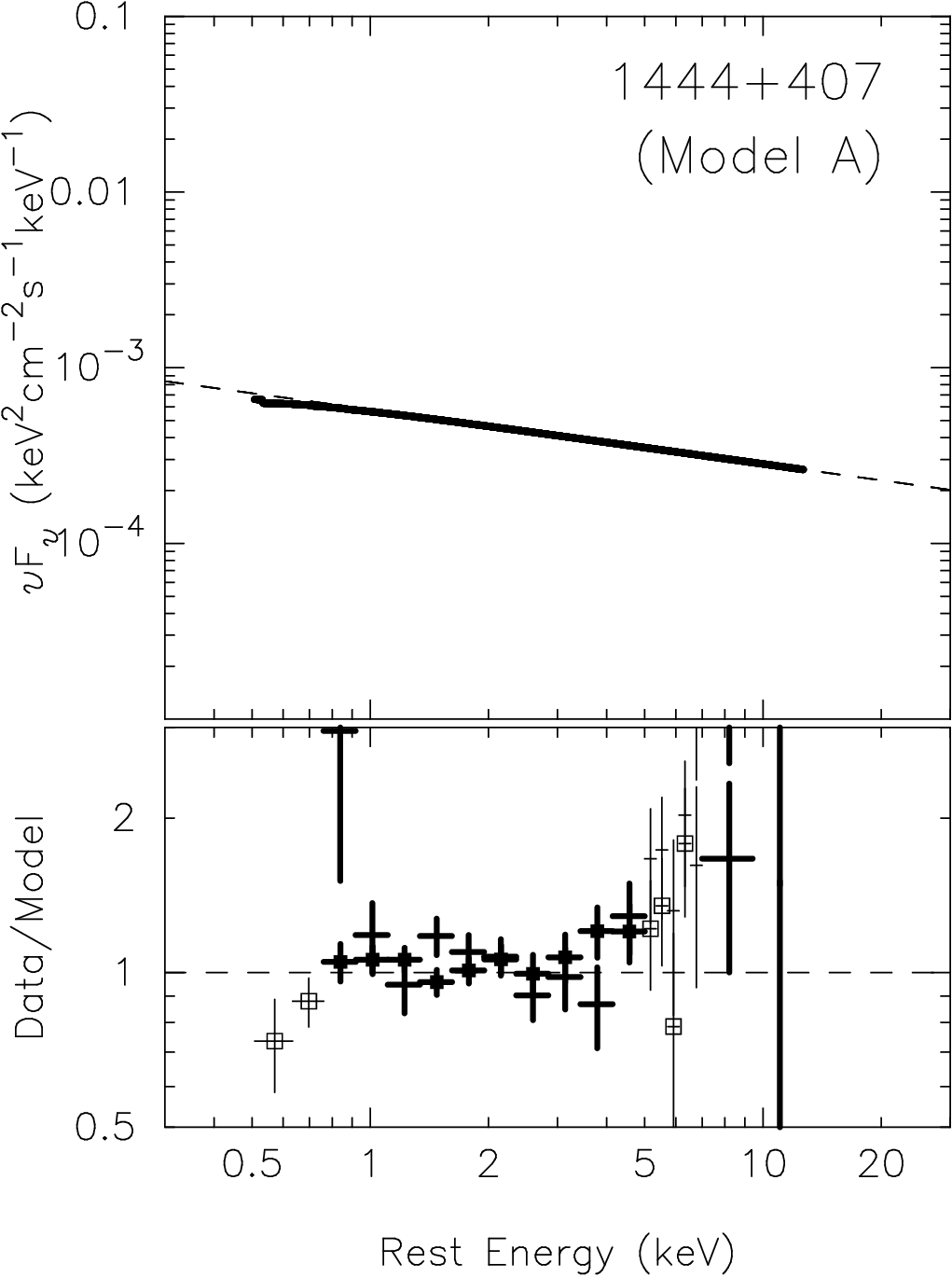}{2cm}{0}{45}{20}{-0}{+75}
\plotfiddle{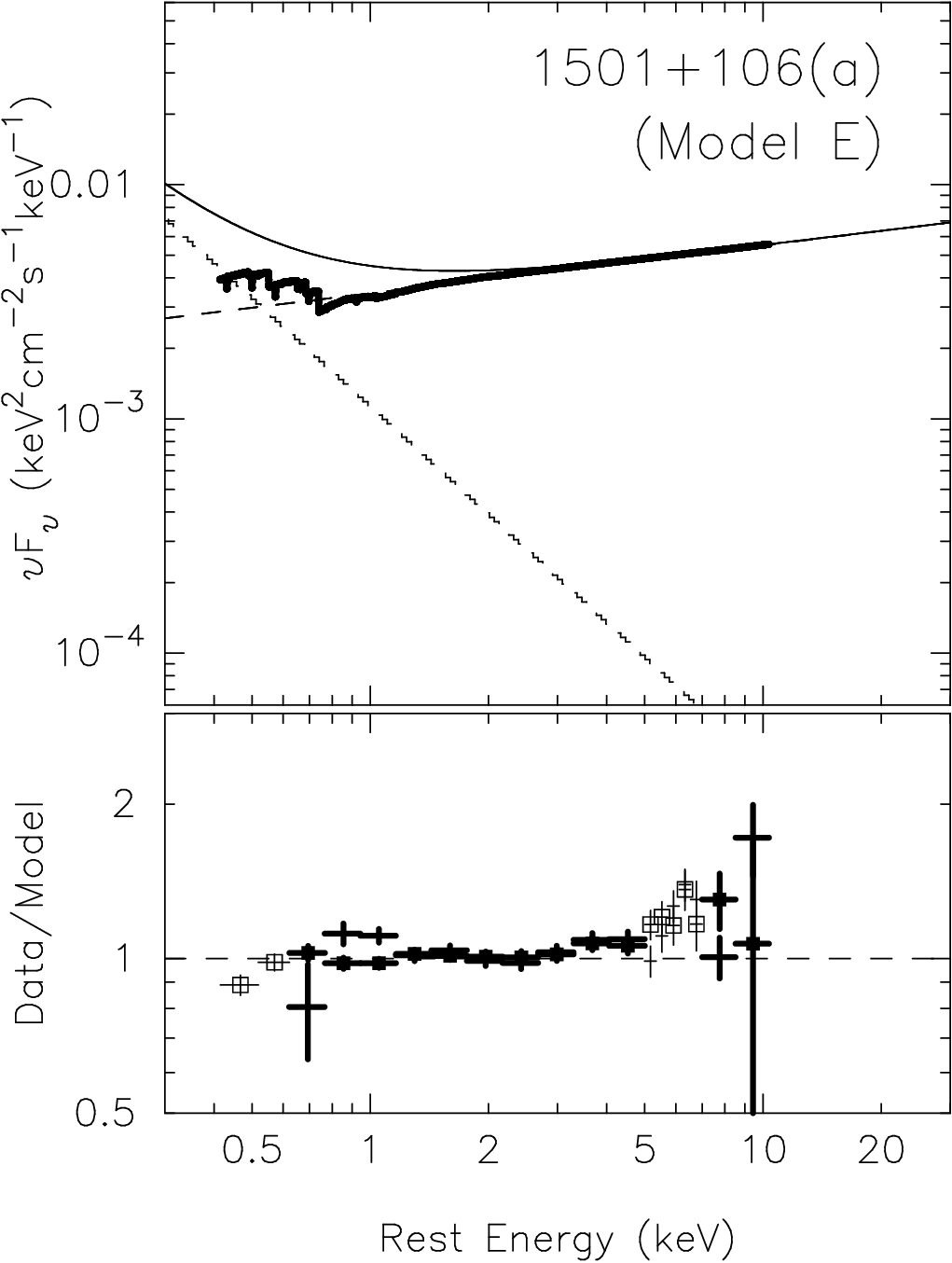}{2cm}{0}{45}{20}{-250}{0}
\plotfiddle{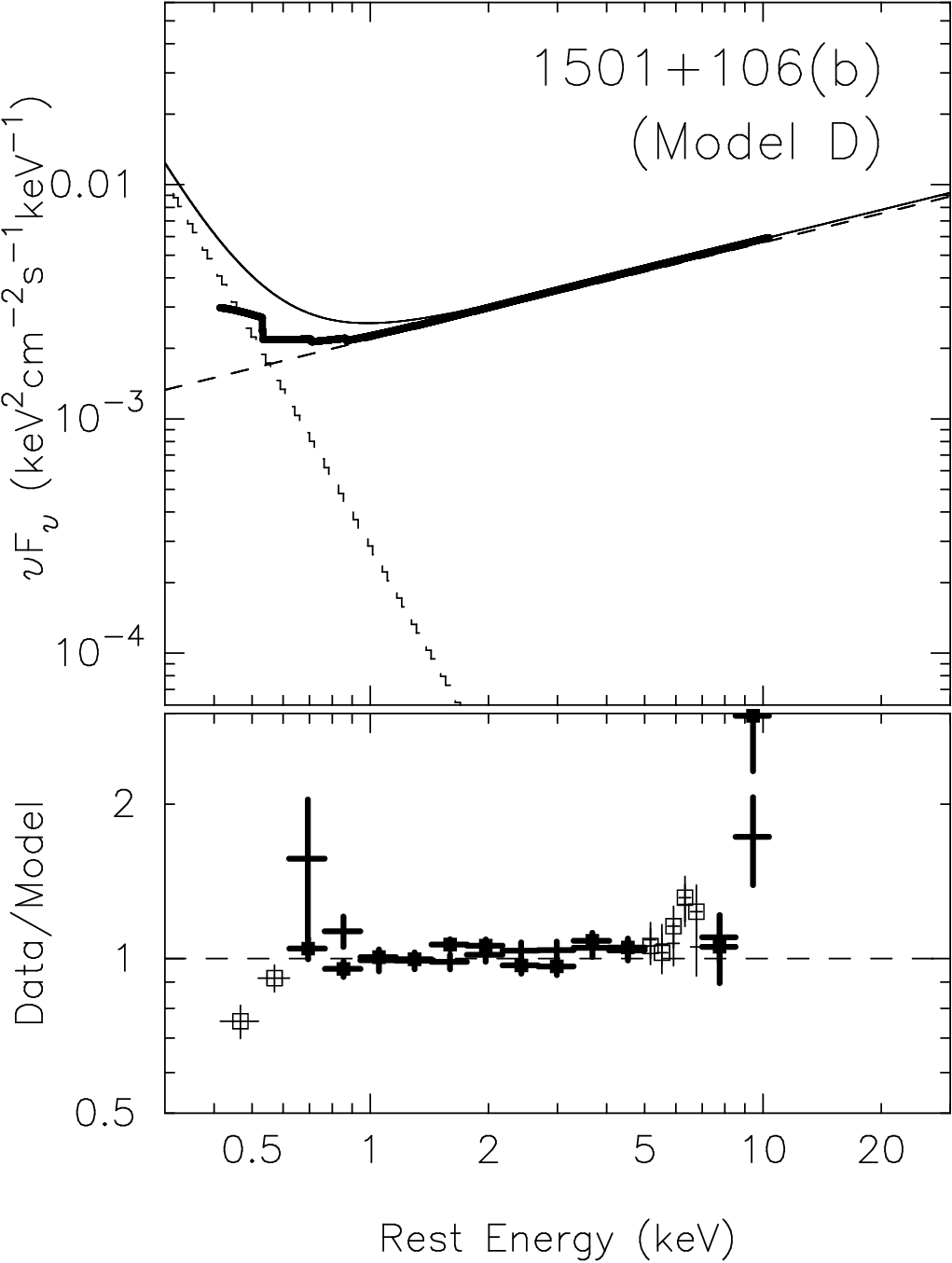}{2cm}{0}{45}{20}{-0}{+75}
\plotfiddle{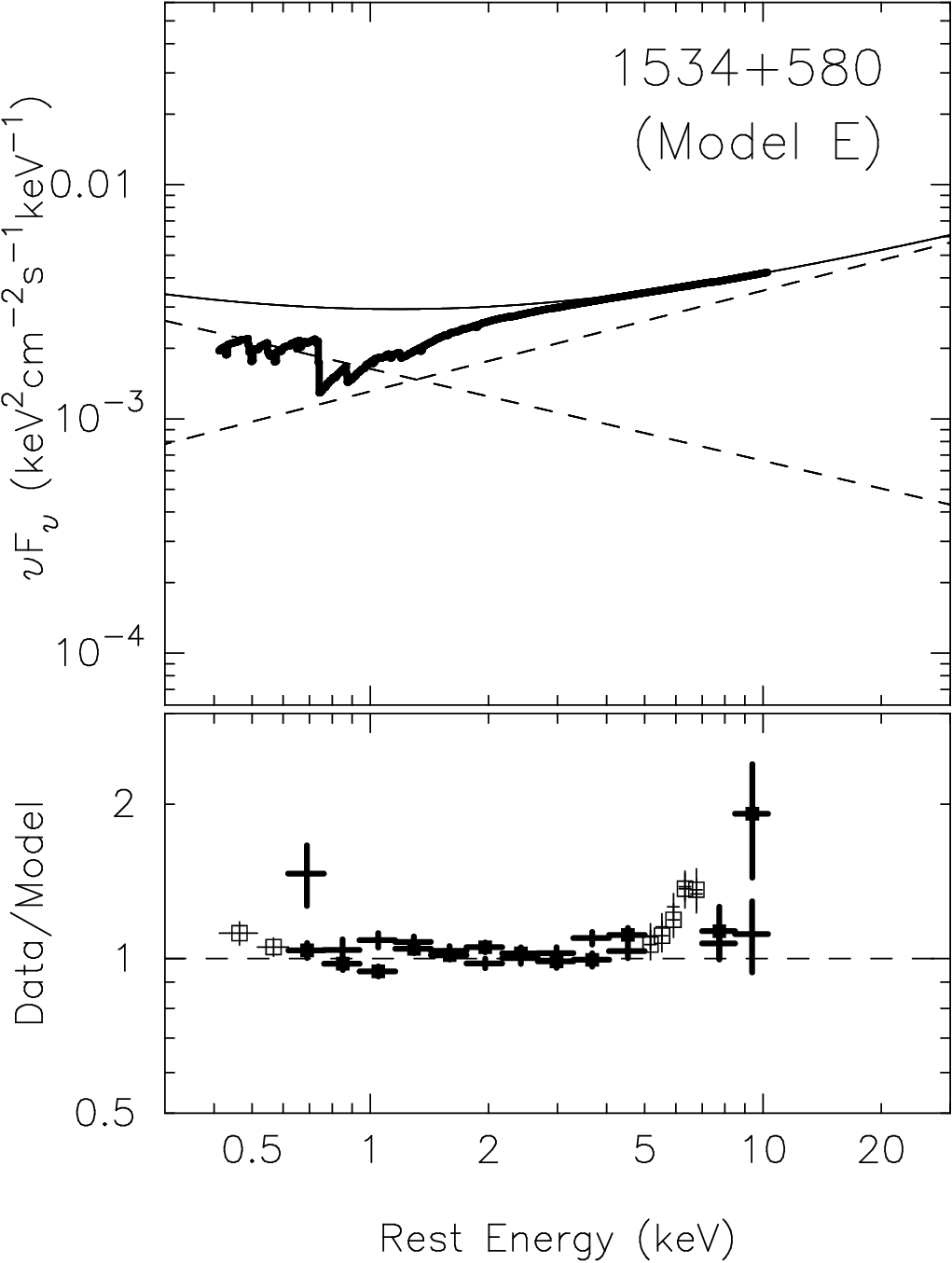}{2cm}{0}{45}{20}{-250}{0}
\plotfiddle{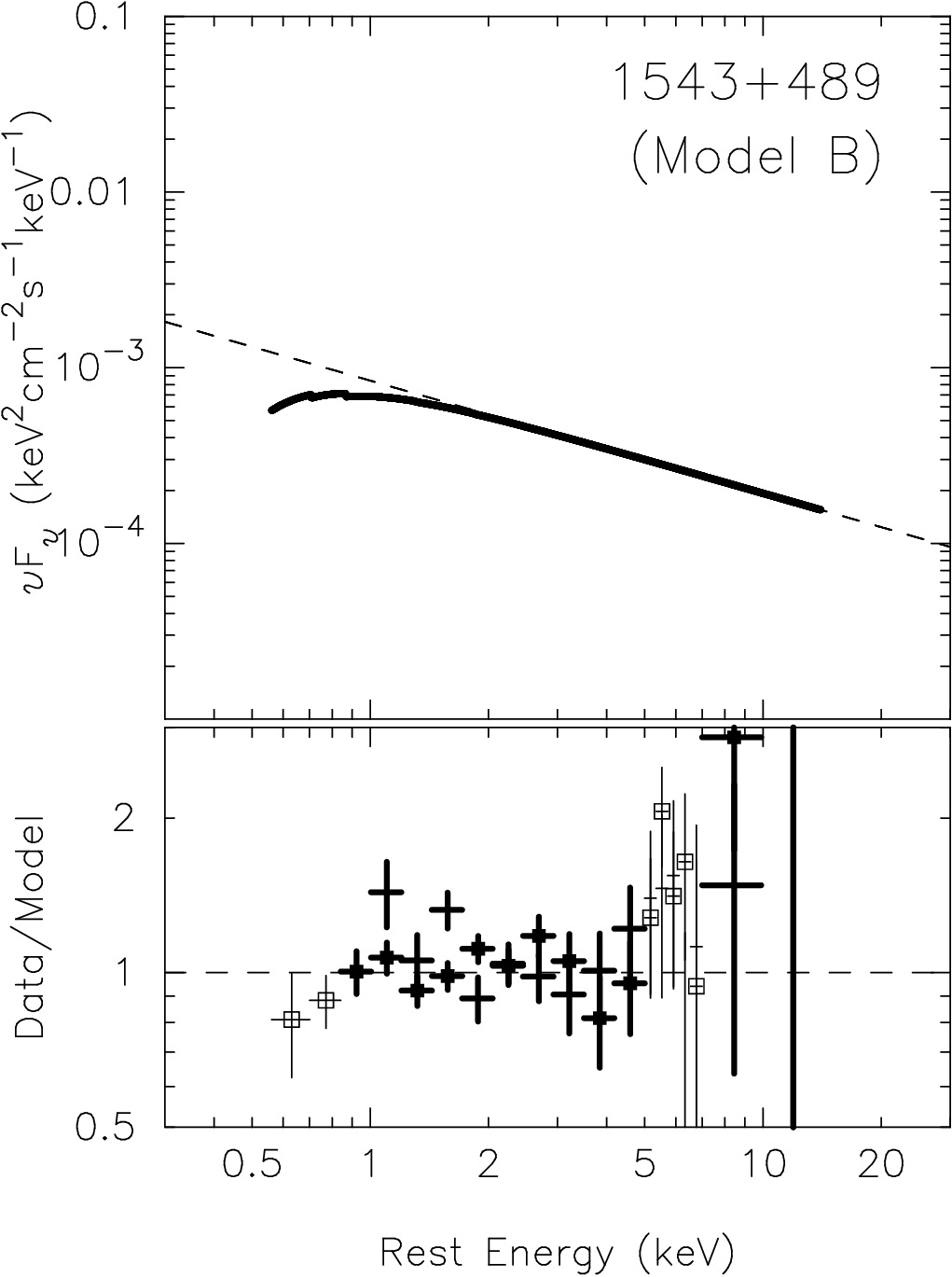}{2cm}{0}{45}{20}{-0}{+75}
\caption{\it continued}
\end{figure}

\addtocounter{figure}{-1}
\begin{figure}
\thispagestyle{empty}
\plotfiddle{fig3y.ps}{2cm}{0}{45}{20}{-250}{0}
\caption{\it continued}
\end{figure}

\clearpage
\begin{figure}
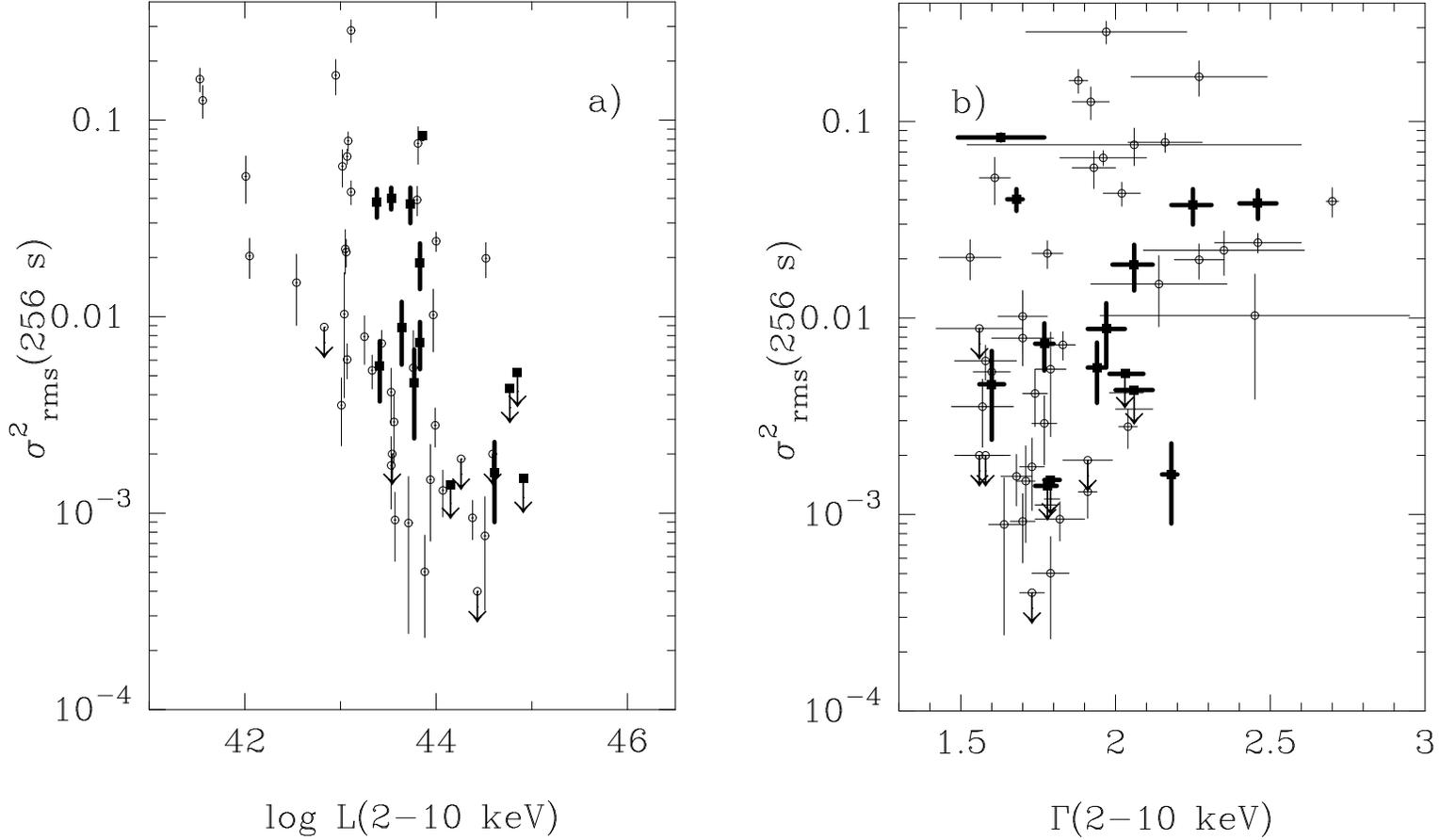

\thispagestyle{empty}
\plotfiddle{fig4a.ps}{2cm}{0}{50}{50}{-300}{-100}
\plotfiddle{fig4b.ps}{2cm}{0}{50}{50}{-0}{-27}
\caption{a) The excess variance, $\sigma_{rms}$(256~s), against 
luminosity in the 2--10~keV band, $L$(2--10~keV) 
in units of ${\rm erg\ s^{-1}}$. The bold squares indicate
the RQQs presented here, and the circles show other RQ AGN
adapted from Turner et al (1999a).
The RQQs behave in a similar way to other AGN despite 
covering only a limited range in luminosity.
b) $\sigma_{rms}$(256~s) against the photon index in the 
2--10~keV band, $\Gamma_{2-10}$, for the same set of objects. 
We find no significant correlation, but do note a trend whereby most 
of the 
sources exhibiting significant variability have either 'soft excesses' 
of steeper than average continuua.
\label{fig:correlations_xray_rms}}
\end{figure}

\clearpage
\begin{figure}
\thispagestyle{empty}
\plotfiddle{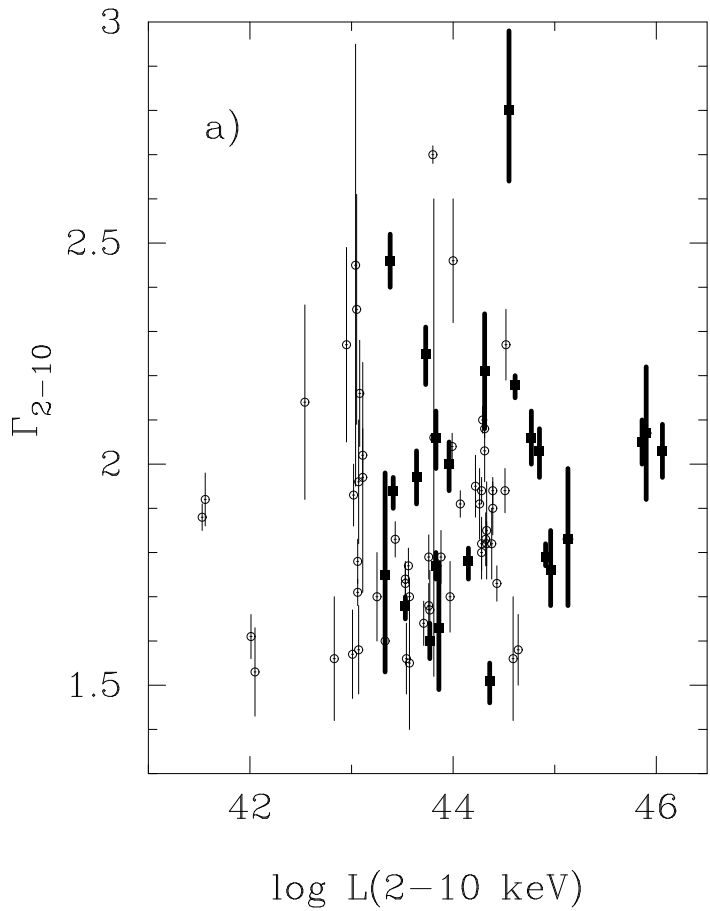}{2cm}{0}{50}{50}{-300}{-100}
\plotfiddle{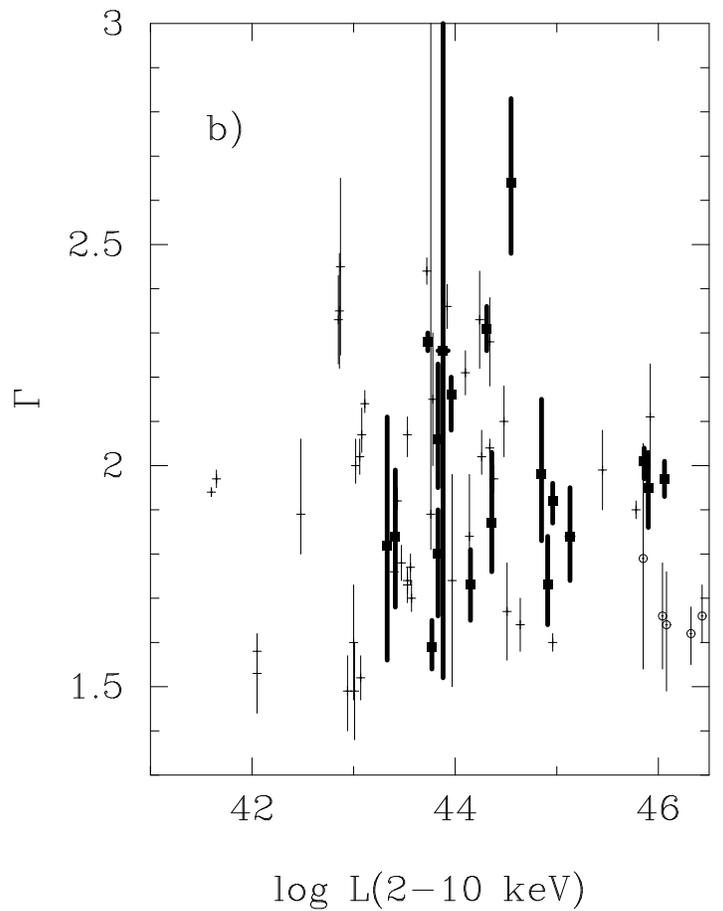}{2cm}{0}{50}{50}{-0}{-27}
\caption{a) 
Luminosity in the 2--10~keV band, $L$(2--10~keV) 
in units of ${\rm erg\ s^{-1}}$, 
versus $\Gamma_{2-10}$
 for the RQQs presented here (bold), 
and the other RQ objects from Turner et al (1999a).
b) $L$(2--10~keV) versus $\Gamma$ 
from the analysis of the 0.6-10~keV (observed frame) 
continuum.
The crosses are the RQQs from Reeves et al (1997) along with 
Seyfert 1 Galaxies published in George et al (1998a,b,c)
and Vaughan et al (1999). The circles 
are the high-$z$ RQQs from Vignali, et al (1999).
\label{fig:correlations_xray_gamma}}
\end{figure}

\clearpage
\begin{figure}
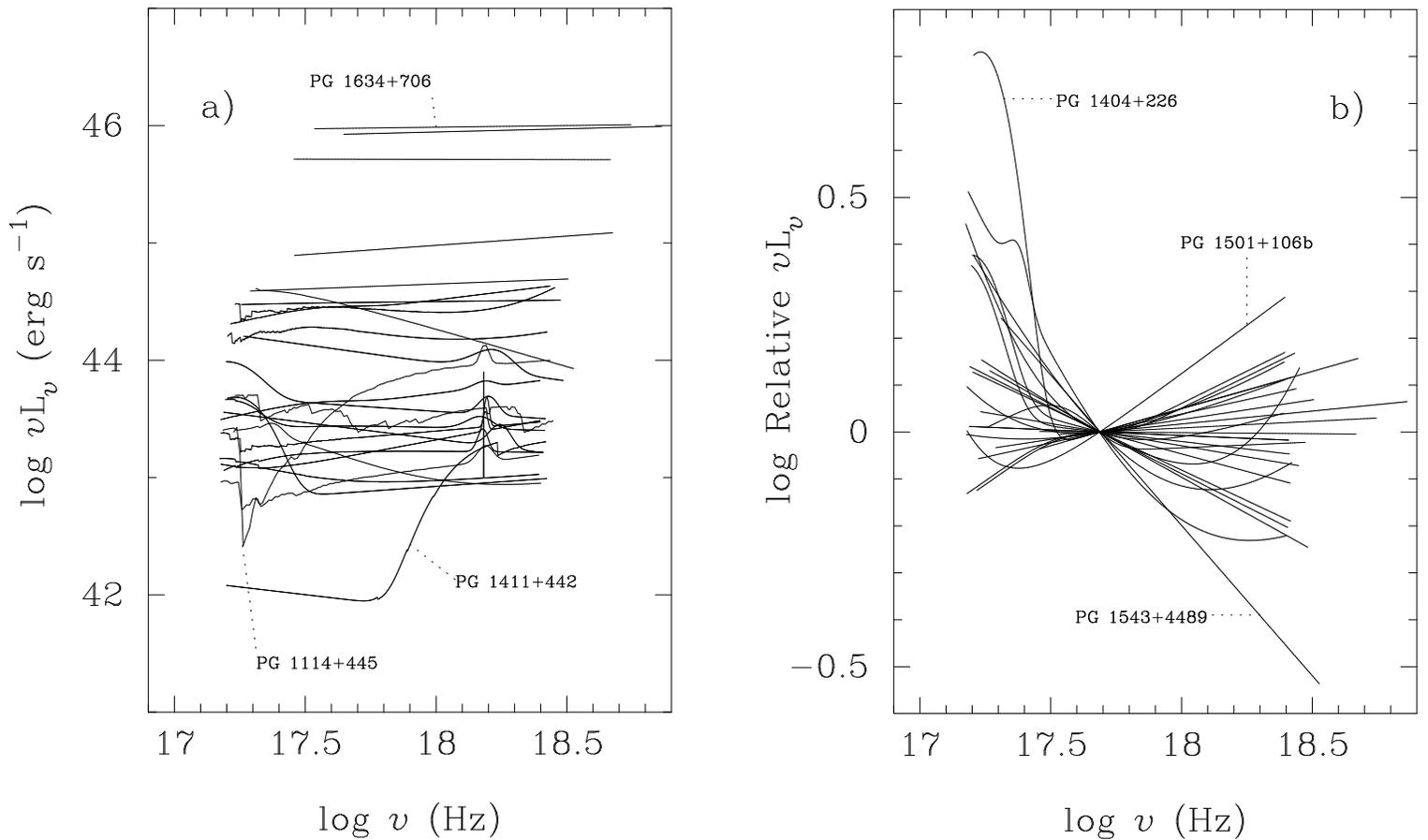

\thispagestyle{empty}
\plotfiddle{fig6a.ps}{2cm}{0}{50}{50}{-300}{-100}
\plotfiddle{fig6b.ps}{2cm}{0}{50}{50}{-0}{-27}
\caption{a) $\nu L_{\nu}$ 
versus frequency 
for the 
best-fitting models. Each curve indicates the 0.6--10~keV (observed-frame) 
band and has been corrected for Galactic absorption but not any absorption 
intrinsic to the RQQ. 
b) Comparison of the underlying continua when corrected for 
all absorption and renormalized to the luminosty at 2~keV in the quasar-frame. 
Any Fe $K$-shell emission feature is not shown for the sake of clarity.
PG~1501+106b has $\Gamma \simeq 1.6$ and 
PG~1543+489 has $\Gamma \simeq 2.6$.
\label{fig:sample_asca_nulnu}}
\end{figure}

\clearpage
\begin{figure}
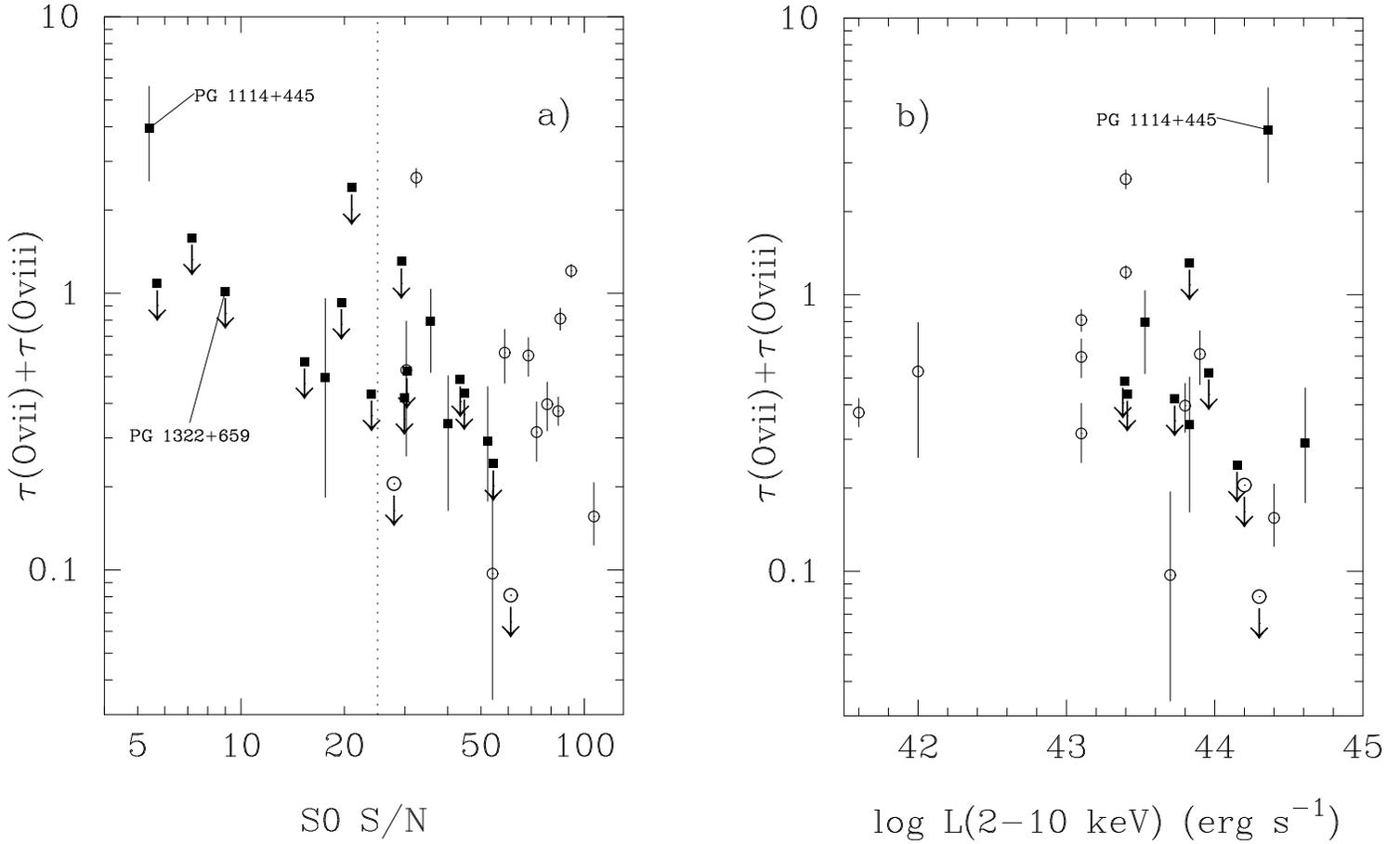

\thispagestyle{empty}
\plotfiddle{fig7a.ps}{2cm}{0}{50}{50}{-300}{-100}
\plotfiddle{fig7b.ps}{2cm}{0}{50}{50}{-0}{-27}
\caption{a) 
Sum of the optical depths of the 
edges due to O{\sc vii} (739~eV)
and O{\sc viii} (871~eV) versus signal to noise ratio
(Table~\ref{tab:2edges}).
The RQQs are shown as filled squares, and the other RQ AGN from 
George et al (1998a) as open circles.
The dotted line indicates S0 S/N $=25$.
The signal to noise ratio ratio for PG~1114+445 is low as a result of the 
deep absorption in this source. 
The lack of significant O{\sc vii} and O{\sc viii} edges in 
PG~1322+659 is discussed in the text.
b) Sum of the optical depths of  O{\sc vii} and O{\sc viii}
versus luminosity in the 2--10~keV band.
Only those sources with S0 S/N $> 25$, plus
PG~1114+445, are plotted.
\label{fig:2edge}}
\end{figure}

\clearpage
\begin{figure}
\thispagestyle{empty}
\plotfiddle{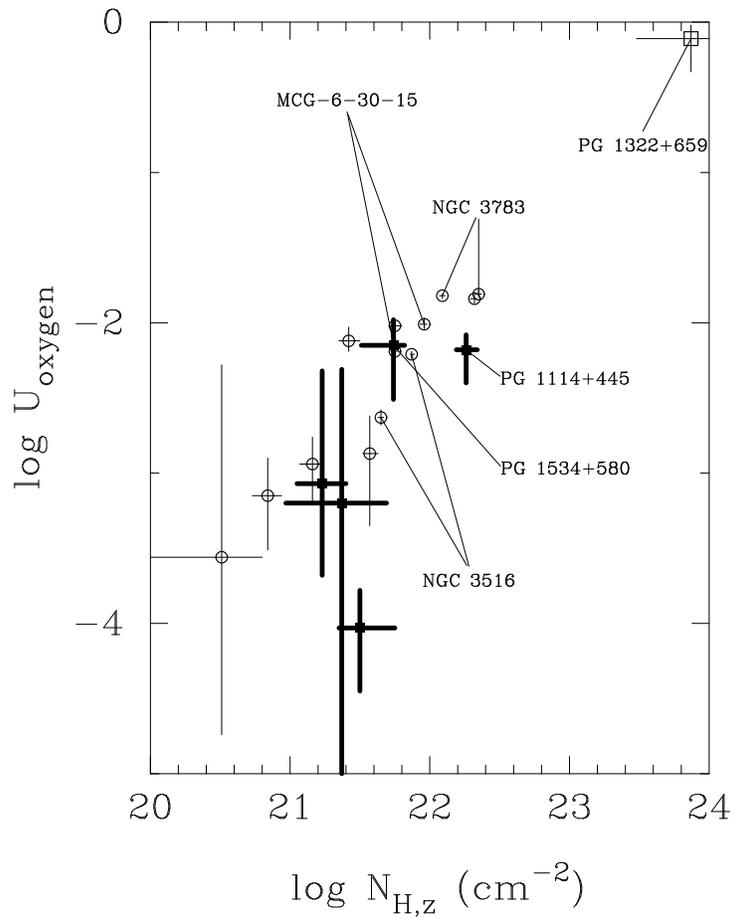}{8cm}{0}{50}{50}{-150}{0}
\caption{A comparison between the characteristics of the 
ionized absorbing material in the RQQs (squares)
to those for the other RQ AGN in George et al (1998a,b,c).
Besides the extreme values in the case of PG~1322+659 
(which should be treated with caution -- see text), we find no 
clear differences between the two samples.
\label{fig:nh_vs_ionox}}
\end{figure}

\clearpage
\begin{figure}
\thispagestyle{empty}
\plotfiddle{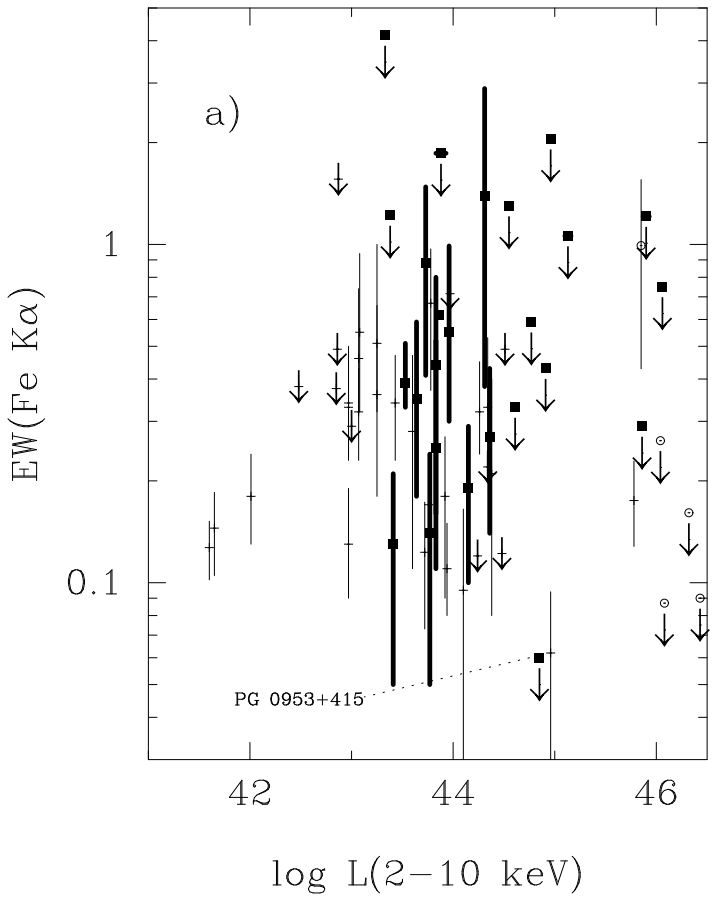}{2cm}{0}{50}{50}{-300}{-100}
\plotfiddle{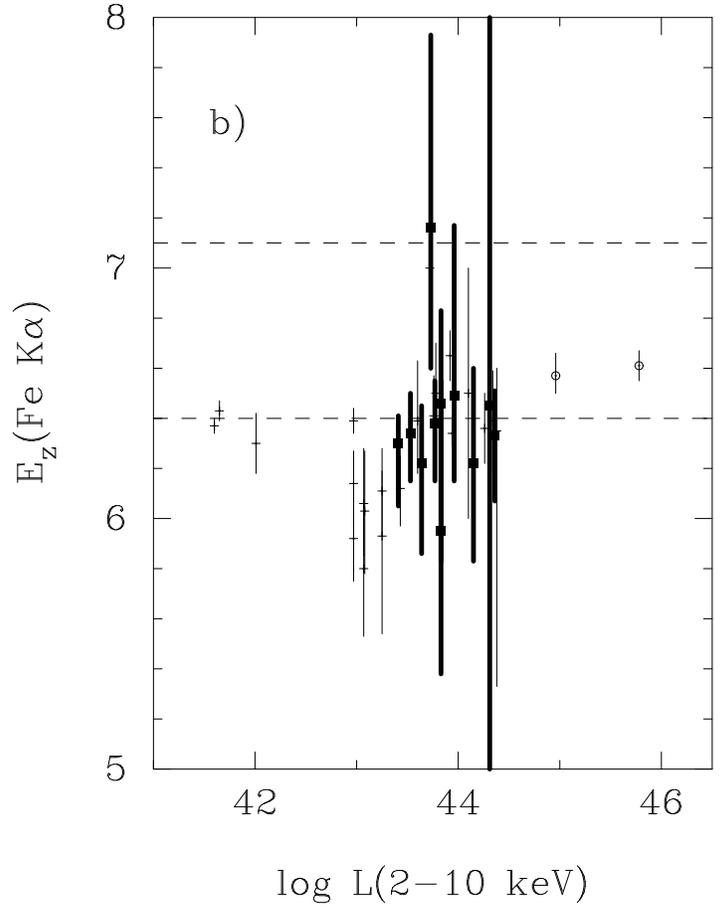}{2cm}{0}{50}{50}{-0}{-27}
\caption{a) 
Luminosity in the 2--10~keV band, $L$(2--10~keV) 
in units of ${\rm erg\ s^{-1}}$,  
versus the equivalent width, EW(Fe K$\alpha$)
in units of eV, of any Fe $K$-shell emission.
The crosses are the RQQs from Reeves et al (1997) along with 
Seyfert 1 Galaxies published in George et al (1998a,b,c)
and Vaughan et al (1999). The circles 
are the high-$z$ RQQs from Vignali, et al (1999).
b) 
$L$(2--10~keV) vesus the 
energy of the line, $E_z$ (in keV), in the quasar frame.
The crosses are for the Seyfert 1 galaxies published in
Nandra et al (1997b) and Vaughan et al (1999).
The circles are the RQQs from Reeves et al (1997).
\label{fig:correlations_xray_feline}}
\end{figure}

\clearpage
\begin{figure}
\thispagestyle{empty}
\plotfiddle{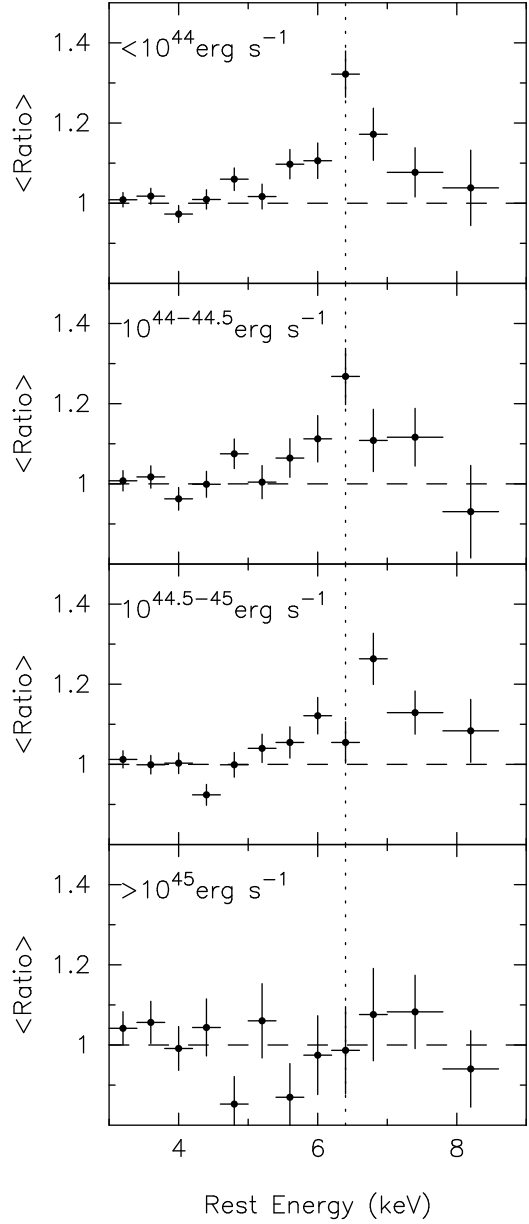}{13.5cm}{0}{70}{70}{-150}{-50}
\caption{Mean SIS data/model ratios in the Fe $K$-shell regime as a function of 
2--10~keV luminosity. Note that a different number of datasets were used in the
construction of these mean profiles 
(from top to bottom: 12, 3, 6 \& 4 datasets).
The vertical dotted line at 6.4~keV 
indicates the energy of $K\alpha$ fluorescence from 
Fe{\sc i}--{\sc xii}
\label{fig:fe-profile}}
\end{figure}

\clearpage

\begin{deluxetable}{llrrrrlrc}

\tablecaption{THE SAMPLE
\label{tab:sample}}

\tablecolumns{9}
\tablewidth{0pt}
\tablehead{
\colhead{PG Name} &
	\colhead{Alt. Name} &
	\colhead{RA} 	&
	\colhead{dec} 	&
	\colhead{$z$}	&
	\colhead{$M_V$}	&
	\colhead{$R_L$}	&
	\multicolumn{2}{c}{$N_{H,0}^{gal}$}
\nl
	&
	&
	(J2000) &
	(J2000)	&
	&
	&
	&
	\multicolumn{2}{c}{$(10^{20}\ {\rm cm^{-2}})$}	
\nl
        \colhead{(1)}   &
        \colhead{(2)}   &
        \colhead{(3)}   &
        \colhead{(4)}   &
        \colhead{(5)}   &
        \colhead{(6)}   &
        \colhead{(7)}   &
        \colhead{(8)}   &
        \colhead{(9)}   
}

\startdata
0003+199	&
	Mrk 335	&
	00~06~19.3 	&
	$+$20~12~12	&
	0.025	&		
	-22.13	&
	0.27	&
	3.70 &	a
	\nl
0043+039	&
	PB~6151	&
	00~45~44.9	&
	$+$04~10~57	&
	0.384	&
	-26.00	&
	0.12	&
	2.99 &	b
	\nl
0050+124	&
	I Zw 1	&
	00~53~34.9 	&
	$+$12~41~35	&
	0.061	&		
	-23.74	&
	0.33	&
	5.07 &	c
	\nl
0804+761	&
	\nodata	&
	08~10~58.5	&
	$+$76~02~42	&
	0.100	&
	-24.43  &
	0.60	&
	3.26	&	b
	\nl
0844+349	&
	Ton 951	&
	08~47~42.5	& 
	$+$34~45~05	&
	0.064	&
	-23.28	&
	0.03	&
	3.32	&	b
	\nl
0921+525	&
	Mrk 110	&
	09~25~12.9	&   
	$+$52~17~17  	&
	0.035	&		
	-21.23  &
	1.49	&
	1.21 &	a
	\nl
0953+415	&
	\nodata	&
	09~56~52.5	&
	$+$41~15~41	&
	0.239	&
	-25.67  &
	0.44	&
	1.12 & b
	\nl
1114+445	&
	\nodata	&
	 11~17~06.4	&
	$+$44~13~33	&
	0.144	&
	-24.00  &	
	0.13	&
	1.93 &	a
	\nl
1116+215	&
	TON~1388	&
	 11~19~08.7	&
	$+$21~19~18	&
	0.177	&
	-25.58  &	
	0.72	&
	1.40 &	b
	\nl
1148+549 	&
	\nodata	&
	11~51~20.5 	&
	$+$54~37~32	&
	0.969	&
	-28.85  &		
	0.59	&
	0.96 &	b
	\nl
1211+143	&
	\nodata	&
	12~14~17.6 	&
	$+$14~03~12	&
	0.085	&
	-24.60  &		
	0.13	&
	2.76 &	b
	\nl
1216+069	&
	\nodata	&
	 12~19~20.3	&
	$+$06~38~40	&
	0.334	&
	-26.33  &
	1.65	&
	1.56 &	b
	\nl
1244+026	&
	\nodata	&
	12~46~35.3	&
	$+$02~22~08	&
	0.048	&
	-21.68  &
	0.53	&
	1.93 &	c
	\nl
1247+267 	&
	\nodata	&
	12~50~05.7	&
	$+$26~31~09	&
	2.038	&
	-30.58	&		
	0.36	&
	0.86 &	a
	\nl
1322+659	&
	\nodata	&
	13~23~49.6	&
	$+$65~41~48	&
	0.168	&
	-24.16	&		
	0.12	&
	0.19 &	a
	\nl
1404+226	&
	\nodata	&
	14~06~21.9	&
	$+$22~23~47	&
	0.098	&
	-22.95	&		
	0.47	&
	2.00 &	c
	\nl
1407+265 	&
	\nodata	&
	14~09~23.8	&
	$+$26~18~21	&
	0.947	&
	-28.43  &		
	3.43	&
	1.36 	& b
	\nl
1411+442	&
	PB~1732	&
	14~13~48.4	&
	$+$44~00~14	&
	0.089	&
	-23.53  &
	0.13	&
	1.05 &	b
	\nl
1416--129 	&
	\nodata	&
	14~19~03.8	&
	$-$13~10~45	&
	0.129	&
	-25.05  &		
	1.15	&
	7.20 &	c
	\nl
1440+356	&
	Mrk~478	&
	 14~42~07.5	&
	$+$35~26~24	&
	0.077	&
	-23.50	&	
        0.37	&
	0.96 	& a
	\nl
1444+407	&
	\nodata		&
	14~46~45.9	&
	$+$40~35~06	&
	0.267	&
	-25.20	&
        0.08$\ddag$ &
	1.08 	& b
	\nl
1501+106 	&
	Mrk 841	&
	15~04~01.3	&
	$+$10~26~17	&
	0.036	&		
	-22.75	&
        0.36	&
	2.17 	& a	
	\nl
1534+580	&
	Mrk 290	&
	15~35~52.9	&
	$+$57~54~09	&
	0.030	&		
	-21.40	&
        0.70	&
	2.32 	& c	
	\nl
1543+489	&
	\nodata		&
	 15~45~30.3	&
	$+$48~46~08	&
	0.400	&
	-25.88	&
        0.63	&
	1.68 	& b
	\nl
1634+706 	&
	\nodata	&
	16~34~29.0	&
	$+$70~31~33	&
	1.334	&		
	-30.75  &
        0.44	&
	5.74 	& c
	\nl
1700+518 	&
	\nodata	&
	17~01~25.0 	&
	$+$51~49~21	&
	0.272	&		
	-26.31  &
        2.40	&
	2.48	& b
	\nl
\tablecomments{
Cols.(1--2): PG name and alternative name of the source in common usage.
Cols.(3--4): Optical position from Green, Schmidt \& Liebert (1986) 
transformed to J2000 coordinates.
Col.(5--7): Redshift, absolute magnitude in the V-band, and 
		radio-loudness (see \S\ref{Sec:sample}).
Cols.(8--9): Galactic column density  from 
(a) Murphy et al (1996),
(b) Lockman \& Savage (1995), and
(c) Elvis, Lockman \& Wilkes (1989).
The measurement errors on these values are typically 
$\lesssim$few$\times10^{19}\ {\rm cm^{-2}}$.
}
\enddata
\end{deluxetable}

\clearpage

\begin{deluxetable}{llcc cccl}

\tablecaption{LOG OF THE {\it ASCA} OBSERVATIONS
\label{tab:asca_obs_log}}

\tablecolumns{8}
\tablewidth{0pt}
\tablehead{
\colhead{Source(obs)} &
	\colhead{Start Date} &
	\colhead{Sequence} &
	\colhead{SIS} &
        \colhead{$t_{exp}$}     &
        \colhead{$t_{dur}$}     &
        \colhead{$Rate$}        &
	\colhead{Refs}
\nl
	&
        &
	&
	\colhead{C.Mode} &
        \colhead{$(10^{3}\ {\rm s})$}   &
        \colhead{$(10^{3}\ {\rm s})$}   &
        \colhead{$({\rm count\ s^{-1}})$}       &
\nl
        \colhead{(1)}   &
        \colhead{(2)}   &
        \colhead{(3)}   &
        \colhead{(4)}   &
        \colhead{(5)}   &
        \colhead{(6)}   &
        \colhead{(7)}   &
        \colhead{(8)}   
}

\startdata
0003+199	&
	1993 Dec 09	&
	71010000	&
	2	&
	18.8	&
	\phn46.8	&
	$0.494\pm0.005$	&
	a,b,c
	\nl
0043+030	&
	1996 Dec 21	&
	74078000	&
	2	&
	25.8		&
	\phn66.5		&
	$<0.003$	&
	d,e
	\nl
0050+124	&
	1995 Jul 15	&
	73042000 &
	1	&
	27.1	&
	\phn92.3	&
	$0.191\pm0.003$	&
	b,c
	\nl
0804+761	&
	1997 Nov 04	&
	75058000	&
	1	&
	41.9	&
	\phn89.3	&
	$0.456\pm0.003$	&
	\nl
0844+349	&
	1998 Apr 05	&
	76059000	&
	1	&
	45.6	&
	133.3	&
	$0.101\pm0.002$	&
	\nl
0921+525	&
	1995 Apr 05 	&
	73091000  	&
	1	&
	20.7	&
	\phn40.5	&
	$1.022\pm0.007$	&
	\nl
0953+415	&
	1997 Nov 19	&
	75060000	&
	1	&
	34.7	&
	\phn83.5	&
	$0.136\pm0.002$	&
	\nl
1114+445	&
	1996 May 05	&
	74072000  &
	1	&
	60.9	&
	151.0	&
	$0.048\pm0.001$	&
	f
	\nl
1116+215	&
	1995 May 19	&
	73064000  &
	2	&
	17.9	&
	\phn40.7	&
	$0.207\pm0.004$	&
	g
	\nl
1148+549 	&
	1995 Dec 07	&
	73037000  &
	1	&
	35.0	&
	\phn76.6 	&
	$0.011\pm0.001$	&
	\nl
1211+143 	&
	1993 Jun 03 	&
	70025000  &
	4	&
	23.6	&
	\phn90.3	&
	$0.167\pm0.003$	&
	b,c,h,i
	\nl
1216+069	&
	1995 Dec 25	&
	74074000  &
	1	&
	21.6	&
	\phn42.9	&
	$0.073\pm0.002$	&
	\nl
1244+026	&
	1996 Jul 01	&
	74070000	&
	2	&
	36.7	&
	103.9	&
	$0.222\pm0.002$	&
	b,g,c
	\nl
1247+267 	&
	1995 Jun 17	&
	73048000  &
	1	&
	35.0	&
	\phn94.3	&
	$0.014\pm0.001$	&
	\nl
1322+659	&
	1999 May 09	&
	77058000 &
	1	&
	33.2	&
	\phn81.3	&
	$0.028\pm0.001$	&
	\nl
1404+226	&
	1994 Jul 13	&
	72021000  &
	1	&
	34.4	&
	\phn96.9	&
	$0.041\pm0.001$	&
	b,c,k,l
\nl
1407+265 	&
	1993 Jul 02	&
	70024000  &
	1	&
	35.3	&
	\phn81.3	&
	$0.059\pm0.002$	&
	i \nl
1411+442	&
	1996 Dec 08	&
	74075000  &
	1	&
	36.0	&
	\phn88.2	&
	$0.008\pm0.001$	&
	d
	\nl
1416--129 	&
	1994 Jul 29	&
	72043000  &
	1 \& 2	&
	0.7 \& 30.1 	&
	\phn75.4	&
	$0.344\pm0.004$	&
	i \nl
1440+356	&
	1995 Jul 02	&
	73067000  &
	1	&
	31.9	&
	\phn84.6	&
	$0.134\pm0.002$	&
	b,c
\nl
1444+407	&
	1997 Jan 25	&
	75061000  &
	1	&
	39.9	&
	\phn89.0	&
	$0.037\pm0.001$	&
	\nl
1501+106(a) 	&
	1993 Aug 22	&
	70009000  &
	2	&
	30.2	&
	\phn78.2	&
	$0.408\pm0.004$	&
	a
	\nl
1501+106(b)        &
        1994 Feb 21	&
	71040000  &
	2	&
	20.6	&
	\phn59.6	&
	$0.329\pm0.004$	&
	a
	\nl
1534+580	&
	1994 Jun 15	&
	72027000  &
	2	&
	41.0	&
	\phn98.3	&
	$0.271\pm0.003$	&
	m
	\nl
1543+489	&
	1997 Jan 29	&
	75059000  &
	1	&
	39.5	&
	\phn89.0	&
	$0.032\pm0.001$	&
	b \nl
1634+706 	&
	1994 May 02	&
	71036000  &
	1 \& 2	&
	21.9 \& 20.5 	&
	\phn47.7	&
	$0.041\pm0.002$	&
	i \nl
1700+518 	&
	1998 Mar 24	&
	76041000  &
	1 	&
	19.0 	&
	532.6	&
	$<0.003$	&
	e
	\nl
\tablecomments{\it see over}
\enddata
\end{deluxetable}

\clearpage
Notes for Table~\ref{tab:asca_obs_log} --- 
Cols.(1--4): Source, Observation date (start of exposure), 
{\it ASCA} sequence number, and
SIS CCD Clocking Mode in use.
Col.(5--6): Total (SIS0) exposure time, $t_{exp}$ (for each 
	SIS clocking mode), and total duration $t_{dur}$ 
	between first and last events after screening.
Cols.(7): Mean background-subtracted SIS0 
	count rate in the 0.6--10.0~keV band 
	(uncorrected for the counts falling outside the extraction cell).
	In the case of PG~1416-129 \& PG~1634+706 exposure-weighted 
	mean values are quoted.
Col.(8): References to other papers discussing the same dataset ---
        (a) Nandra et al (1997a,b), George et al (1998a);
        (b) Vaughan et al (1999); 
	(c) Leighly (1999);
        (d) Brinkmann et al (1999);
        (e) Gallagher et al (1999);
        (f) George et al (1997);
        (g) Nandra et al (1996);
        (h) Yaqoob et al (1994);
	(i) Reeves et al (1997);
        (j) Fiore et al (1998);
        (k) Leighly et al (1997);
	(l) Ulrich et al (1999);
	(m) Turner et al (1996).

\clearpage

\begin{deluxetable}{lcc}

\tablecaption{VARIABILITY AMPLITUDE
$\sigma^2_{rms}$(256~s)
\label{tab:xs_rms_128}}

\tablecolumns{3}
\tablewidth{0pt}
\tablehead{
\colhead{Source(obs)} &
        \colhead{SIS}   &
        \colhead{GIS}   
\nl
        &
        \colhead{(0.5--10~keV)} &
        \colhead{(2--10~keV)}   
\nl
        \colhead{(1)}   &
        \colhead{(2)}   &
        \colhead{(3)}   
}

\startdata
0003+199	&
	$0.56\pm0.19$	&
	$0.72\pm0.29$
\nl
0050+124	&
	$3.76\pm0.77$	&
	$3.12\pm0.68$
\nl
0804+761	&
	$0.16\pm0.07$	&
	$<0.26$
\nl
0844+349	&
	$0.88\pm0.31$	&
	$0.91\pm0.42$
\nl
0921+525	&
	$<0.14$	&
	$<0.19$
\nl
0953+415	&
	$<0.52$	&
	$<0.40$
\nl
1116+215	&
	$<0.43$	&
	$<0.57$
\nl
1211+143 	&
	\nodata	&
	$<1.12$
\nl
1244+026	&
	$3.83\pm0.64$		&
	$4.26\pm0.78$	
\nl
1322+659	&
	$8.30\pm0.29$	&
	\nodata	
\nl
1416--129 	&
	$<0.15$	&
	$<0.46$
\nl
1440+356	&
	$1.87\pm0.49$		&
	$1.91\pm0.54$
\nl
1501+106(a) 	&
	$0.74\pm0.20$		&
	$0.58\pm0.18$
\nl
1501+106(b) 	&
	$0.46\pm0.22$		&
	$<0.30$
\nl
1534+580	&
	$4.02\pm0.51$		&
	$3.50\pm0.48$
\nl
\tablecomments{Normalized excess variance, $\sigma^2_{rms}$(256~s)
in units of $10^{-2}$. Upper limits are at 90\% confidence. Lack of an entry
indicates the criteria described in \S\ref{Sec:temporal} were not fulfilled.}
\enddata
\end{deluxetable}

\clearpage

\begin{deluxetable}{lrllr}

\tablecaption{MODEL A APPLIED TO THE 2-10~keV (QUASAR-FRAME) CONTINUUM
(see \S4.1.1) 
\label{tab:zpo_ngal_210_noline_fits}}

\tablecolumns{5}
\tablewidth{0pt}
\tablehead{
\colhead{Source(obs)} &
	\colhead{$N_{pts}$}	&
	\colhead{$\Gamma_{2-10}$}	&
	\colhead{$\chi^2_{\nu}$}	&
	\colhead{$P(\chi^2)$}	
\nl
        \colhead{(1)}   &
        \colhead{(2)}   &
        \colhead{(3)}   &
        \colhead{(4)}   &
        \colhead{(5)}   
}
\startdata
0003+199	&
	331	&
	\phs$1.94^{+0.03}_{-0.04}$	&
	0.98	& 0.41	
	\nl
0050+124	&
	202	&
	\phs$2.25^{+0.06}_{-0.07}$	&
	1.02	&	0.60	
	\nl
0804+761	&
	529	&
	\phs$2.18^{+0.02}_{-0.03}$	&
	1.02	&	0.65	
	\nl
0844+349	&
	236	&
	\phs$1.97^{+0.06}_{-0.06}$	&
	0.93	&	0.23	
	\nl
0921+525	&
	618	&
	\phs$1.78^{+0.03}_{-0.04}$ 	&
	1.05	&	0.79	
	\nl
0953+415	&
	230	&
	\phs$2.03^{+0.06}_{-0.05}$	&
	0.98	&	0.44	
	\nl
1114+445	&
	289	&
	\phs$1.51^{+0.04}_{-0.05}$	&
	1.08	&	0.82	
	\nl
1116+215	&
	188	&
	\phs$2.06^{+0.06}_{-0.06}$	&
	1.11	& 0.86	
	\nl
1148+549	&
	73	&
	\phs$1.83^{+0.16}_{-0.15}$	&
	0.96	&	0.43	
	\nl
1211+143	&
	223	&
	\phs$2.00^{+0.05}_{-0.06}$	&
	0.82	&	0.02	
	\nl
1216+069 	&
	120	&
	\phs$1.76^{+0.09}_{-0.08}$		&
	0.83	& 	0.10	
	\nl
1244+026	&
	225	&
	\phs$2.46^{+0.06}_{-0.06}$	&
	0.92	&	0.20	
	\nl
1247+267	&
	68	&
	\phs$2.07^{+0.15}_{-0.15}$	&
	0.82	&	0.16	
	\nl
1322+659	&
	90	&
	\phs$1.63^{+0.14}_{-0.14}$	&
	0.91	&	0.28	
	\nl
1404+226	&
	64	&
	\phs$1.75^{+0.23}_{-0.22}$	&
	1.13	&	0.76	
	\nl
1407+265	&
	196	&
	\phs$2.05^{+0.05}_{-0.05}$	&
	0.86	&	0.08	
	\nl
1411+442	&
	58	&
	$-0.42^{+0.24}_{-0.26}$	&
	1.49	&	0.99	
	\nl
1416--129	&
	531	&
	\phs$1.79^{+0.03}_{-0.02}$	&
	1.02	&	0.65	
	\nl
1440+356	&
	193	&
	\phs$2.06^{+0.06}_{-0.07}$	&
	0.92	&	0.21	
	\nl
1444+407	&
	111	&
	\phs$2.21^{+0.13}_{-0.13}$	&
	1.07	&	0.71	
	\nl
1501+106(a)	&
	449	&
	\phs$1.77^{+0.03}_{-0.03}$	&
	1.03	&	0.66	
	\nl
1501+106(b) 	&
	255	&
	\phs$1.60^{+0.04}_{-0.04}$	&
	0.97	&	0.36	
	\nl
1534+580	&
	492	&
	\phs$1.68^{+0.02}_{-0.03}$	&
	0.94	&	0.17	
	\nl
1543+489	&
	93	&
	\phs$2.80^{+0.18}_{-0.16}$	&
	0.78	&	0.06	
	\nl
1634+706	&
	175	&
	\phs$2.03^{+0.06}_{-0.06}$	&
	1.18	&	0.95	
	\nl
\tablecomments{\it see over}
\enddata
\end{deluxetable}

\clearpage
Notes to Table~\ref{tab:zpo_ngal_210_noline_fits} ---
Results of fitting a single power-law with galactic absorption to the 
2--10~keV band, excluding the 5--7~keV band. 
Col.(1--2): Dataset and 
number of data points used in the spectral analysis
	(sum of all instruments).
Col.(3): Best-fitting value of the photon index of the underlying power law 
	with errors quoted at 68\% confidence for 1 interesting parameter.
Cols.(4--5): Reduced-$\chi^2$ statistic ($\chi^2_{\nu}$) and 
	probability ($P(\chi^2)$) that the $\chi^2$-statistic will be less
	than the observed value for the number of $dof$.
The model is considered an acceptable representation of the data 
($P(\chi^2) \leq 0.95$) in all cases except PG~1411+442.

\clearpage

\begin{deluxetable}{lccccl}

\tablecaption{SUMMARY OF THE PREFERRED MODEL
FOR THE FULL {\it ASCA} CONTINUUM
\label{tab:fit_summary2}}

\tablecolumns{6}
\tablewidth{0pt}
\tablehead{
\colhead{Source(obs)} &
	\colhead{Preferred}	&
	\colhead{$F_{2-10}$}	&
	\colhead{$F_{\nu}$(2~keV)}	&
	\colhead{$\log L$(2-10~keV)} 	&
	\colhead{Notes}	
\nl
	&
	\colhead{Model}	&
	&
	&
	&
\nl
        \colhead{(1)}   &
        \colhead{(2)}   &
        \colhead{(3)}   &
        \colhead{(4)}   &
        \colhead{(5)}   &
        \colhead{(6)}   
}

\startdata
0003+199	&
	D	&
	\phn$9.44\pm0.10$	&
	$3.00\pm0.03$		&
	$43.41\pm0.02$		&
	Var., SXS, Fe
\nl
0043+039	&
	$\ddag$	&
	$<0.07$	&
	$<0.03$		&
	$<43.78$		&
\nl    
0050+124	&
	A	&
	\phn$3.22\pm0.04$	&
	$1.35\pm0.02$		&
	$43.73\pm0.01$		&
	Var., Fe (?)
\nl    
0804+761	&
	E	&
	\phn$8.79\pm0.07$	& 
	$3.84\pm0.04$	& 
	$44.61\pm0.01$	&
	Abs
\nl
0844+349	&
	B	&
	\phn$2.36\pm0.02$	& 
	$0.72\pm0.01$	& 
	$43.64\pm0.01$	&
	Var., Abs, Fe
\nl
0921+525	&
	F	&
        $26.85\pm0.19$	&
	$7.21\pm0.05$	&
        $44.15\pm0.01$	&
	Fe
\nl
0953+415	&
	F	&
	\phn$2.59\pm0.04$        &
	$1.08\pm0.02$	&
	$44.85\pm0.01$        &
\nl
1114+445	&
	C	&
	\phn$2.13\pm0.05$	&
	$0.72\pm0.02$	& 
	$44.36\pm0.01$	&
	Abs, Fe 
\nl
1116+215	&
	E	&
        \phn$3.94\pm0.07$	&
	$1.68\pm0.03$	&
        $44.77\pm0.01$	&
	Var. ? Abs ?
\nl
1148+549	&
	A	&
	\phn$0.27\pm0.02$	&	
	$0.13\pm0.01$	&
	$45.13\pm0.03$	&
	SXS (?) $\dag$
\nl
1211+143	&
	F	&
	\phn$2.78\pm0.04$	&
	$1.08\pm0.02$	&
	$43.96\pm0.01$	&
	Var. ? SXS, Fe
\nl
1216+069	&
	A	&
	\phn$1.72\pm0.05$	&
	$0.65\pm0.02$	&
	$44.96\pm0.02$	&
\nl
1244+026	&
	D+Ga	&
	\phn$2.32\pm0.03$	&
	$1.22\pm0.01$	&
	$43.38\pm0.01$	&
	Var., 1~keV line
	$\dag$
\nl
1247+267	&
	A	&
	\phn$0.29\pm0.02$	&	
	$0.25\pm0.02$	&
	$45.90\pm0.03$	&	
\nl
1322+659	&
	C 	&
        \phn$0.54\pm0.02$	&
	$0.14\pm0.01$	&
        $43.86\pm0.02$	&
	Abs
\nl
1404+226	&
	F	&
        \phn$0.51\pm0.02$	&
	$0.15\pm0.01$	&
        $43.33\pm0.02$	&
	SXS, Var.
\nl
1407+265	&
	A	&
	\phn$1.38\pm0.04$	&	
	$0.84\pm0.02$	&
	$45.86\pm0.01$	&	
\nl
1411+442        &
	B$_{\rm pc}$	&
	\phn$0.76\pm0.09$		&
        $0.90\pm0.11$		&
	$43.88\pm0.05$		&
	Abs $\dag$
\nl
1416--129	&
	F	&
        $11.08\pm0.24$	&
	$0.33\pm0.01$	&
        $44.91\pm0.01$	&
\nl
1440+356	&
	F	&
        \phn$2.54\pm0.04$	&
	$0.91\pm0.02$	&
        $43.83\pm0.01$	&
	Var., SXS, Fe
\nl
1444+407	&
	A	&
	\phn$0.55\pm0.02$	&	
	$0.29\pm0.01$	&
	$44.31\pm0.01$	&	
	Fe (?)
\nl
1501+106(a)	&
	E	&
	$11.90\pm0.12$	& 
	$3.46\pm0.03$	& 
	$43.83\pm0.01$	&
	Var., SXS, Abs, Fe
\nl
1501+106(b)	&
	D	&
	$10.41\pm0.14$	& 
	$2.36\pm0.03$	&
	$43.77\pm0.01$	&
	SXS, Fe
\nl
1534+580	&
	E	&
	\phn$8.49\pm0.08$	& 
	$2.40\pm0.02$	& 
	$43.53\pm0.01$	&
	Var., Abs, Fe
\nl
1543+489	&
	B	&
	\phn$0.36\pm0.01$	&
	$0.31\pm0.01$	&
	$44.55\pm0.02$	&
\nl
1634+706	&
	A	&
	\phn$1.04\pm0.04$	&	
	$0.72\pm0.03$	&
	$46.06\pm0.01$	&
	$\dag$
\nl
1700+518	&
	$\ddag$	&
	$<0.08$	&
	$<0.03$		&
	$<43.51$		&
\nl    

\tablecomments{\it see over}
\enddata
\end{deluxetable}

\clearpage
Notes to Table~\ref{tab:fit_summary2} ---
Cols.(1--2): Dataset, and
preferred spectral model for the continuum 
	(see \S\ref{Sec:spect_all}).
For the two sources not detected ($\ddag$), upper limits are calculated 
assuming Model~A with $\Gamma=2.0$.
Col.(3): Flux in the {\it observed} 2--10~keV band in 
	units of $10^{-12}\ {\rm erg\ cm^{-2}\ s^{-1}}$.
Col.(4): Specific intensity, corrected for absorption,
	at 2~keV in the {\it quasar-frame}
	in units of $10^{-12}\ {\rm erg\ cm^{-2}\ s^{-1}\ keV^{-1}}$
	($\simeq 0.41\times10^{-6}$~Jy).
Col.(5): Luminosity, corrected for absorption, in the 2--10~keV band 
	({\it quasar-frame}) in units of ${\rm erg\ s^{-1}}$
	assuming $H_0 = 50\ {\rm km\ s^{-1}\ Mpc^{-1}}$ and 
	$q_0 = 0.5$.
Col.(6): Var. -- Variable source (Table~\ref{tab:xs_rms_128}); 
	SXS -- soft X-ray component;
	Abs -- intrinsic absorption, Fe -- Fe $K$-shell emission
	(Table~\ref{tab:feline}), $\dag$ -- see text.

\clearpage

\begin{deluxetable}{llllrcll}

\tablecaption{BEST-FITTING PARAMETERS OF RQQs FOR WHICH 
MODELS A, B or C IS PREFERRED 
\label{tab:prefer_abc}}

\tablecolumns{7}
\tablewidth{0pt}
\tablehead{
\colhead{Source(obs)} &
	\colhead{$N_{H,z}$}	&
	\colhead{$\log U_{oxygen}$}	&
	\colhead{$\Gamma$}	&
	\colhead{$N_{pts}$}	&
	\colhead{$\chi^2_{\nu}$}	&
        \colhead{$P(\chi^2)$} 
&
\nl
        \colhead{(1)}   &
        \colhead{(2)}   &
        \colhead{(3)}   &
        \colhead{(4)}   &
        \colhead{(5)}   &
        \colhead{(6)}   &
        \colhead{(7)}   
}
\startdata
0050+124	&
	\nodata	&
	\nodata	&
	$2.28^{+0.02}_{-0.02}$	&
	420	&
	0.95	&	0.26	
	\nl
0844+349	&
	\phn\phn$0.68^{+0.28}_{-0.28}$	&
	\nodata	&
	$2.02^{+0.07}_{-0.06}$	&
	457	&
	1.07	&	0.87	
	\nl
1114+445	&
	\phn$18.20^{+4.45}_{-3.54}$	&
	$-2.184^{+0.126}_{-0.275}$	&
	$1.87^{+0.16}_{-0.11}$	&
	439	&
	1.02	&	0.61	
	\nl
1148+549	&
	\nodata	&
	\nodata	&
	$1.84^{+0.11}_{-0.10}$	&
	114	&
	0.81	&	0.07	
	\nl
1216+069	&
	\nodata &
	\nodata &
	$1.92^{+0.04}_{-0.05}$		&
	219	&
	0.86	&	0.06	
	\nl
1247+267	&
	\nodata &
	\nodata &
	$1.95^{+0.08}_{-0.09}$	&
	114	&
	0.86	&	0.14	
	\nl
1322+659	&
	$758^{+241\ (p)}_{-522}$	&
	$-0.105^{+0.089}_{-0.273}$	&
	$2.10^{+0.12}_{-0.11}$	&
	214	&
	0.90	&	0.16	
	\nl
1407+265	&
	\nodata &
	\nodata &
	$2.01^{+0.03}_{-0.04}$	&
	272	&
	0.89	&	0.11	
	\nl
1411+442 $\ddag$	&
        $244^{+113}_{-83}$		&
	\nodata &
        $2.26^{+0.78}_{-0.74}$	&
	103	&
        1.16	&
        0.87	
\nl
1444+407	&
	\nodata &
	\nodata &
	$2.31^{+0.05}_{-0.05}$	&
	229	&
	1.03	&	0.65	
	\nl
1543+489	&
	\phn\phn$0.69^{+1.14}_{-0.69\ (p)}$	&
	\nodata &
	$2.64^{+0.19}_{-0.16}$	&
	203	&
	0.93	&	0.25	
	\nl
1634+706	&
	\nodata &
	\nodata &
	$1.97^{+0.04}_{-0.04}$	&
	240	&
	1.29	&	0.98	
	$\dag$
	\nl
\tablecomments{
Fits undertaken in the {\it observed} 0.6--10.0~keV band, but excluding 
	the 5--7~keV band in the {\it quasar-frame}.
	All models include
	Galactic absorption of column density $N_{H,0}^{gal}$ 
	as given in Table~\ref{tab:sample}.
	Errors are at 68\% confidence for the relevant number of 
	interesting parameters.
$\ddag$ The model for PG~1411+442 includes a fraction 
$D_f = 0.02^{+0.06}_{-0.01}$ of the continuum which does not 
suffer attenuation.
Col.(1): Dataset.
Col.(2): Column density of absorbing material intrinsic to the 
	quasar in units of $10^{21}\ {\rm cm^{-2}}$
	(Models~B \& C only).
Col.(3): Ionization parameter of the absorbing material as defined 
	in \S\ref{Sec:UX-definition}
	(Model~C only). 
Col.(4): Photon index of the power law continuum.
Cols.(5--7): Number of data points used in the spectral analysis ($N_pts$), 
	reduced-$\chi^2$ statistic ($\chi^2_{\nu}$) and
        probability ($P(\chi^2)$) that the $\chi^2$-statistic will be less
        than the observed value for the number of $dof$.
$\dag$ see text.}
\enddata
\end{deluxetable}

\clearpage

\addtolength{\oddsidemargin}{-2cm}
\addtolength{\evensidemargin}{-2cm}
\scriptsize
\begin{deluxetable}{lll lll rcc}

\tablecaption{BEST-FITTING PARAMETERS OF RQQs FOR WHICH 
MODELS D or E IS PREFERRED 
\label{tab:prefer_de}}

\tablecolumns{9}
\tablewidth{0pt}
\tablehead{
\colhead{Source(obs)} &
        \colhead{$N_{H,z}$} &
	\colhead{$\log U_{oxygen}$}	&
        \colhead{$\Gamma_s$} &
        \colhead{$\log R_{s/h}$} &
        \colhead{$\Gamma_h$} &
	\colhead{$N_{pts}$}	&
        \colhead{$\chi^2_{\nu}$} &
        \colhead{$P(\chi^2)$} 
\nl
        \colhead{(1)}   &
        \colhead{(2)}   &
        \colhead{(3)}   &
        \colhead{(4)}   &
        \colhead{(5)}   &
        \colhead{(6)}   &
        \colhead{(7)}   &
	\colhead{(8)}	&
	\colhead{(9)}	
}

\startdata
0003+199	&
        \phn$0.90^{+1.31}_{-0.90\ (p)}$     &
	\nodata	&
        $4.18^{+0.82\ (p)}_{-1.11}$        &
        $-0.01^{+0.44}_{-0.41}$   &
        $1.84^{+0.15}_{-0.16}$        &
	627	&
        0.97 	&	0.28	
	\nl
0804+761	&
        \phn$3.16^{+3.74}_{-1.94}$     &
	$-4.04^{+0.42}_{-1.40}$	&
        $3.00^{+1.71}_{-0.58}$        &
        $+0.90^{+1.62}_{-0.61}$   &
        $1.43^{+0.77}_{-0.72}$        &
	803	&
        1.04	&	0.78		
	\nl
1116+215	&
        \phn$2.34^{+4.58}_{-1.88}$     &
	$-3.20^{+1.24}_{-2.11\ (p)}$	&
        $2.36^{+1.57}_{-0.26}$        &
        $+2.30^{+0.62}_{-1.42}$   &
        $0.00^{+0.00\ (p)}_{-1.25}$        &
	354	&
        1.05	&	0.75		
	\nl
1244+026 $\ddag$	&
        \phn$1.01^{+2.19}_{-0.99}$     &
	\nodata	&
        $3.24^{+1.76\ (p)}_{-0.61}$        &
        $+0.95^{+2.10}_{-5.00\ (p)}$   &
        $1.60^{+0.89}_{-1.60\ (p)}$        &
	504	&
        1.08	&	0.89		
	\nl
1501+106(a) 	&
        \phn$1.70^{+3.87}_{-0.93}$     &
	$-3.07^{+1.07}_{-2.24}$	&
        $3.61^{+1.49\ (p)}_{-2.11\ (p)}$        &
        $-0.51^{+1.41}_{-1.07}$   &
        $1.80^{+0.10}_{-0.14}$        &
	742	&
        1.02	& 0.78		
	\nl
1501+106(b)	&
        \phn$0.30^{+0.16}_{-0.30\ (p)}$     &
	\nodata	&
        $5.00^{+0.00\ (p)}_{-1.58}$        &
        $-0.88^{+0.55}_{-0.24}$   &
        $1.59^{+0.06}_{-0.05}$        &
	421	&
        0.93	&	0.16	
	\nl
1534+580	&
        \phn$5.50^{+1.98}_{-2.64}$     &
	$-2.15^{+0.26}_{-1.78}$	&
        $2.39^{+2.61\ (p)}_{-0.89\ (p)}$        &
        $+0.10^{+4.90\ (p)}_{-2.66}$   &
        $1.57^{+0.93\ (p)}_{-1.26}$        &
	780	&
        1.07	& 0.90 		
	\nl
\tablecomments{\normalsize
Symbols and units as for Table~\ref{tab:prefer_abc}, except
Cols.(4--6) contain the best-fitting values for 
	the 
        photon indices for the 'soft' and 'hard' power laws
        ($\Gamma_s$ \& $\Gamma_h$) and $R_{s/h}$, 
        the ratio  of the 
normalizations of the soft
        to hard power laws at 1~keV (in the quasar-frame).
$(p)$ indicates the 
parameter 'pegged' 
at the specified value.
$\ddag$ The model for PG~1244+026 includes a Gaussian 
emission line 
with $E_z = 0.96^{+0.07}_{-0.31}$~keV,
$\sigma_z = 0.09^{+0.16}_{-0.09}$~keV,
and
$L_{line} \simeq 3\times10^{42}\ {\rm erg\ s^{-1}}$ 
(see text).
}
\enddata
\end{deluxetable}
\addtolength{\oddsidemargin}{+2cm}
\addtolength{\evensidemargin}{+2cm}

\clearpage

\begin{deluxetable}{llllll r ccl}

\tablecaption{BEST-FITTING PARAMETERS OF RQQs FOR WHICH 
MODEL~F IS PREFERRED
\label{tab:prefer_f}}

\small
\tablecolumns{12}
\tablewidth{0pt}
\tablehead{
\colhead{Source(obs)} &
        \colhead{$N_{H,z}$} &
        \colhead{$\Gamma$} &
        \colhead{$E_z$} &
        \colhead{$\sigma_z$} &
        \colhead{$\log L_{line}$} &
	\colhead{$N_{pts}$}	&
        \colhead{$\chi^2_{\nu}$} &
        \colhead{$P(\chi^2)$} 
\nl
        \colhead{(1)}   &
        \colhead{(2)}   &
        \colhead{(3)}   &
        \colhead{(4)}   &
        \colhead{(5)}   &
        \colhead{(6)}   &
        \colhead{(7)}   &
	\colhead{(8)}	&
	\colhead{(9)}	
}

\startdata
0921+525	&
        $0.10^{+0.30}_{-0.10\ (p)}$	&
        $1.73^{+0.08}_{-0.08}$	&
        $0.36^{+0.87}_{-0.06\ (p)}$	&
        $0.76^{+0.13}_{-0.43}$	&
        $43.04^{+0.16}_{-0.94}$	&
	914	&
        1.04	&
        0.80	
\nl
0953+415	&
	$0.01^{+1.09}_{-0.01\ (p)}$        &
	$1.98^{+0.17}_{-0.15}$        &
	$0.71^{+0.80}_{-0.28}$        &
	$0.60^{+0.21}_{-0.22}$        &
	$44.10^{+0.37}_{-0.54}$        &
	403	&
	0.97        &
	0.36        
\nl
1211+143 $\ddag$	&
	$0.00^{+0.28}_{-0.00\ (p)}$	&
	$2.16^{+0.04}_{-0.08}$	&
	$0.38^{+0.13}_{-0.01}$	&
	$0.24^{+0.01}_{-0.05}$	&
	$43.99^{+0.13}_{-0.25}$	&
	480	&
	0.93	&
	0.15	
\nl
1404+026 $\ddag$	&
        $0.00^{+0.49}_{-0.00\ (p)}$	&
        $1.82^{+0.29}_{-0.26}$		&
        $0.39^{+0.10}_{-0.01}$		&
        $0.28^{+0.02}_{-0.04}$		&
        $43.73^{+0.13}_{-0.17}$	&
	166	&
        1.01	&
        0.53	
\nl
1416--129	&
        $0.12^{+0.58}_{-0.12\ (p)}$	&
        $1.73^{+0.11}_{-0.09}$		&
        $0.40^{+0.95}_{-0.10\ (p)}$	&
        $0.81^{+0.17}_{-0.54}$	&
        $43.83^{+0.19}_{-1.20}$	&
	867	&
 	0.98    &
        0.37	
\nl
1440+356 $\ddag$	&
        $0.54^{+1.53}_{-0.54\ (p)}$	&
        $2.06^{+0.17}_{-0.11}$		&
        $0.43^{+0.71}_{-0.06}$		&
        $0.22^{+0.05}_{-0.13}$		&
        $43.67^{+0.33}_{-1.03}$	&
	384	&
        0.92	&
        0.13	
\nl
\tablecomments{\normalsize
Symbols and units as for Table~\ref{tab:prefer_abc}, 
except
cols.(4--6) which contain the energy, width and equivalent width
of the Gaussian component (all in keV).
$\ddag$ In the case of PG~1211+143, PG~1404+026 and PG1440+356 the
Gaussian represents a {\it parameterization} of a 
'soft-excess'. The Gaussian parameterizes only 
subtle curvature of the underlying continuum in 
the other cases.
}
\enddata
\end{deluxetable}

\clearpage
\normalsize

\begin{deluxetable}{lrrrllll}

\tablecaption{FE EMISSION LINE PARAMETERS
\label{tab:feline}}

\tablecolumns{8}
\tablewidth{0pt}
\tablehead{
\colhead{Source(obs)} &
        \colhead{$\Delta N_{pts}$} &
        \colhead{$\frac{\Delta \chi^2}{\Delta N_{pts}}$} &
        \colhead{$F$} &
        \colhead{$E_z$} &
        \colhead{$\sigma_z$} &
        \colhead{$EW$(Fe-{\it K})} &
        \colhead{$\log L$(Fe-{\it K})} 
\nl
        &
        &
        &
        &
        \colhead{(keV)}   &
        \colhead{(keV)}   &
	\colhead{(keV)}	&
        \colhead{(${\rm erg\ s^{-1}}$)}   
\nl
        \colhead{(1)}   &
        \colhead{(2)}   &
        \colhead{(3)}   &
        \colhead{(4)}   &
        \colhead{(5)}   &
        \colhead{(6)}   &
        \colhead{(7)}   &
	\colhead{(8)}	
}

\startdata
0003+199	&  
	73 &  0.93 &  
	3.90	& $6.30^{+0.11}_{-0.25}$	&
		$< 1.20$	& 
		$0.13^{+0.08}_{-0.08}$	&
		$41.52^{+0.21}_{-0.42}$	
\nl
0050+124 $\ddag$	&  
	38 &  1.47 &  
	4.86	& $7.16^{+0.77}_{-0.56}$	&
		$0.63^{+0.53}_{-0.33}$	& 
		$0.88^{+0.60}_{-0.47}$	&
		$42.57^{+0.23}_{-0.33}$	
\nl
0804+761	&  
	137 &  0.93 &
	1.32	& \nodata & \nodata & 
		$<0.33$	& 
		$<43.14$	
\nl
0844+349	&  
	58 &  1.30 &
	15.12	& $6.22^{+0.23}_{-0.36}$ & 
		$<1.96$	 & 
		$0.35^{+0.24}_{-0.17}$	& 
		$42.21^{+0.23}_{-0.29}$	
\nl
0921+525	& 
	172 & 0.97 & 
	7.33	& $6.22^{+0.38}_{-0.39}$	&
		$0.76^{+0.30}_{-0.55}$	& 
		$0.19^{+0.10}_{-0.09}$	&
		$42.48^{+0.19}_{-0.30}$	
\nl
0953+415	&  
	54 &  0.91 &  
	1.40   & \nodata & \nodata   &
                $<0.06$	& 
		$<42.69$	
\nl
1114+445	&  
	86 &  1.02 &  
	5.61	& $6.33^{+0.18}_{-0.26}$	&
		$< 0.85$	& 
		$0.27^{+0.16}_{-0.13}$	&
		$42.80^{+0.20}_{-0.29}$ 	
\nl
1116+215	&  
	43 &  1.02 &  
	2.30 & \nodata & \nodata   & 
		$<0.59$ 	&
		$<43.58$ 
\nl
1148+549 	&  
	21 &  0.68 &  
	0.87 & \nodata & \nodata   &
                $<1.06$	&
		$< 44.32$	
\nl
1211+143 	&  
	50 &  1.66 &  
	7.74	& $6.49^{+0.68}_{-0.34}$	& 
		$0.52^{+1.16}_{-0.28}$		& 
                $0.55^{+0.44}_{-0.25} $	&
		$42.69^{+0.25}_{-0.25}$	
\nl
1216+069	&  
	31 &  1.38 &  
	1.48    & \nodata & \nodata   &
                $<2.05$	&
		$<44.26$	
\nl
1244+026	&  
	45 &  1.00 &  
	2.88 & \nodata 	&
		\nodata & 
		$< 1.22$	&
                $< 42.41$	
\nl
1247+267 	&  
	31 &  0.71 &  
	0.40 & \nodata & \nodata   &
                $<1.21$	&
		$<45.18$	
\nl
1322+659	&  
	22 &  1.36 &  
	0.01	& \nodata & \nodata   &
                $<0.62$	&
		$<42.91$	
\nl
1404+226	&  
	21 &  0.90 &  
	1.79 & \nodata & \nodata   &
                $<4.15$	&
		$<42.95$	
\nl
1407+265 	&  
	44 &  1.00 &  
	1.49 & \nodata & \nodata   &
                $<0.29$	&
		$<44.50$	
\nl
1411+442	&  
	26 &  1.04 &  
	1.23 & \nodata & \nodata   &
                $<1.86$ &
		$<42.80$	
\nl
1416--129 	& 
	145 & 1.17 & 
	1.67 & \nodata & \nodata   &
                $<0.43$ &
		$< 43.60$	
\nl
1440+356	&  
	43 &  1.53 &  
	4.25 & $5.95^{+0.88}_{-0.57}$ 	&
		\nodata	& 
		$0.44^{+0.36}_{-0.33}$	&
		$42.50^{+0.26}_{-0.61}$	
\nl
1444+407 $\ddag$	&  
	28 &  1.32 &  
	3.27 & $6.45^{+1.55\ (p)}_{-1.47}$	&
		\nodata	&
		$1.39^{+1.50}_{-1.01}$	&
		$43.44^{+0.32}_{-0.57}$	
\nl
1501+106(a) 	&  
	114 &  1.18 &  
	11.00 & $6.46^{+0.09}_{-0.63}$	&
		$0.13^{+0.70}_{-0.09}$	&
		$0.25^{+0.27}_{-0.09}$	&
		$42.24^{+0.32}_{-0.19}$	
\nl
1501+106(b) 	&  
	69 &  1.04 &  
	3.70	& $6.38^{+0.17}_{-0.23}$	&
		$< 1.03$	& 
		$0.14^{+0.10}_{-0.09}$ &
		$41.96^{+0.23}_{-0.44}$	
\nl
1534+580	&  
	134 &  1.12 &  
	17.16 	& $6.34^{+0.16}_{-0.19}$	&
		$0.41^{+0.21}_{-0.15}$	& 
		$0.38^{+0.12}_{-0.06}$	&
		$42.14^{+0.12}_{-0.07}$	
\nl
1543+489	&  
	25 &  1.08 &  
	1.83 & \nodata & \nodata   &
                $<1.30$ &
		$< 43.74$	
\nl
1634+706 	&  
	46 &  0.63 &  
	1.16 & \nodata & \nodata   &
                $<0.75$ &
		$< 45.25$	
\nl
\tablecomments{
Col.(1): Dataset.
Col.(2,3): Number of additional spectral bins ($\Delta N_{pts}$) and
fractional increase in the $\chi^2$-statistic per additional bin 
($\Delta\chi^2$/$\Delta N_{pts}$) when the preferred model from
\S\ref{Sec:0.6-10-observed} is applied to the data in the 5--7~keV
band (quasar-frame).
Col.(4): $F$-statistic when a Gaussian emission component is added to
	the spectral analysis.
Cols.(5--8): Energy ($E_z$), width ($\sigma_z$), equivalent width 
	($EW$), and luminosity of the Gaussian component. 
$\ddag$ Line parameters considered dubious (see text).}
\enddata
\end{deluxetable}

\clearpage


\begin{deluxetable}{lccc}

\tablecaption{CONSTRAINTS ON O{\sc vii} AND O{\sc viii} EDGES
\label{tab:2edges}}

\tablecolumns{4}
\tablewidth{0pt}
\tablehead{
\colhead{Source(obs)} &
        \colhead{S0 S/N} &
        \colhead{$\tau$(O{\sc vii})} &
        \colhead{$\tau$(O{\sc viii})}
\nl
        \colhead{(1)}   &
        \colhead{(2)}   &
        \colhead{(3)}   &
        \colhead{(4)}  
}

\startdata
0003+199	&  
	45 &
	$0.16^{+0.12}_{-0.12}$	&
	$<0.07$
\nl
0050+124 	&  
	30	&
	$< 0.29$	&
	$< 0.06$
\nl
0804+761	&  
	52	&
	$0.29^{+0.13}_{-0.11}$	&
	$< 0.03$
\nl
0844+349	&  
	24	&
	$< 0.32$	&
	$< 0.11$
\nl
0921+525	& 
	54	&
	$< 0.09$	&
	$< 0.10$	
\nl
0953+415	&  
	20	&
	$< 0.41$	&
	$<0.29$	
\nl
1114+445	&  
	\phn5	&
	$2.56^{+1.18}_{-0.65}$	&
	$1.39^{+0.48}_{-0.76}$
\nl
1116+215	&  
	18	&
	$0.44^{+0.27}_{-0.26}$	&
	$< 0.26$
\nl
1211+143 	&  
	30	&
	$< 0.10$	&
	$< 0.15$
\nl
1216+069	&  
	\phn6	&
	$< 0.27$	&
	$< 0.14$
\nl
1244+026	&  
	44	&
	$< 0.24$	&
	$< 0.09$
\nl
1322+659	&  
	\phn9	&
	$< 0.53$	&
	$< 0.24$
	
\nl
1404+226	&  
	21	&
	$< 0.93$	&
	$0.40^{+0.60}_{-0.30}$
\nl
1411+442	&  
	\phn3	&
	$< 1.14$	&
	$< 1.65$
\nl
1416--129 	& 
	15	&
	$< 0.32$	&
	$< 0.11$
\nl
1440+356	&  
	29	&
	$< 0.82$	&
	$< 0.25$
\nl
1444+407 	&  
	10	&
	$< 0.37$	&
	$< 0.19$
\nl
1501+106(a) 	&  
	40	&
	$0.28^{+0.11}_{-0.12}$	&
	$< 0.14$
\nl
1501+106(b) 	&  
	28	&
	$< 0.28$	&
	$< 0.14$
\nl
1534+580	&  
	36	&
	$0.48^{+0.16}_{-0.15}$	&
	$0.31^{+0.09}_{-0.13}$	
\nl
1543+489	&  
	\phn7	&
	$< 0.63$	&
	$< 0.47$
\nl
\tablecomments{
Col.(1): Dataset (sources with $z > 0.45$ are excluded).
Col.(2): The signal-to-noise ratio of the SIS0 data between 0.6~keV
         (observed frame) and 1~keV (quasar frame).
Cols.(3,4): Constraints on the optical depth of the two edges. 
}
\enddata
\end{deluxetable}

\clearpage

\begin{deluxetable}{lrrclll}

\tablecaption{SERENDIPTOUS SOURCES WITHIN THE FIELDS OF VIEW
\label{tab:asca_serendip}}

\tablecolumns{7}
\tablewidth{0pt}
\tablehead{
\colhead{PG Field} &
	\colhead{RA} 	&
	\colhead{dec} 	&
	\colhead{GIS2 Rate} 	&
	\multicolumn{3}{c}{Possible Identification} 	
\nl
	&
        \colhead{J2000}   &
        \colhead{J2000}   &
	\colhead{($10^{-2}\ {\rm ct\ s^{-1}}$)}&
	\colhead{Name}	&
	\colhead{Class}	&
	\colhead{Refs}	
\nl
        \colhead{(1)}   &
        \colhead{(2)}   &
        \colhead{(3)}   &
        \colhead{(4)}   &
        \colhead{(5)}   &
        \colhead{(6)}	&
        \colhead{(7)}	
}
\startdata
0804+761	&
        08~12~48.8	&
        +76~06~12	&
	$1.35\pm0.07$	&
	1RXS J081241.7+760615	&
	\nodata	&
	\nodata	        
\nl
1216+069	&
        12~20~16.3	&
        +06~41~20	&
	$1.42\pm0.09$	&
	1RXS~J122019.0+064126 	&
	\nodata	&
	\nodata	        
\nl
1216+069	&
        12~19~28.1	&
        +06~42~59	&
	$0.99\pm0.08$	&
	2E~1216.9+0700 	&
	AGN	&
	St91	        
\nl
1247+267	&
        12~49~46.7	&
        +26~29~07	&
	$0.34\pm0.04$	&
	1RXP~J124946+262	&
	\nodata	&
	\nodata	        
\nl
1404+226	&
        14~05~29.0	&
        +22~23~50	&
	$0.71\pm0.06$	&
	1RXP~J140528+2223.4	&
	\nodata	&
	\nodata	        
\nl
1440+356	&
       14~41~07.4	&
        +35~20~29	&
	$0.47\pm0.05$		&
	1WGA~J1441.1+3519	&
	\nodata	&
	\nodata	        
\nl
1501+106		&
        15~03~36.9	&
        +10~16~44	&
	$0.69\pm0.08$		&
	1RXP~J150340+1016.3	&
	\nodata	&
	\nodata	        
\nl
1501+106		&
        15~04~25.6	&
        +10~29~33	&
	$0.16\pm0.04$		&
	PKS~1502+106	&
	AGN	&
	Ge94	        
\nl
\tablecomments{
Col.(1--3): Target field and position of serendipitous 
source as determined from the GIS data after correcting for 
any offset (typically $\lesssim 1$~arcmin) between the target 
source and catalogued position listed in Table~\ref{tab:sample}.
Cols.(4): GIS2 count rates in the 1--10~keV (observer's frame) band.
The extraction cells varied, but were typically circular 
of radius 1.5--2.5~arcmin. 
Cols.(5--6): Tentative identification from a cross-correlation with
previously detected X-ray sources and (if known) 
the class of the object.
Col.(8): References discussing X-ray observations of the 
serendipitous source: St91 - Stocke et al (1991), 
Ge94 - George et al (1994).
Two additional sources were detected:
tentatively identified as the cluster of galaxies ACO~1885
(in the PG~1411+442 field) and the AGN 1E~0803.3+7557
(in the PG~0804+761 field). 
However both objects lie 
close to the edge of the GIS field--of--view and an accurate 
position or count rate are not possible using 
these {\it ASCA}\ data.}
\enddata
\end{deluxetable}
\clearpage


\begin{references}

\reference{Ar96}	Arnaud, K.A., 
		1996, in ASP Conf. Proc. 101, 
		{\it Astronomical Data Analysis Software and Systems V}, 
		eds. G. Jacoby, J.Barnes (San Fransico: ASP), p17
\reference{Ba86}	Barr, P., Mushotzky, R.F., 
		1986, Nature, 320, 421
\reference{Bo96}	Boller, T., Brandt, W.N., Fink, H.H., 
		1996, A\&A, 305, 53
\reference{Bo92} 	Boroson, T.A., Green, R.F., 
		1992, \apjs, 80, 109
\reference{Br99}	Brinkmann, W., Wang, T., Matsuoka, M., Yuan, W. 
		1999, A\&A, 345, 43
\reference{Ca89}	Canizares, C.R., White, J.L., 
		1989, \apj, 339, 27
\reference{Ca87}   	Carleton, N.P., Elvis, M., Fabbiano, G., Willner, S.P.,
			Lawrence, A., Ward, M.,
		1987, \apj, 318, 595 
\reference{Co 92}	Comastri, A., Setti, G., Zamorani, G., Elvis, M., 
 			Wilkes, B.J., McDowell, J.C., Giommi, P.,
		1992, \apj, 384, 62
\reference{El89}	Elvis, M., Lockman, F.J., Wilkes, B.J., 
		1989, \aj, 97, 777 
\reference{FU96}	Fabian, D., Usher, P.D.,
		1996, \aj, 111, 645
\reference{Fi98a}	Fiore, F., et al., 
		1998a, \mnras, 298, 103
\reference{Fi98b}	Fiore, F., Laor, A., Elvis, M., Nicastro, F., 
			Giallongo, E., 
		1998b, \apj, 503, 607
\reference{Ga99}	Gallagher, S.C., Brandt, W.N., Sambruna, R.M., 
			Mathur, S., Yamasaki, N., 
		1999, \apj, 519, 549
\reference{fgf}		George, I.M., Fabian, A.C., 
		1991, \mnras, 249, 352
\reference{Ge94} 	George, I.M., Nandra, K., Turner, T.J., Celotti, A., 
		1994, \apj, 436, L59
\reference{Ge97} 	George, I.M., Nandra, K., Laor, A., Turner, T.J.,
			Netzer, H., Mushotzky, R.F., 
		1997, \apj, 491, 508
\reference{Ge98a}	George, I.M., Turner, T.J., Netzer, H., Nandra, K., 
			Mushotzky, R.F., Yaqoob, T., 
		1998a, \apjs, 114, 73
\reference{Ge98b}	George, I.M., Turner, T.J., Mushotzky, R.F., Nandra, K.,
			Netzer, H.,
		1998b, \apj, 503, 174
\reference{Ge98c}       George, I.M.,  Mushotzky, R.F., Turner, T.J., 
			Yaqoob, T., Ptak, A., Nandra, K., Netzer, H.,
     		1998c, \apj, 509, 146
\reference{Ge99}       George, I.M.,  
			Netzer, H., Laor, A., Turner, T.J., 
			Yaqoob, T., Mushotzky, R.F., Nandra, K., 
			Takahashi, T., 
     		1999, in preparation (Paper~II)
\reference{Go97}	Gotthelf, E.V., Ishibashi, K., 
		1997, in 
		{\it X-Ray Imaging and Spectroscopy of Cosmic Hot Plasmas}, 
		eds. F.Makino, K.Misuda 
		(Tokyo: Universal Academy Press), p631
\reference{Gr93}	Green, A.R., McHardy, I.M., Lehto, H.J., 
		1993, \mnras, 265, 664
\reference{Gr96}	Green, P.J., Mathur, S., 
		1996, \apj, 462, 637
\reference{Gr86}	Green, R.F., Schmidt, M., Liebert, J.,
		1986, \apjs, 61, 305
\reference{Iw93}	Iwasawa, K., Taniguchi, Y., 
		1993, \apj, 370, L61
\reference{Ke89}	Kellermann, K.I., Saramek, R., Schmidt, M., 
			Shaffer, D.B., Green, R., 
		1989, \aj, 98, 1195
\reference{Ke94}	Kellermann, K.I., Saramek, R., Schmidt, M., 
			Green, R.F., Shaffer, D.B., 
		1994, \aj, 108, 1163
\reference{Ko97}	 K\"{o}nig, M., Staubert, R., Wilms, J.,
		1997, A\&A, 326, L25
\reference{La97}		Laor, A., Fiore, F., Elvis, M., Wilkes, B.J.,
			 McDowell, J.C., 
		1997, \apj, 477, 93
\reference{La93}	Lawrence, A., Papadakis, I., 
		1993, \apj, 414, L85
\reference{La97}        Lawson, A.J., Turner, M.J.L., 
                1997, \mnras, 288, 920
\reference{La92}	Lawson, A.J., Turner, M.J.L., Williams, O.R., 
			Stewart, G.C., Saxton, R.D.,
		1992, \mnras, 259, 743
\reference{Le99}	Leighly, K., 
		1999, \apj, in press
\reference{Le97} 	Leighly, K.M., Mushotzky, R.F., Nandra, K., 
		Forster, K., 
		1997, \apj, 489, L25
\reference{Lo95}	Lockman, F.J.,Savarge, B.D., 
		1995, \apjs, 97, 1
\reference{Ma88}	Maccacaro, T., et al., 
		1988, \apj, 326, 680
\reference{Ma96}	Makishima, K., et al., 
		1996, \pasj, 48, 171
\reference{M+M83}	Morrison, R., McCammon, D., 
		1983, \apj, 270, 119
\reference{Mu96}	Murphy, E.M., Lockman, F.J., Laor, A., Elvis, M., 
		1996, \apjs, 105, 369
\reference{Na96}	Nandra, K., George, I.M., Turner, T.J., Fukazawa, Y. 
       		1996, ApJ, 464, 165 
\reference{Na97a} Nandra, K., George, I.M., Mushotzky, R.F., Turner, T.J.,
                 Yaqoob, T.,
                   1997a, \apj, 476, 70
\reference{Na97b} Nandra, K., George, I.M., Mushotzky, R.F., Turner, T.J.,
                 Yaqoob, T.,
                   1997b, \apj, 477, 602
\reference{Na97c} Nandra, K., George, I.M., Mushotzky, R.F., Turner, T.J.,
                 Yaqoob, T.,
                   1997c, \apj, 488, L91
\reference{HN93}	Netzer, H., 
		1993, \apj, 411, 594
\reference{HN96}	Netzer, H., 
		1996, \apj, 473, 781
\reference{Ne87}	Neugebauer, G., Green, R.F., Matthews, K., Schmidt, M.,
			Soifer, B.T., Bennett, J., 
		1987, \apjs,63, 615
\reference{Orr98}	Orr, A., Yaqoob, T., Parmar, A.N., Piro, L., 
			White, N.E., Grandi, P., 
		1998, A\&A, 337, 685
\reference{Ree97}	Reeves, J.N., Turner, M.J.L., Ohashi, T., 
			Kii, T., 
		1997, \mnras, 292, 468
\reference{Re97}	Reynolds, C.S., 
		1997, \mnras, 286, 513
\reference{Sa99}	Sambruna, R.M., Eracleous, M., Mushotzky, R.F., 
		1999, \apj, in press
\reference{Sa89}	Sanders, D.B.,  Phinney, E.S., Neugebauer, G.,
			Soifer, B.T., Matthews, K.,   
		1989, \apj, 347, 29
\reference{SG83}	Schmidt, M., Green, R.F., 
		1983, \apj, 269
\reference{St91}	Stocke et al (1991) \apjs, 76, 813
\reference{Tan94}	Tanaka, Y., Inoue, H., Holt, S.S., 
		1994, PASJ, 46, L37
\reference{Tan95}	Tanaka, Y., et al., 
		1995, Nature, 375, 659
\reference{Tu96}	Turner, T.J. , George, I.M., Kallman, T., Yaqoob, T., 
			Zycki, P.T.,
       		1996, ApJ, 472, 571. 
\reference{Tu98}	Turner, T.J., George, I.M., Nandra, K.,
		1998, \apj, 508, 648
\reference{Tu99a}	Turner, T.J., Nandra, K., George, I.M., Turcan, D., 
		1999a, \apj, in press.
\reference{Tu99b}	Turner, T.J., George, I.M., Netzer, H.,
		1999b, \apj, in press
\reference{Turn}	Turnshek, D.A., Monier, E.M., Sirola, C.J., 
			Espey, B.R. 
   		1997, \apj, 476, 40
\reference{Ul99}	Ulrich, M.-H., Comastri, A., Komossa, S., 
			Crane, P., 
		1999, A\&A, in press
\reference{Si99}	Vaughan, S., Reeves. J., Warwick, R.S., Edelson, R., 
		1999, \mnras, in press
\reference{Vi99}	Vignali, C., Comastri, A., Cappi, M., Palumbo, G.G.C., 
			Matsuoka, M., Kubo, H., 
		1999, \apj, 516, 582
\reference{wi92}	Williams, O.R., et al., 
		1992, \apj, 389, 157
\reference{wi87}	Wilkes, B.J., Elvis, M.,
		1987, \apj, 323, 243
\reference{Ya94}	Yaqoob, T., Serlemitsos, P.J., Mushotzky, R.F., 
			Madejski, G.M., Turner, T.J., Kunieda, H.,
		1994, PASJ, 46. 173
\reference{ze64}  	Zel'dovich, Ya.B., Noovikov, I.D., 
		1964, Sov. Phys. Dokl., 158, 811
\end{references}
\end{document}